\documentclass{emulateapj}
\usepackage{multirow,lscape}
\begin{document}
\newcommand{\Oii}{[\textsc{O ii}]}
\newcommand{\Oiii}{[\textsc{O iii}]}
\newcommand{\Ha}{H$\alpha$}

\newcommand{\Hb}{H$\beta$}
\newcommand{\Lya}{Ly$\alpha$}
\newcommand{\Rc}{R_{\rm C}}
\submitted{2006 March 28}
\accepted{2006 October 25}
\title{The Luminosity Function and Star Formation Rate between Redshifts of 0.07 and 1.47 for Narrow-band Emitters
  in the Subaru Deep Field\altaffilmark{1}}
\author{Chun Ly,\altaffilmark{2} Matt A. Malkan,\altaffilmark{2} Nobunari Kashikawa,\altaffilmark{3,4}
  Kazuhiro Shimasaku,\altaffilmark{5,6} Mamoru Doi,\altaffilmark{7} Tohru Nagao,\altaffilmark{3,8}
  Masanori Iye,\altaffilmark{3,4} Tadayuki Kodama,\altaffilmark{3,4} Tomoki Morokuma,\altaffilmark{7} and
  Kentaro Motohara\altaffilmark{7}}
\shorttitle{LF and SFR for NB Emitters in the SDF}
\shortauthors{Ly et al.}

\email{chun@astro.ucla.edu}
\altaffiltext{1}{Based in part on data collected at Subaru Telescope, which is operated by the National Astronomical
  Observatory of Japan.}
\altaffiltext{2}{Department of Astronomy, University of California at Los Angeles, Los Angeles, CA 90095-1547.}
\altaffiltext{3}{Optical and Infrared Astronomy Division, National Astronomical Observatory, Mitaka, Tokyo
  181-8588, Japan.}
\altaffiltext{4}{Department of Astronomy, School of Science, Graduate University for Advanced Studies, Mitaka,
  Tokyo 181-8588, Japan.}
\altaffiltext{5}{Department of Astronomy, School of Science, The University of Tokyo, Bunkyo, Tokyo 113-0033, Japan.}
\altaffiltext{6}{Research Center for the Early Universe, School of Science, The University of Tokyo, Tokyo 113-0033, Japan.}
\altaffiltext{7}{Institute of Astronomy, University of Tokyo, Mitaka, Tokyo 181-8588, Japan.}
\altaffiltext{8}{INAF-Osservatorio Astrofisico di Arcetri, Largo E. Fermi 5, 50125 Florence, Italy.}

\begin{abstract}
  Subaru Deep Field line-emitting galaxies in four narrow-band filters (NB704, NB711, NB816, and NB921) at low and
  intermediate redshifts are presented. Broad-band colors, follow-up optical spectroscopy, and multiple narrow-band filters
  are used to distinguish \Ha, \Oii, and \Oiii~emitters between redshifts of 0.07 and 1.47 to construct their averaged
  rest-frame optical-to-UV spectral energy distributions and luminosity functions. These luminosity functions are derived
  down to faint magnitudes, which allows for a more accurate determination of the faint end slope. With a large (N $\sim$
  200 to 900) sample for each redshift interval, a Schechter profile is fitted to each luminosity function. Prior to dust
  extinction corrections, the \Oiii~and \Oii~luminosity functions reported in this paper agree reasonably well with those
  of Hippelein et al. The $z=0.08$ \Ha~LF, which reaches two orders of magnitude fainter than Gallego et al., is steeper
  by 25\%. This indicates that there are more low luminosity star-forming galaxies for $z<0.1$. The faint end slope $\alpha$
  and $\phi_{\star}$ show a strong evolution with redshift while $L_{\star}$ show little evolution. The evolution in $\alpha$
  indicates that low-luminosity galaxies have a stronger evolution compared to brighter ones. These results can only be
  achieved with deep NB observations over a wide range in redshift. Integrated star formation rate densities are derived
  via \Ha~for $0.07<z<0.40$, \Oiii~for $0.40<z<0.84$, and \Oii~for $0.89<z<1.47$. A steep increase in the star-formation
  rate density, as a function of redshift, is seen for $0.4\lesssim z\lesssim0.9$. For $z\gtrsim1$, the star-formation rate
  densities are more or less constant. The latter is consistent with previous UV and \Oii~measurements. Below $z\lesssim0.4$,
  the SFR densities are consistent with several \Ha, \Oii, and UV measurements, but other measurements are a factor of
  two higher. For example, the $z=0.066-0.092$ \Ha~LF agrees with those of Jones \& Bland-Hawthorn, but at $z=0.24$
  and 0.40, their number density is higher by a factor of two. This discrepancy can be explained by cosmic variance.
\end{abstract}

\keywords{
  galaxies: photometry ---
  galaxies: emission lines ---
  galaxies: distances and redshifts ---
  galaxies: luminosity function ---
  galaxies: evolution
}

\section{INTRODUCTION}
\label{1}
Over the past decade, deep spectroscopic surveys have utilized emission lines to measure the cosmic star-formation rate
(SFR). Estimates of the SFR can be obtained from the \Ha~emission line in star-forming galaxies \citep{kennicutt83}.
However, \Ha~is no longer visible (in optical spectrographs) beyond $z\,\sim\,0.4$. To study the SFR at higher redshifts,
one must obtain infrared spectroscopy of the \Ha~line or detect bluer emission lines where optical spectrographs are used.
Although the former has been successful
\citep[e.g.,][]{MTM95,glazebrook99,M99,yan99,hopkins00,moorwood00,tresse02,doherty06,rodriguez06}, difficulties, such as a
bright background (for ground-based observations) and smaller areal coverage limit IR searches to small samples, mostly of
the brightest galaxies.

The latter has been attempted by measuring the \textsc{O\,ii} doublet (\Oii~$\lambda\lambda$3726, 3729). It has been used
to determine the SFR out to $z$ = 1.6 \citep{hogg98,hicks02,hippelein03,teplitz03,drozdovsky05}, but its measurements are
more affected by internal extinction and metallicity uncertainties \citep{kennicutt92,kewley04}. Studies have shown that
the comoving SFR density increases by a factor of 10 from $z\sim0$ to 1-1.5 and declines or flattens out at higher redshifts
\citep{hopkins04}. The behavior above a redshift of 3 is not well known for two reasons: ($i$) the amount of UV extinction
is questionable, and ($ii$) the shallowness of recent Lyman Break Galaxies studies at $z>5$ has resulted in an extrapolation
of the faint-end slope for a SFR estimate.

Since past studies identified galaxies and redshifts via spectra, the measured SFRs are biased toward the small selected
sample of bright objects, and spectroscopy requires a greater demand of allocated telescope time, as opposed to the
approach of using deep narrow-band (NB) imaging with large fields-of-view. The NB imaging method has proven to be quite
effective in finding many emission line galaxies with the appropriate redshift for a strong line (e.g., \Lya, \Ha,
\Oii~$\lambda$3727, \Hb, and \Oiii~$\lambda$5007) to fall within the NB filter. For example, \citet{hu02},
\citet{ajiki03}, \citet{kodaira03}, \citet{taniguchi03}, \citet{hu04}, \citet{kashik06}, and \citet{shima06} have
confirmed candidate \Lya~emitters (LAEs) at $z=5.7$ and 6.6 with follow-up spectroscopy. These NB emitters (identified
when their NB magnitude is substantially brighter than that of the broad-band continuum) provide an opportunity to study
the cosmic evolution of star formation. \citet{fujita03}, \citet{kodama04}, and \citet{umeda04} have measured the
\Ha~luminosity function (the latter two are for clusters) at $z$ = 0.24 or 0.40 by identifying NB emitters and then
using their broad-band colors to distinguish a few hundred \Ha~emitters from other line emitters. \citet{ajiki06}
also examined the same field as \citet{fujita03} for other strong emission line galaxies such as \Oiii~and \Oii.
Fabry-Perot (FP) interferometers have also been used to find emission line galaxies \citep{jones01,hippelein03,glazebrook04},
but the comoving volume or limiting flux of past surveys is not comparable to that of the NB imaging technique, and
their surveys currently lack broad-band colors. Other work, such as COMBO-17 \citep{combo02} that uses intermediate-band
filters is capable of selecting emission line galaxies, but these wider filters (compared to NBs) will only detect very
strong line emitting galaxies.

In this paper, the luminosity function (LF) and SFR in almost a dozen redshift windows between $z$ = 0.07 and 1.47 are
presented from line-emitting galaxies in the Subaru Deep Field \citep[SDF;][]{kashik04}. The approach of using
broad-band colors to separate NB emitters will be considered. However, with spectra of some of our NB emitters, galaxies
(with appropriate redshifts) from the Hawaii HDF-N (a deep spectroscopic survey), and multiple NB filters to
cover two different lines at similar redshifts, the LF for line emitters other than the typical \Ha~and \Lya~can be
studied. Broad-band (BB) multi-color selection of \Oii~and \Oiii~emitters has yet to be done at these intermediate
redshifts. The combination of deep, wide imaging with multiple broad- and narrow-band filters makes the SDF scientifically
unique. In \S~\ref{2}, deep broad- and narrow-band imaging are presented. Selection criteria for different NB emitters are
described and follow-up spectroscopy of the brightest line-emitting galaxies are also presented in \S~\ref{2}.
Section~\ref{3} will discuss our methods of distinguishing different line emitters, derive emission line fluxes from NB
photometry, calculate the luminosity function, and derive SFRs at 11 redshift windows. Comparisons with previous studies
will be made in \S~\ref{4}, and a discussion of the evolution of the luminosity function and star formation rate density,
and suggestions for future work are given in \S~\ref{5}. Concluding remarks are made in the final section.

A flat cosmology with H$_{\circ}$ = 70~km~s$^{-1}$~Mpc$^{-1}$, $\Omega_{\Lambda}$ = 0.7, and $\Omega_M$ = 0.3 is
adopted for consistency with recent papers related to this topic and cosmological measurements \citep{spergel03,spergel06}.
Throughout this paper, all magnitudes are given in the AB system: $m_{\rm AB} = -2.5\log{f_{\nu}} - 48.60$, where
$f_{\nu}$ is the flux density in ergs s$^{-1}$~cm$^{-2}$~Hz$^{-1}$.

\section{OBSERVATIONS}
\label{2}
\subsection{Optical Imaging}
Deep optical imaging of the SDF (centered at 13$^{\rm h}$24$^{\rm m}$38\fs9, +27\arcdeg29\arcmin25\farcs9) has been
obtained with Suprime-Cam on the 8.2-m Subaru Telescope \citep{kaifu98,iye04}. Five broad-band ($B$, $V$, $\Rc$,
$i$\arcmin, and $z$\arcmin) and four narrow-band (NB704, NB711, NB816, and NB921\footnotemark[9]) images were obtained
with a total integration time of 595, 340, 600, 801, 504, 198, 162, 600, and 899 minutes, respectively.
\footnotetext[9]{NB704, NB711, NB816, and NB921 are centered at 7046, 7126, 8150, and 9196\AA~with FWHM of 100, 73, 120,
  and 132\AA, respectively.}
The NB704 and NB711 images were part of a LAE study at $z\,\simeq\,5$ taken in 2001 March-June and 2002 May before the SDF
project began \citep{ouchi03,shima03,shima04}. The remaining data were obtained as part of the SDF project. The limiting
magnitudes (3$\sigma$ with a 2\arcsec-aperture) for each 27\arcmin~$\times$~34\arcmin~image are ($B$) 28.45, ($V$) 27.74,
($\Rc$) 27.80, ($i$\arcmin) 27.43, ($z$\arcmin) 26.62, (NB704) 26.67, (NB711) 25.99, (NB816) 26.63, and (NB921) 26.54.
The correction for galactic reddenning is small, $E(B-V)$ = 0.017 \citep{schlegel98}. Each image contains over 100,000
objects. After removing regions of low quality (the edges of the CCD and saturated regions around foreground stars), the
effective field-of-view is about 868 sq. arcmin. Catalogs for each bandpass were constructed using Source Extractor
v2.1.6 \citep[SExtractor;][]{bertin96}.

This paper will only discuss low and intermediate redshift NB704, NB711, NB816, and NB921 emitters. High redshift LAEs
in the SDF are discussed in \citet{kodaira03}, \citet{ouchi03}, \citet{shima03,shima04}, \citet{taniguchi05},
\citet{kashik06}, and \citet{shima06}.

\subsubsection{NB704, NB711, NB816, and NB921 Line Emitters}
\label{2.1.1}
\begin{figure}
  \plottwo{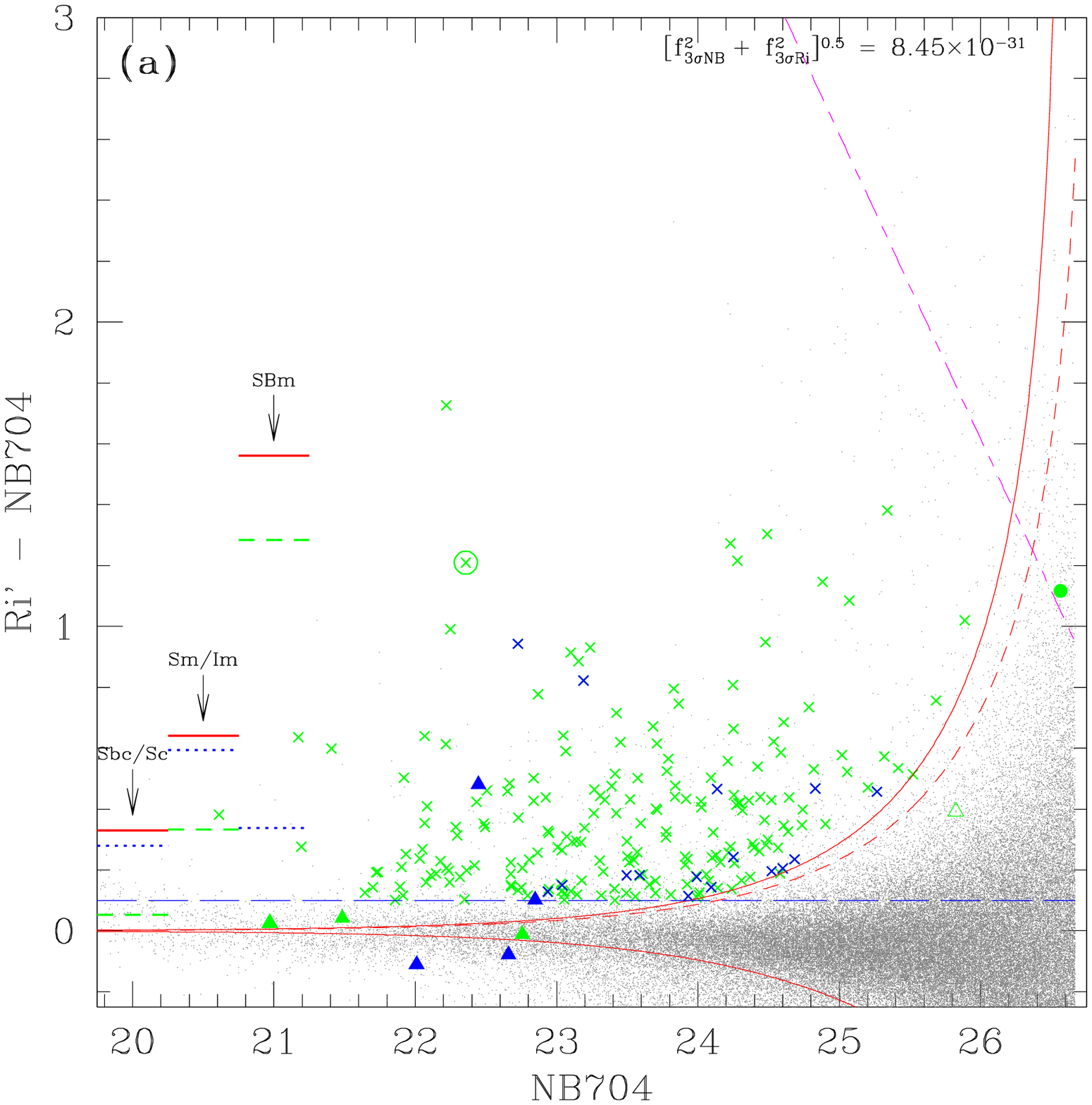}{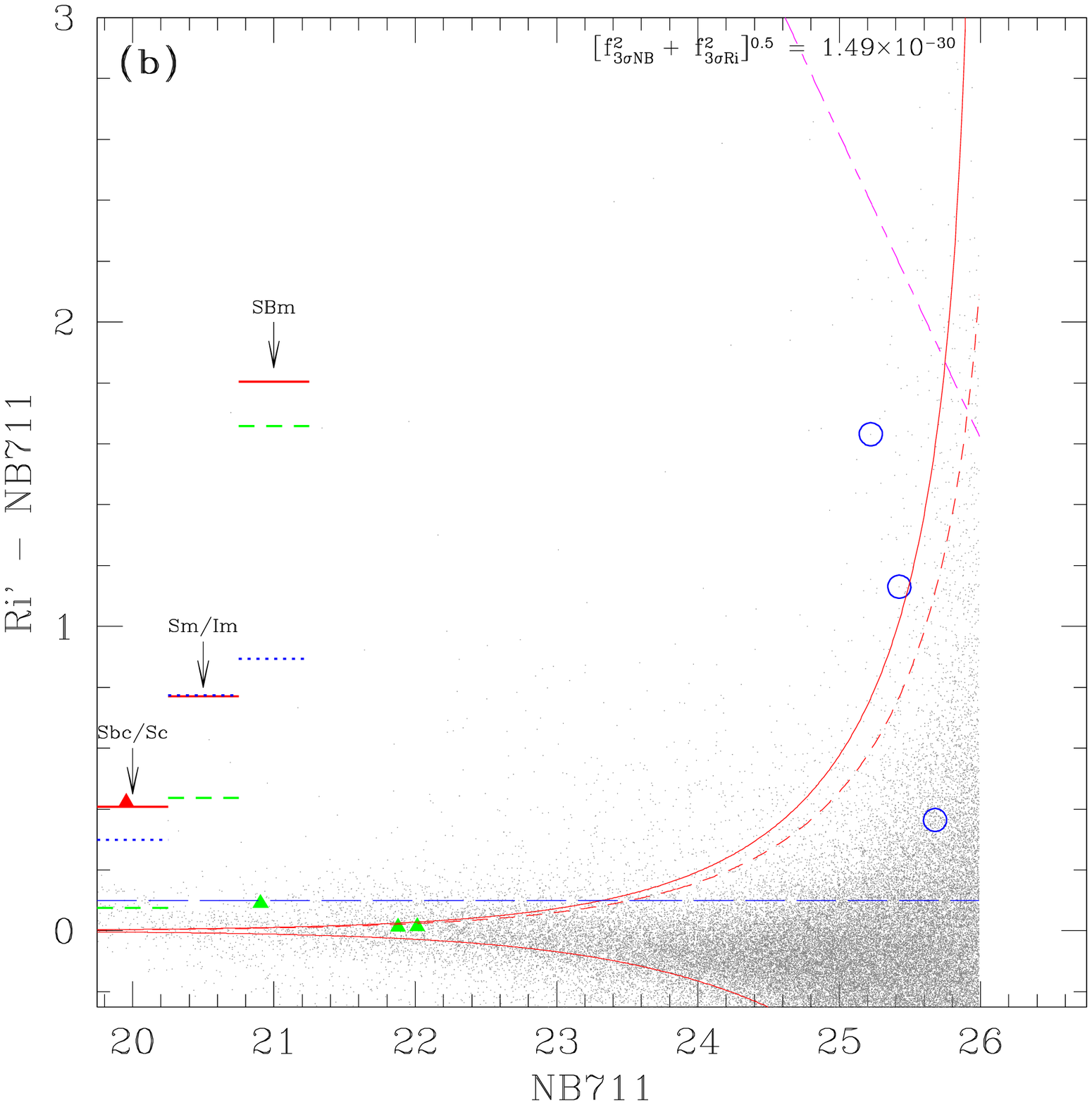}
  \plottwo{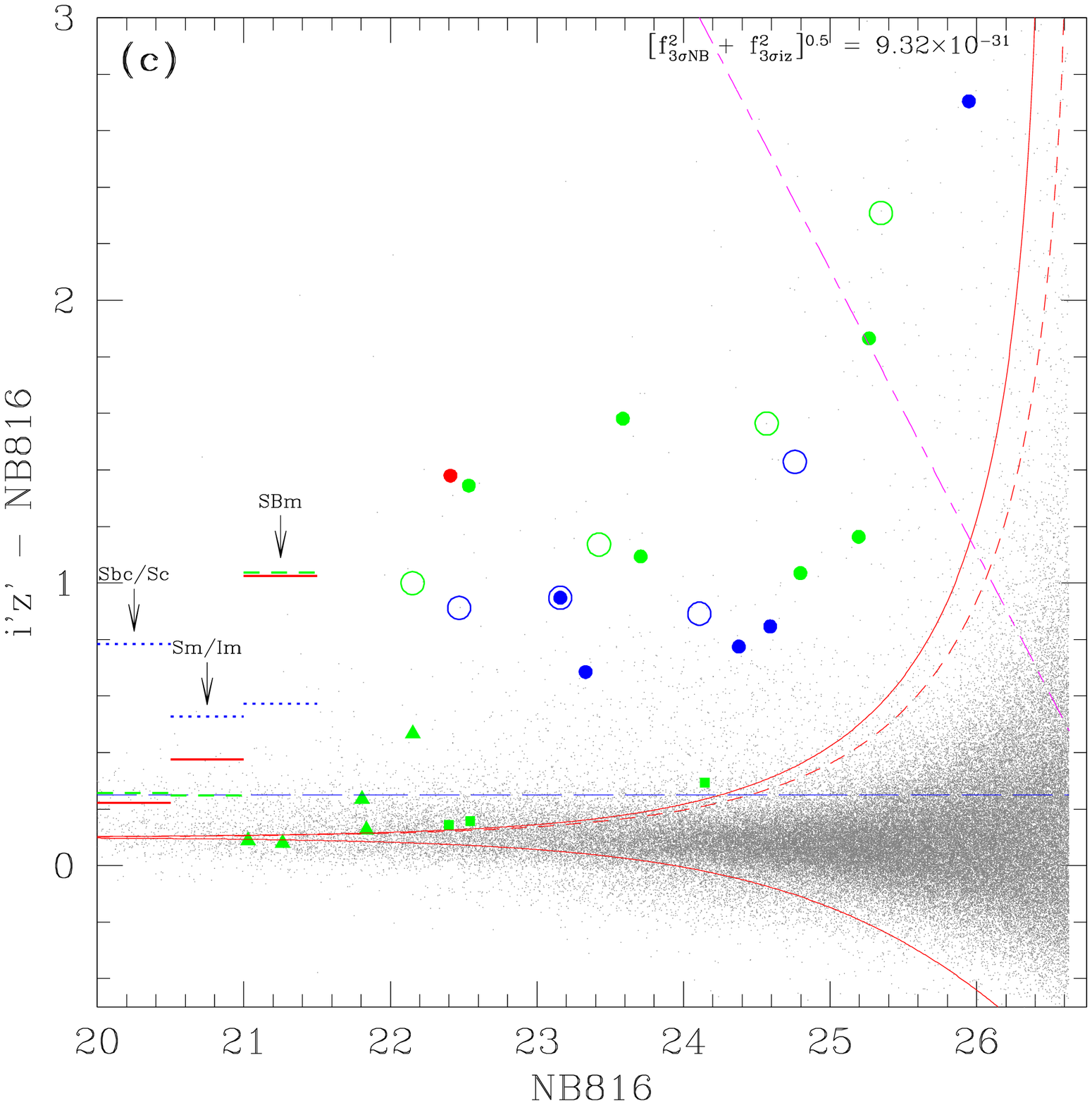}{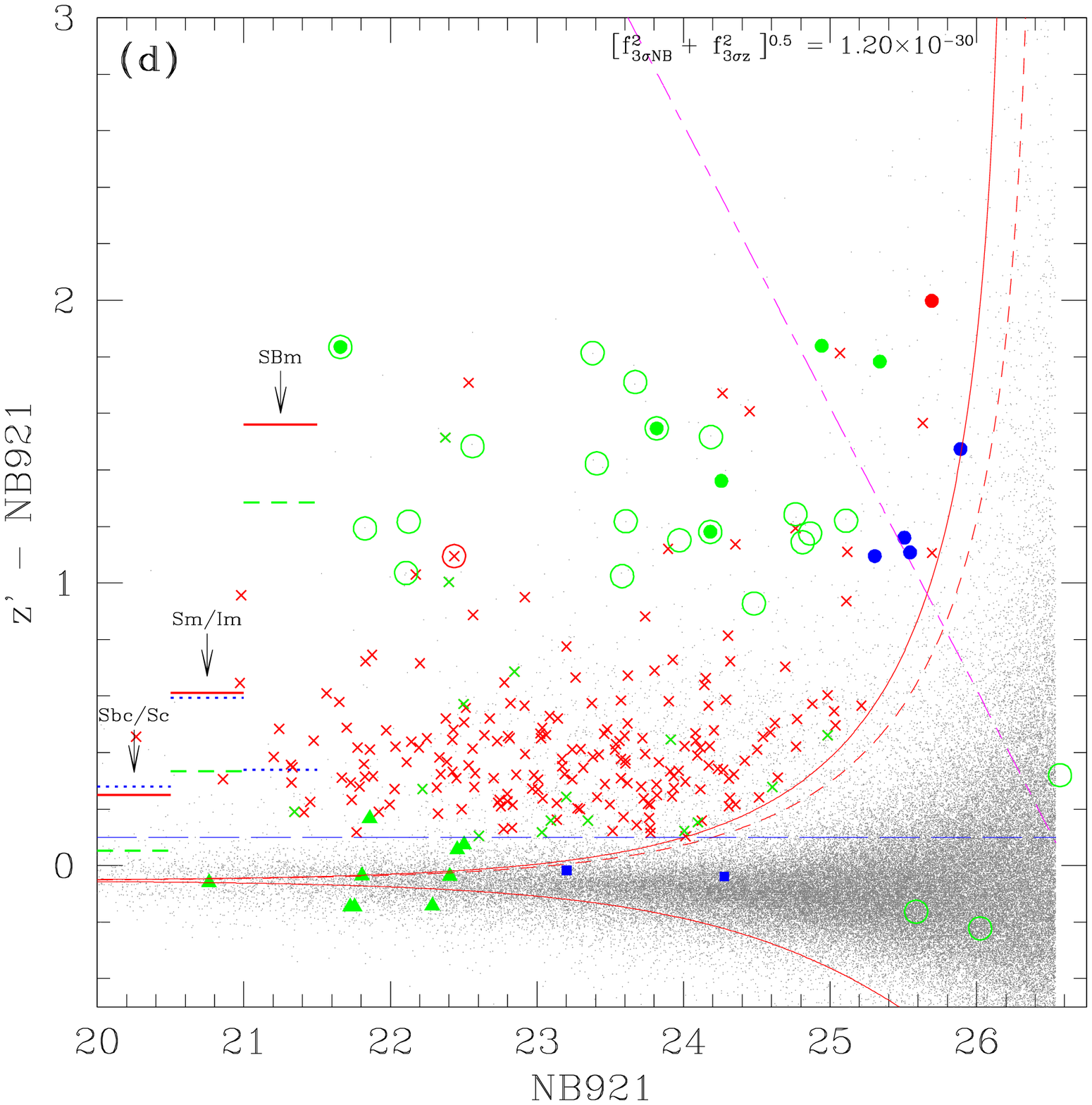}
  \caption{NB excess plots for the ({\it a}) NB704, ({\it b}) NB711, ({\it c}) NB816, and ({\it d}) NB921 catalog. The
    absiccas show the NB magnitude with a 2\arcsec-aperture, and the ordinates show the line excess relative to the
    broad-band continuum. \Oii, \Oiii/\Hb, and \Ha~emitters are identified as blue, green, and red, respectively
    in the electronic edition. Open circles are FOCAS sources, filled circles are DEIMOS targets, filled squares are
    serendipitous objects, crosses are NB704+921 emitters, and filled triangles are fortuitous sources. The solid and
    dashed lines (colored red in the electronic edition) are the excess of BB-NB for $\pm3\sigma$ and 2.5$\sigma$,
    respectively. Long-dashed lines (colored blue in the electronic edition) represent an excess of (a) 0.10,
    (b) 0.10, (c) 0.25, and (d) 0.10 mag. Points above the short-long dashed lines (colored magenta in the electronic
    edition) have their broad-band continuum fainter than 3$\sigma$. Small horizontal lines on the left-hand side
    of the figures are the predicted excess for late type galaxies from the SDSS \citep{yip04}. Solid lines are for
    \Ha, short-dashed lines are for \Oiii, and dotted lines are for \Oii. The three columns from left to right are for
    Sbc/Sc, Sm/Im, and SBm galaxies.}
  \label{fig1}
\end{figure}

BB-NB excess diagrams for the NB704, NB711, NB816, and NB921 catalogs are shown in Figures~\ref{fig1}a-d for NB magnitudes
up to the 3$\sigma$ limiting magnitude. The NB704, NB711, NB816, and NB921 excesses are described by $Ri$\arcmin\,-\,NB704,
$Ri$\arcmin\,-\,NB711, $i$\arcmin$z$\arcmin\,-\,NB816, and $z$\arcmin\,-\,NB921, respectively, where
$Ri$\arcmin\,=\,\case{1}{2}($\Rc+i$\arcmin) and $i$\arcmin$z$\arcmin\,=\,0.6$i$\arcmin~+ 0.4$z$\arcmin. The limiting
magnitude of $Ri$\arcmin~is 27.62 and 27.11 for $i$\arcmin$z$\arcmin. Objects above the short-long dashed magenta lines in
Figure~\ref{fig1}a-d are fainter than 3$\sigma$ of their broad-band flux ($Ri$\arcmin, $i$\arcmin$z$\arcmin, or $z$\arcmin).
The median (i.e., featureless spectra) for the NB816 and NB921 excesses are 0.10 and -0.05 mag, respectively. NB line
emitters are identified as points above the long-dashed blue (a minimum NB excess) and solid red lines in Figure~\ref{fig1}.
The solid red lines represent the $3\sigma$ excess:
$\pm3\sigma_{\rm BB-NB} = -2.5\log{[1\mp(f_{3\sigma NB}^2+f_{3\sigma BB}^2)^{0.5}/f_{NB}]}$, where the error
$[f_{3\sigma NB}^2+f_{3\sigma BB}^2]^{0.5}$ is shown in the upper right-hand corners of Figures~\ref{fig1}a-d. The minimum
NB excesses were chosen `by eye' to be above the NB-BB scatter around NB of 23 mag.

These selection criteria yield 1135 NB704, 1068 NB711, 1916 NB816, and 2135 NB921 line emitters. These values are
reduced to 1000, 986, 1563, and 1942 with good photometric errors ($\Delta m < 0.1$) for broad-band filters used in the
color selections (see \S\S\S~\ref{3.1.1}-\ref{3.1.2}). These line excess limits reach similar equivalent widths
(see \S\S~\ref{3.3}) as \citet{fujita03} and \citet{umeda04}.

\subsection{Spectral Identification of NB Emitters}
\Ha~and \Oiii~NB emitters are the easiest to be identified in an optical spectrum. \Ha~emitters can be confirmed from
detection of other strong lines, \Oiii~and \Hb. And in cases (NB816 emitters) where the spectrum is truncated, the
[\textsc{N ii}] $\lambda\lambda$6548, 6583 and [\textsc{S ii}] $\lambda\lambda$6718, 6732 doublet can be used.
\Oiii~emitters are easily confirmed by the presence of its doublet feature, and \Hb~for some objects. Also, for some
NB921 \Oiii~emitters, the \Oii~doublet appears on the blue side of the spectrum. \Lya~and \Oii~emitters are difficult to
distinguish since the nearest lines are either weak ([Ne \textsc{iii}] $\lambda$3869, H$\delta$, and H$\gamma$) or are
AGN lines (\textsc{C iv}, [Ne \textsc{iii}] $\lambda$3869), and low spectral resolution cannot resolve the very close
\Oii~doublet (e.g., FOCAS; DEIMOS can resolve the doublet). However, \Lya~appears asymmetric at high redshifts, and are
undetected in $B$ and $V$ for NB704, NB711, and NB816 emitters, and $B$, $V$, $R_{\rm C}$, and $i\arcmin$ for NB921
emitters. Therefore, the asymmetry of the line and broad-band detection can be used to distinguish
\Oii~and \Lya~\citep{kashik06,shima06}.

\subsection{Previous Subaru/FOCAS Spectroscopy}
\begin{figure}
  \epsscale{0.5}
  \plotone{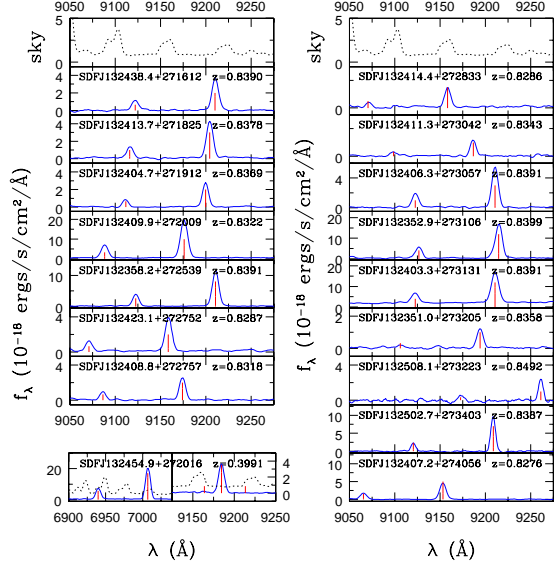}
  \caption{Spectra of NB921 emitters from FOCAS. Vertical lines (colored red in the electronic edition) identify the
    location of emission lines in the spectral window. For \Oiii~emitters, the lines are 4959\AA~and 5007\AA. \Oiii~lines
    for the \Ha~emitter are shown in the adjacent panel. The lines blue-ward and red-ward of \Ha~are
    [\textsc{N ii}] $\lambda\lambda$6548, 6583. The spectrum of the sky is shown in the top panels with arbitrary units
    and overlaid for the \Ha~emitter as a dashed lines.}
  \label{fig2}
\end{figure}

Faint Object Camera and Spectrograph \citep[FOCAS;][]{kashik02} observations primarily targeting NB816 and NB921 emitters
were made on 2002 June 7-10, 2003 June 5-8, and 2004 April 24-27. The description of these observations can be found in
\citet{kodaira03}, \citet{taniguchi05}, \citet{kashik06}, and \citet{shima06}. A total of 24 LAEs, 4 \Oii, and 4
\Oiii~NB816 emitters were identified with FOCAS. For NB921 emitters, 11 LAEs, 19 \Oiii, and 1 \Ha~were identified.
These observations were intended to target LAEs, but a range in broad-band colors were allowed to determine the selection
effects of a color-selected sample. The photometric and redshift information for non-LAEs are provided in
Table~\ref{table1}, and Figure~\ref{fig2} shows the spectrum of NB921 emitters with line fluxes (ordinate) plotted
in ergs s$^{-1}$~cm$^{-2}$~\AA$^{-1}$. The sky's spectrum is plotted at the top with arbitrary units. The spectra of
NB816 emitters can be found in \citet{shima06}, so they are not reproduced. In addition, these NB emitters are plotted
as open circles in Figure~\ref{fig1} and other figures. The convention throughout this paper is that red, green, 
and blue points are \Ha, \Oiii/\Hb, and \Oii~emitters, respectively. Eight NB816 emitters and eight NB921 emitters
remain unclear (either \Lya~or \Oii).

Moreover, deeper broad-band observations have revealed that three NB711 emitters published in \citet{shima03} appear to
be \Oii~emitters by detection in the $B$ and $V$ filters, but a chance projection of a foreground object cannot be ruled
out. These sources are listed at the end of Table~\ref{table1}.

\subsection{DEIMOS Spectroscopy of NB816 and NB921 Emitters}
Deep Imaging Multi-Object Spectrograph \citep[DEIMOS;][]{faber03} observations were made on 2004 April 23 and 24
on Keck II. A total of four masks were used with an 830 lines~mm$^{-1}$ grating and a GG495 order-cut filter. Each
mask had an integration time of 7 - 9 kiloseconds, and had about 100 slits with widths of 1\farcs0 (0.47\AA~pix$^{-1}$,
R $\sim3600$ at 8500\AA). The typical seeing for these observations was 0\farcs5 - 1\farcs0. Standard stars
BD +28\arcdeg~4211 and Feige 110 \citep{oke90} were observed for the flux calibration. The second mask was flux
calibrated with BD +28\arcdeg~4211, and the other three masks were calibrated with Feige 110. All DEIMOS observations
were reduced in the standard manner with the spec2d pipeline. A total of 33 NB816 and 21 NB921 known line emitters
(including LAEs) were targeted with DEIMOS. NB816 emitters were selected for having $i$\arcmin\,-\,NB816\,$\geq$\,1.0 and
20.0\,$\leq$\,NB816\,$\leq$\,25.5 (8.5$\sigma$), and NB921 emitters were selected for $z$\arcmin\,-\,NB921\,$\geq$\,1.0
and 20.0\,$\leq$\,NB921\,$\leq$\,25.5 (7.8$\sigma$). These criteria were used to identify the brightest line emitters in
the sample.

\begin{figure}
  \epsscale{1.17}
  \plottwo{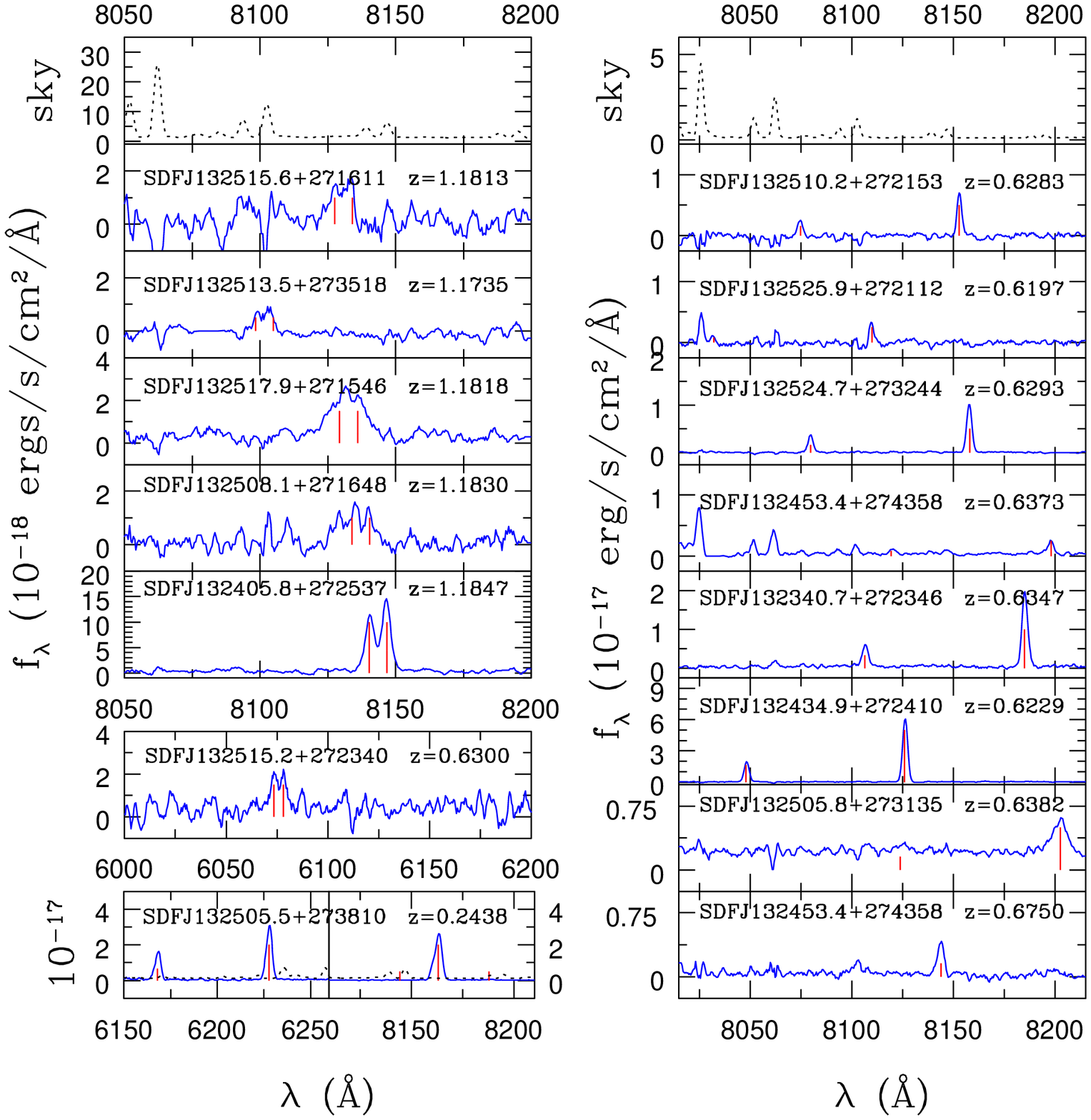}{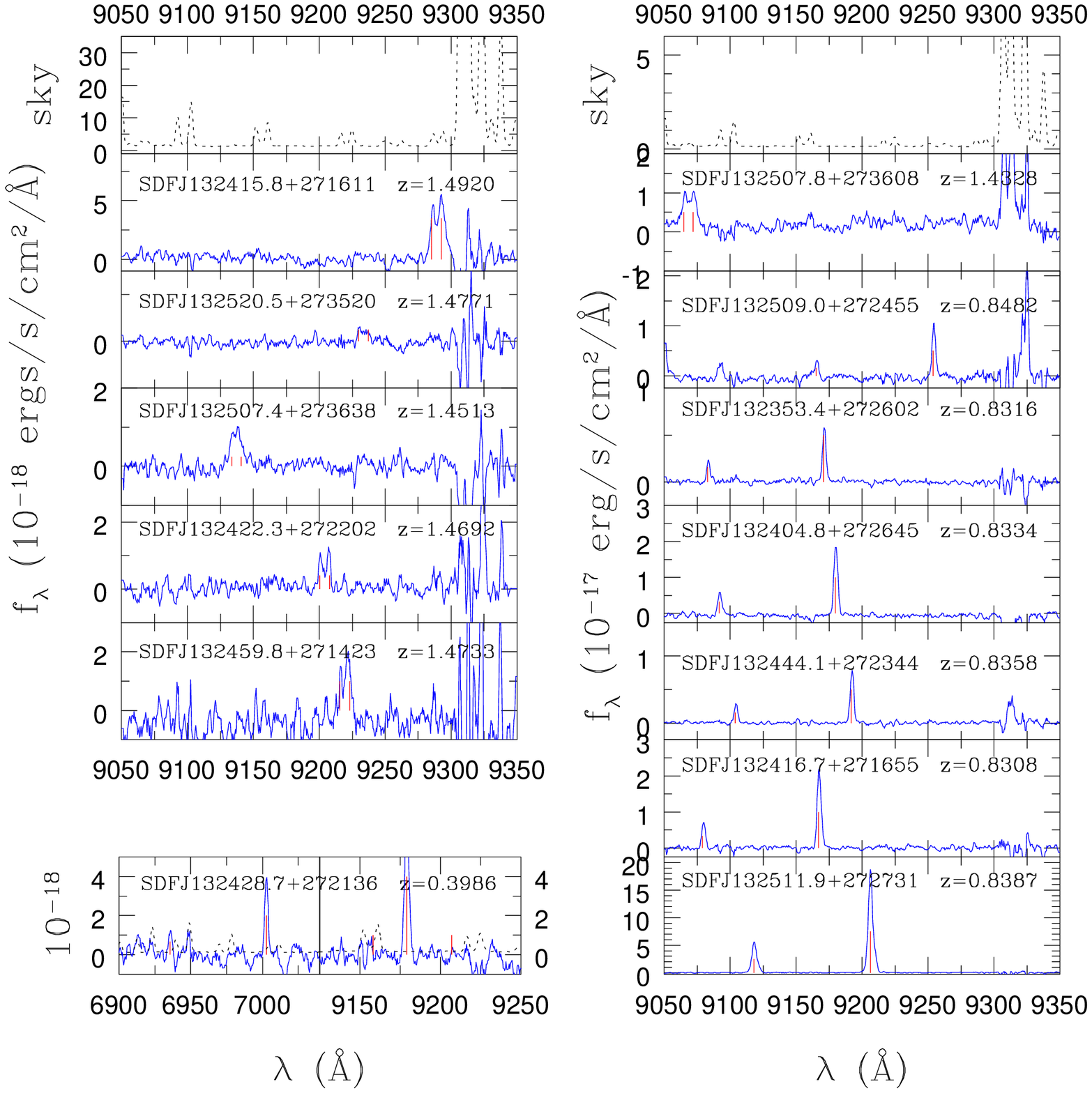}
  \caption{DEIMOS spectra of our identified line-emitting galaxies (including serendipitous sources). NB816 emitters are
    shown on the {\it left}, and NB921 emitters are shown on the {\it right}. Vertical grey lines (colored red in the
    electronic edition) identify the location of emission lines in the spectral window. For \Oii~emitters, the lines are
    3726\AA~and 3729\AA, and 4959\AA~and 5007\AA~for \Oiii. \Oiii~lines for \Ha~emitters are shown in the adjacent panel.
    The lines blue-ward and red-ward of \Ha~are [\textsc{N ii}] $\lambda\lambda$6548, 6583. The spectrum of the sky is
    shown in the top panels with units of 10$^{-18}$ and 10$^{-17}$ ergs s$^{-1}$~cm$^{-2}$~\AA$^{-1}$ for \Oii~and \Oiii,
    respectively. For \Ha~emitters, the sky's spectrum is overlaid on the source's as a dashed line. For the $z=0.675$
    galaxy at the bottom right of the left figure, the \Hb~line is identified in the NB816 filter.}
  \label{fig3}
\end{figure}

Among the NB816 and NB921 line emitters that have been targeted, 4 \Oii~$z$ = 1.47, 4 \Oii~$z$ = 1.19, 3 \Oiii~$z$ = 0.84,
6 \Oiii~$z$ = 0.63, 1 \Ha~$z$ = 0.40, and 1 \Ha~emitter at $z$ = 0.24 have been {\it newly} identified. Their redshift and
photometric information are also provided in Table~\ref{table1}. The flux-calibrated spectra of these sources are shown
in Figure~\ref{fig3}. For Figure~\ref{fig2} and \ref{fig3}, vertical red lines represent the location of emission lines
at the given redshift. For \Oii~emitters, the red lines are for a rest wavelength of 3726\AA~and 3729\AA. While for
\Oiii~emitters, the lines are 4959\AA~and 5007\AA. In the case of the \Ha~emitters, the bluer part (adjacent panel to the
left) of the spectrum has been included to show the \Oiii~lines. Red lines near \Ha~are the expected location of the
[\textsc{N ii}] doublet. The number of new LAEs at $z$ = 5.70 $\pm$ 0.05 and $z$ = 6.56 $\pm$ 0.05 is 10 and 5,
respectively. They are published in \citet{shima06} and \citet{kashik06}.

\subsubsection{Spectroscopy of `Serendipitous' and `Fortuitous' Sources}
\begin{figure}[htp]
  \epsscale{1.17}
  \plottwo{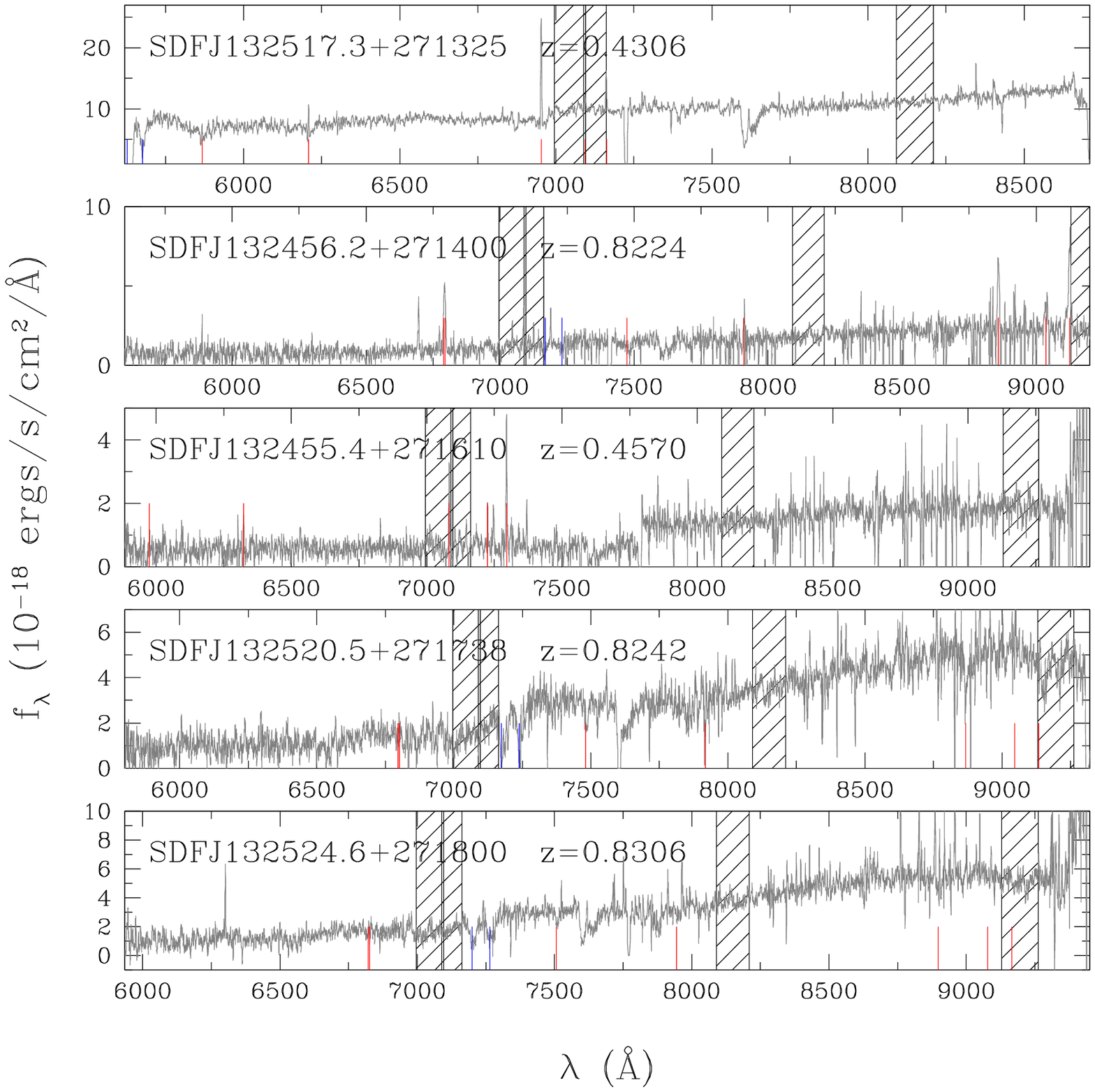}{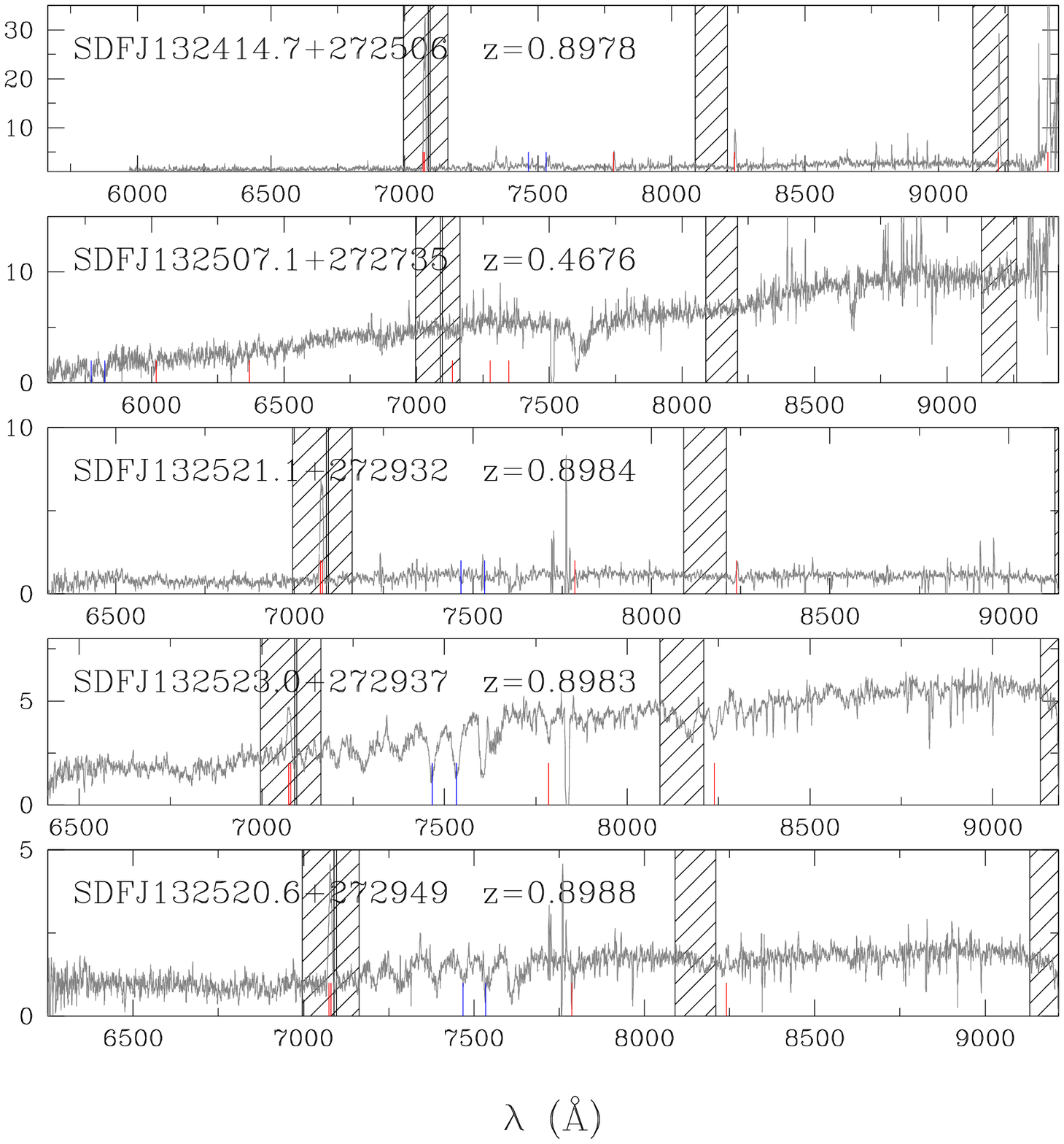}
  \plottwo{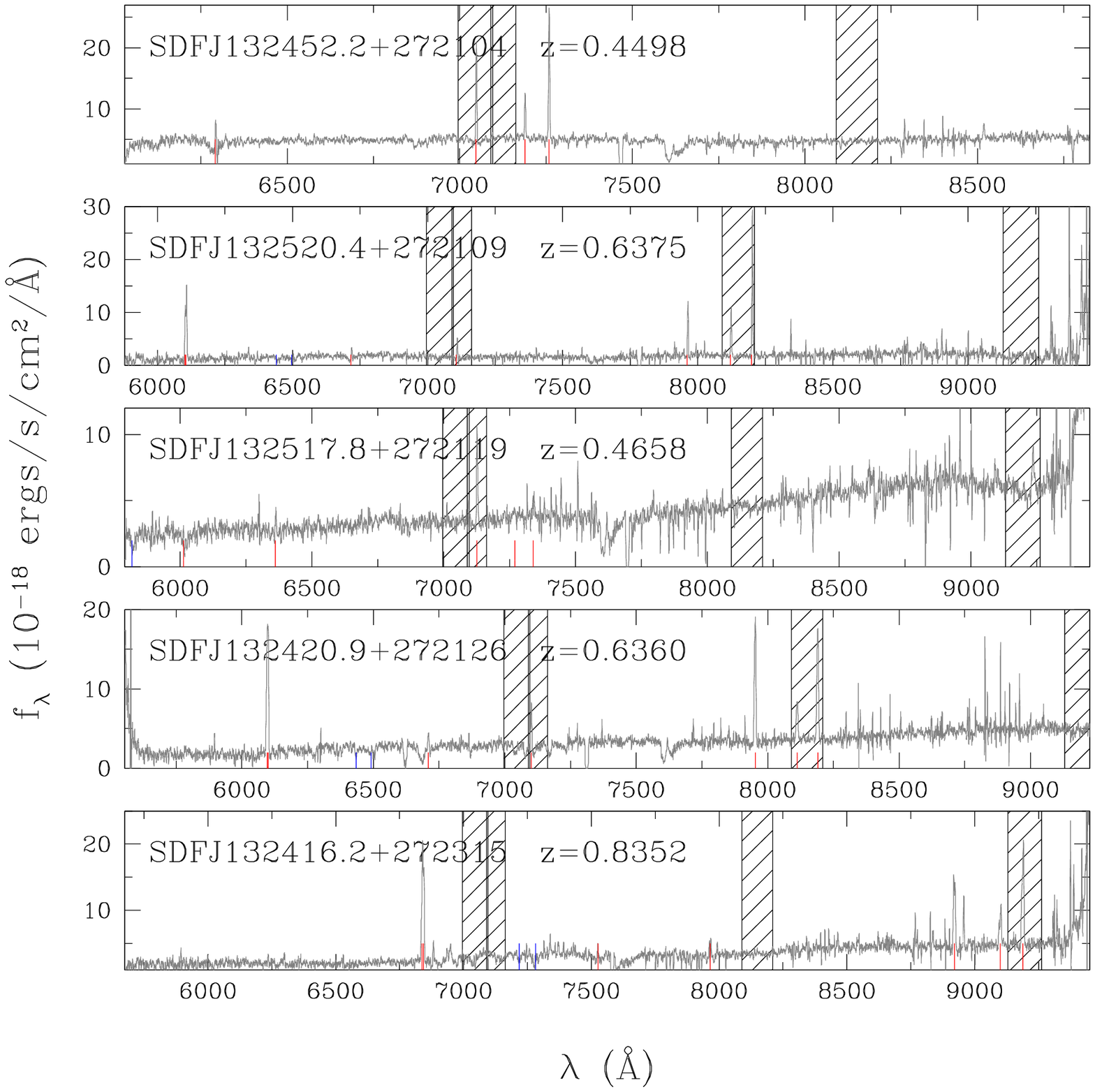}{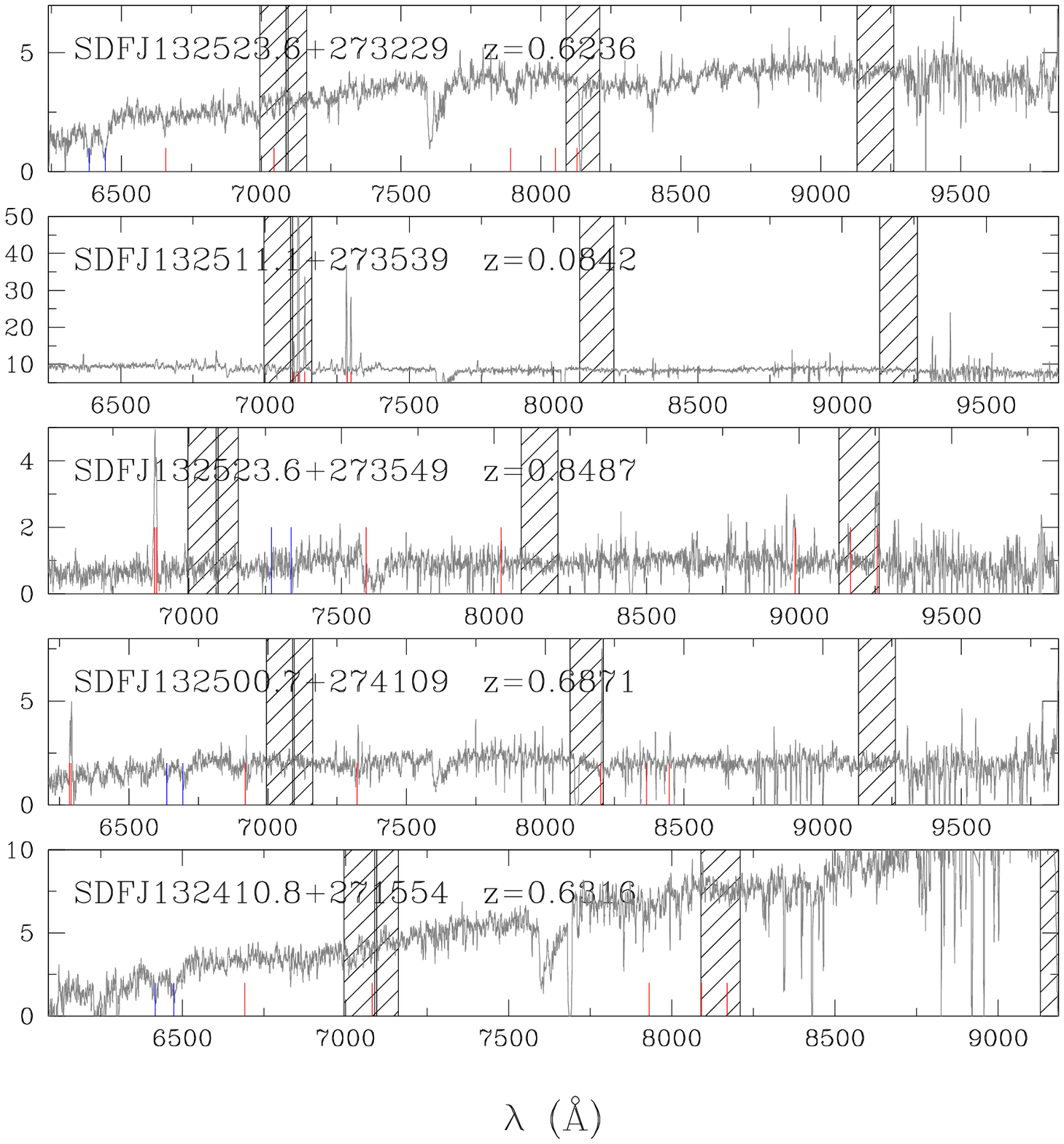}
  \caption{Spectra of the twenty fortuitous sources. The ordinates are given in
    10$^{-18}$ ergs s$^{-1}$~cm$^{-2}$~\AA$^{-1}$. Vertical lines (colored red in the electronic edition) identify the
    location of emission lines in the spectral window. They are from short to long wavelengths: \Oii, H$\delta$, H$\gamma$,
    \Hb, \Oiii~doublet, \Ha, [\textsc{N ii}] doublet, and [\textsc{S ii}] doublet. The blue lines (in the electronic
    edition) identify the location of Ca \textsc{ii} K and H at 3933 and 3968\AA. The FWHMs of the NB filters are shown by
    the shaded regions.}
  \label{fig4}
\end{figure}

Because of the long (up to 10\arcsec) DEIMOS slits, other galaxies falling within the slits are identified by the
reduction pipeline as `serendipitous'. Moreover, other lower priority sources targeted with DEIMOS yield redshifts in the
same range as those of the NB filters. These `fortuitous' and serendipitous sources may not satisfy our NB excess
selection criterion in \S\S~\ref{2.1.1} or have any emission lines (some are identified via absorption features), but
they provide important information about the broad-band colors at these redshifts. There are five serendipitous sources
with relevant redshifts: three $z\approx0.83$ \Oiii~and two $z\approx1.46$ \Oii~that are included in this paper. They
are plotted as filled squares in Figure~\ref{fig1} and subsequent figures, and are listed in Table~\ref{table1}. Twenty
fortuitous sources are identified, and are included in Table~\ref{table1}. The spectra of the fortuitous sources are
shown in Figure~\ref{fig4}, and are identified in NB-excess, and two-color figures as filled triangles.

Therefore, the total (including serendipitous and fortuitous sources) number of spectra that will be used in our line
classification scheme is 75. Table~\ref{table2} summarizes the number of spectroscopically-identified sources within
different redshift windows.

\subsection{NB Excess Predictions from Sloan Digital Sky Survey}
Mean spectra for six galaxy types (from early to late) were obtained from \citet{yip04} and were then redshifted
for either \Ha, \Oiii, or \Oii~to fall within the four NB filters and then convolved with the BB and NB filters.
The spectra were averaged over 100 to 20000 Sloan Digital Sky Survey (SDSS) sources. This procedure tests whether or
not typical galaxies detected in the SDSS are capable of being detected in the NB filters due to strong emission lines.
The BB-NB excesses for the three latest types (Sbc/Sc, Sm/Im, and SBm) are shown as horizontal lines on the left-hand
side of Figure~\ref{1}a-d. The BB-NB excess method shows that the two latest types (Sm/Im and SBm) can easily be detected
with NB filters due to their very strong emission lines, and Sbc/Sc can be detected in some cases. The Sm/Im and SBm
galaxies make up 0.5\% of the entire sample of \citet{yip04}, and the Sbc/Sc sample consists of 23\%. Thus at low
redshift, for example, our NB imaging would detect about a quarter of the SDSS galaxies.

\section{RESULTS}
\label{3}
Multiple possibilities exist for the identification of a detected emission line in a NB filter. \Lya, \Oii~$\lambda$3727,
\Hb, \Oiii~$\lambda\lambda$4959, 5007, and \Ha~are the strongest lines that most likely will be identified.
The spectra of NB strong emitters show that they are either \Ha, \Oiii, or \Oii; neither of them are \Hb. Other
objects (serendipitous and fortuitous) with \Hb~in the filter have very low NB-BB excess. It is therefore assumed that
these NB emitters are more likely to have \Oiii~rather than \Hb. The redshift range, comoving volume, and luminosity
distance ($d_L$) are listed in Table~\ref{table3} for all four NB filters.

\subsection{Broad-band Color Selection}
\label{3.1}
Past studies \citep[e.g.,][]{fujita03,kodama04,umeda04} that have used multi-color spectral energy distributions (SEDs)
of NB emitters, relied on theoretical population synthesis models to identify photometric \Ha~emitters. However, without
spectra of a sample of bright galaxies, the identification of these emitters cannot be confirmed. Since spectra have
been obtained for a few to over two dozen objects in each redshift bin, the multi-color classification of different
(\Oii~and \Oiii) line emitters in this study is more reliable than previous studies. With five broad bandpasses,
distinguishing different line emitters is more feasible in a multidimensional color space, as previous studies were
limited to two or three broad bandpasses. Many of the colors that will be used rely on the Balmer break falling in a
particular bandpass.

The NB704 filter provides the special advantage of determining the redshift of NB921 emitters into two possible
intervals. This is almost equivalent to obtaining a spectra, as a line-emitting galaxy in the NB704 and NB921 filters
correspond to either a redshift of 0.397-0.411 or 0.878-0.904. The former occurs when the \Oiii~$\lambda$5007 line
falls within the NB704 filter, and the \Ha~line is within the NB921 filter. The latter is for \Oii~$\lambda$3727 and
\Hb~(see Table~\ref{table3}). Coincidentally, the FOCAS spectra of an \Ha~emitter (SDFJ132354.9+272016) is one of these
NB704+921 line emitters, which shows that multiple NB filters can be used to select sources. The total number of
NB704+921 line emitters is 212. As a comparison, other sets of filters were investigated. For NB704 and NB816, only 11
objects are emitters in both filters, and 7 objects for NB711 and NB816. NB816 and NB921 filters yielded 99 objects.

\begin{figure}
  \plottwo{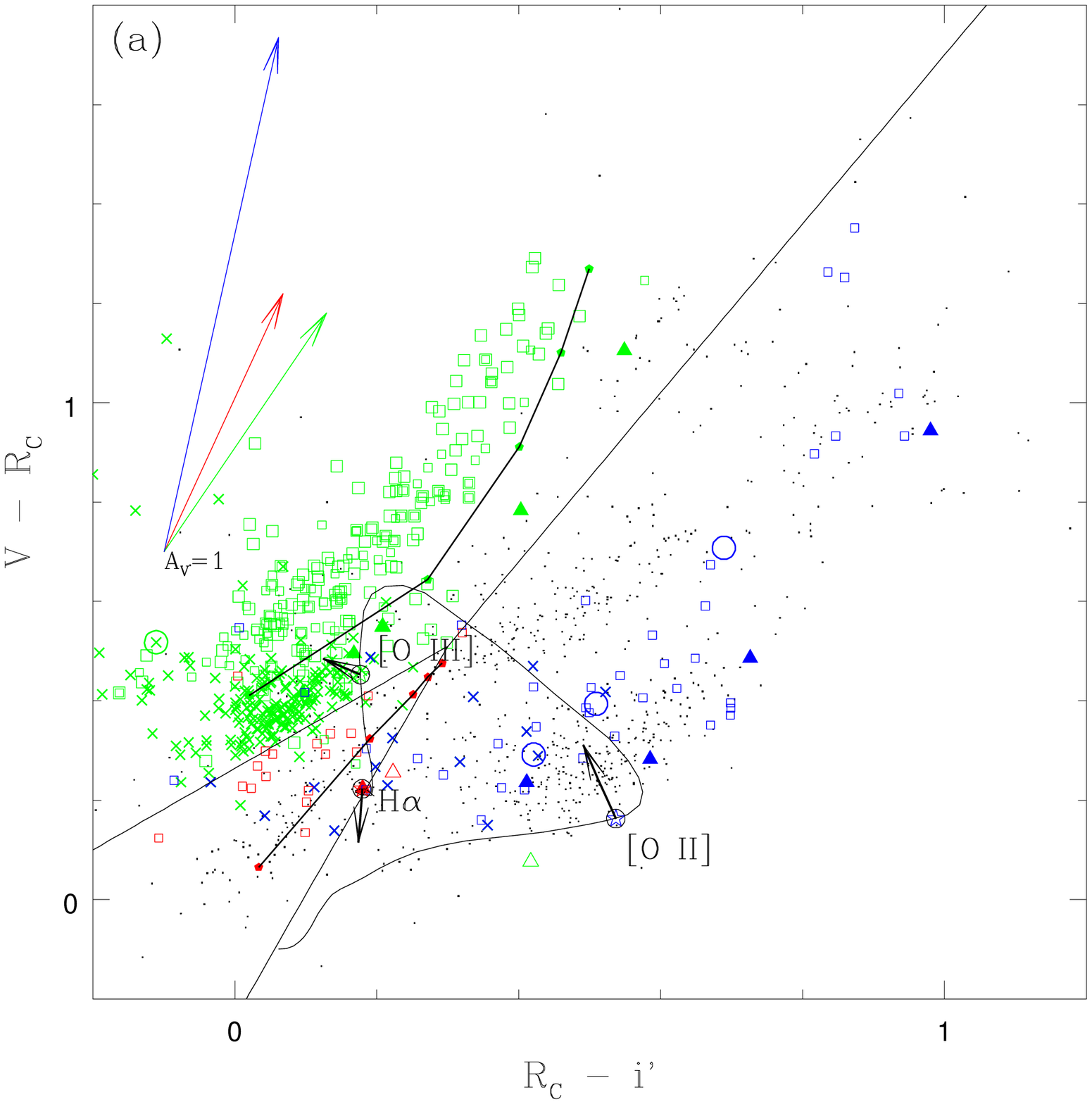}{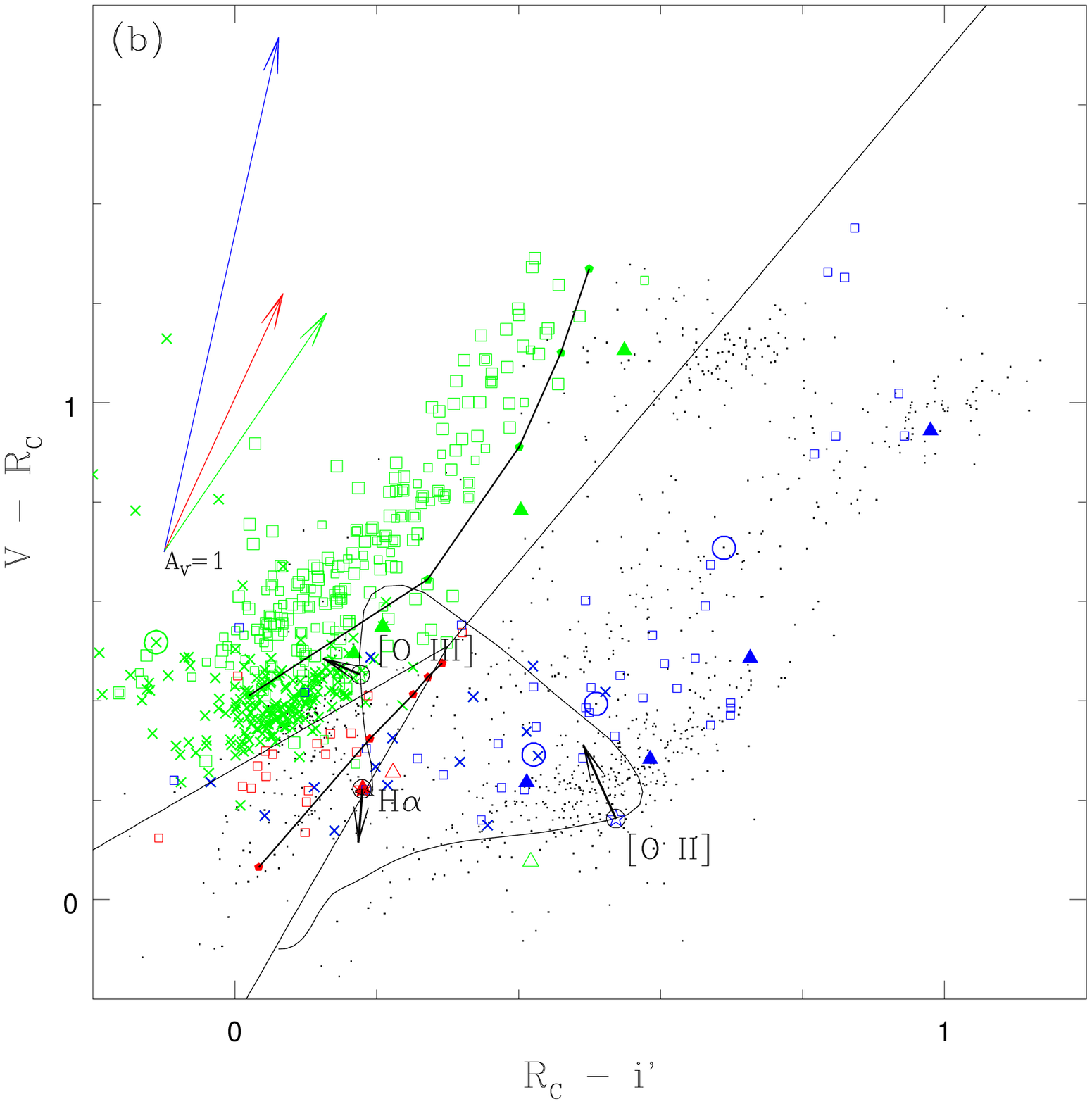}
  \caption{Two-color diagrams in $V-\Rc$ vs. $\Rc-i\arcmin$ for ({\it a}) NB704 and ({\it b}) NB711 emitters. All
    spectroscopic NB704 and NB711 emitters are plotted on both diagrams. In the electronic edition, red points are \Ha,
    blue points are \Oii, and green points are \Oiii~and \Hb~emitters identified by FOCAS (open circles) and DEIMOS
    (filled circles). Open squares are galaxies from the Hawaii HDF-N, and triangles are fortuitous sources (filled) and
    from the NED database (open). NB704 and NB921 dual emitters are shown as crosses. The solid lines are
    $V-\Rc = 1.70(\Rc-i\arcmin)$, $V-\Rc = 0.82(\Rc-i\arcmin) + 0.26$, and $V-\Rc = 2.5(\Rc-i\arcmin) - 0.24$.
    A theoretical model from \citet{bc03} with constant star-formation (without dust extinction) is shown by the
    solid black curve for $z=0$ to 1.5. Along this curve, the broad-band colors at specific redshifts of 0.07, 0.40,
    and 0.89 are shown by starred symbols with a circle surrounding it. Black vectors at these points indicate the
    direction that the colors follow with different emission line strengths. The vectors in the upper left-hand corner
    correspond to 1 magnitude of V extinction using the reddenning curve of \citet{cardelli89}. Filled pentagons and
    thick black lines represent the colors of \citet{yip04}'s spectra from early to late type SDSS galaxies for
    $z=0.07$ and 0.40.}
  \label{fig5}
\end{figure}

To better distinguish different line emitters, galaxies from the Hawaii HDF-N with known redshifts from either
LRIS \citep{oke95} or DEIMOS have been analyzed \citep{cowie04}. $B$, $V$, $\Rc$, $I_{\rm C}$, and $z\arcmin$ photometry
have been obtained by \citet{capak04} using Suprime-Cam. Currently, no transformation between $I_{\rm C}$ and $i\arcmin$
exists for a sample of galaxies. However, SDSS studies\footnotemark[10] of stars have shown that the
transformation between the $I_{\rm C}$ and $i\arcmin$ bandpasses is $I_{\rm C} \approx i\arcmin - 0.4(i\arcmin-z\arcmin)$.
\footnotetext[10]{\url{http://www.sdss.org/dr4/algorithms/sdssUBVRITransform.html}.}
This formula is used to compute the $i\arcmin$ magnitude for these Hawaii HDF-N galaxies. The number of sources within the
NB704 and NB711 redshift intervals of 0.064 - 0.093 (\Ha), 0.395 - 0.475 (\Oiii~and \Hb), and 0.875 - 0.924 (\Oii) is
20, 200, 39, respectively. And the number of sources for NB816 and NB921 intervals of 0.231 - 0.253 (\Ha), 0.614 - 0.658
(\Oiii), 0.662 - 0.691 (\Hb), 1.169 - 1.205 (\Oii), 0.389 - 0.413 (\Ha), 0.821 - 0.870 (\Oiii), 0.876 - 0.904 (\Hb), and
1.448 - 1.487 (\Oii) is 19, 74, 58, 7, 46, 157, 21, and 8, respectively. Hawaii HDF-N galaxies are plotted as open squares
in the two-color diagrams (see below) with the same color conventions used for the SDF spectroscopic sample. Also, there
are two sources within the NASA/IPAC Extragalactic Database (NED) at redshifts of 0.0718 and 0.45, which fall within the
redshift windows. These sources are plotted as open triangles in the color-color diagrams.
The SDSS spectra of \citet{yip04} have been redshifted and convolved with the broad-band filters to obtain the colors.
They are overlayed as thick solid black lines on Figures~\ref{fig5} and \ref{fig6}. Because of the limited coverage
(3500-7000\AA) of these spectra, the desired colors could only be determined at $z=0.07$ (NB704 and NB711 \Ha), 0.25
(NB816 \Ha), and 0.40 (NB704 and NB711 \Oiii). There is good agreement between the SDSS predicted broad-band colors and
those of the NB emitters.

For additional comparison, a stellar population model from GALAXEV \citep{bc03} with constant star-formation is
overlayed on these two-color diagrams. To correct the broad-band colors for strong nebular emission lines, we adopt
the emission line ratios of Sm/Im galaxies from \citet{yip04}. They are \Oiii/\Ha+[\textsc{N ii}] = 1.33,
\Oii/\Ha+[\textsc{N ii}] = 1.05, and \Hb/\Ha+[\textsc{N ii}] = 0.43. The large \Oiii/\Ha~ratio is valid as a subsample
of our data has a large ratio compared to local measurements (see \S\S~\ref{3.6}). Other lines
(e.g., H$\gamma$, [\textsc{S ii}] $\lambda\lambda$6718, 6732) while present in the spectrum do not affect the colors
significantly compared to the strong emission of \Oiii, \Oii, \Hb, and \Ha. Vectors are drawn on
Figures~\ref{fig5}-\ref{fig7} for \Ha+[\textsc{N ii}] line strengths from 0 to 200\AA~EW. These vectors do pass through
the majority of NB line emitters.
  
\subsubsection{NB704 and NB711 Emitters}
\label{3.1.1}
To distinguish NB704 and NB711 \Ha, \Oiii, or \Oii~emitters, $V-\Rc$ and $\Rc-i$\arcmin~colors are plotted in
Figure~\ref{fig5}. \Oiii~emitters are selected by $V-\Rc \geq 1.70(\Rc-i\arcmin)$ and
$V-\Rc \geq 0.82(\Rc - i\arcmin) + 0.26$. \Oii~emitters are selected by $V-\Rc \leq 1.70(\Rc-i\arcmin)$ and
$V-\Rc \leq 2.50(\Rc - i\arcmin) - 0.24$. The remaining sources are identified as \Ha. The total number of identified
NB704 \Ha, \Oiii, and \Oii~emitters is 120, 303, and 580, respectively, and 114, 158, and 713 for NB711 \Ha, \Oiii, and
\Oii~emitters.

The contamination rate$-$percentage for a type of source (e.g., \Oiii~or \Oii) to fall within another type's selection
criteria$-$can be determined from available spectra (including Hawaii HDF-N data). In the selection of \Ha, the
contamination from \Oii~is 5/46 (11\%) and 3/208 (1\%) for \Oiii~or \Hb~emitters. For \Oiii, there is 4/46 (9\%)
contamination from \Oii~and 1/22 (5\%) from \Ha. Finally, for \Oii, \Ha~and \Oiii~or \Hb~contribute 2/22 (9\%) and 1/208
($<1$\%) contamination, respectively.

\subsubsection{NB816 Emitters}
\begin{figure}
  \plottwo{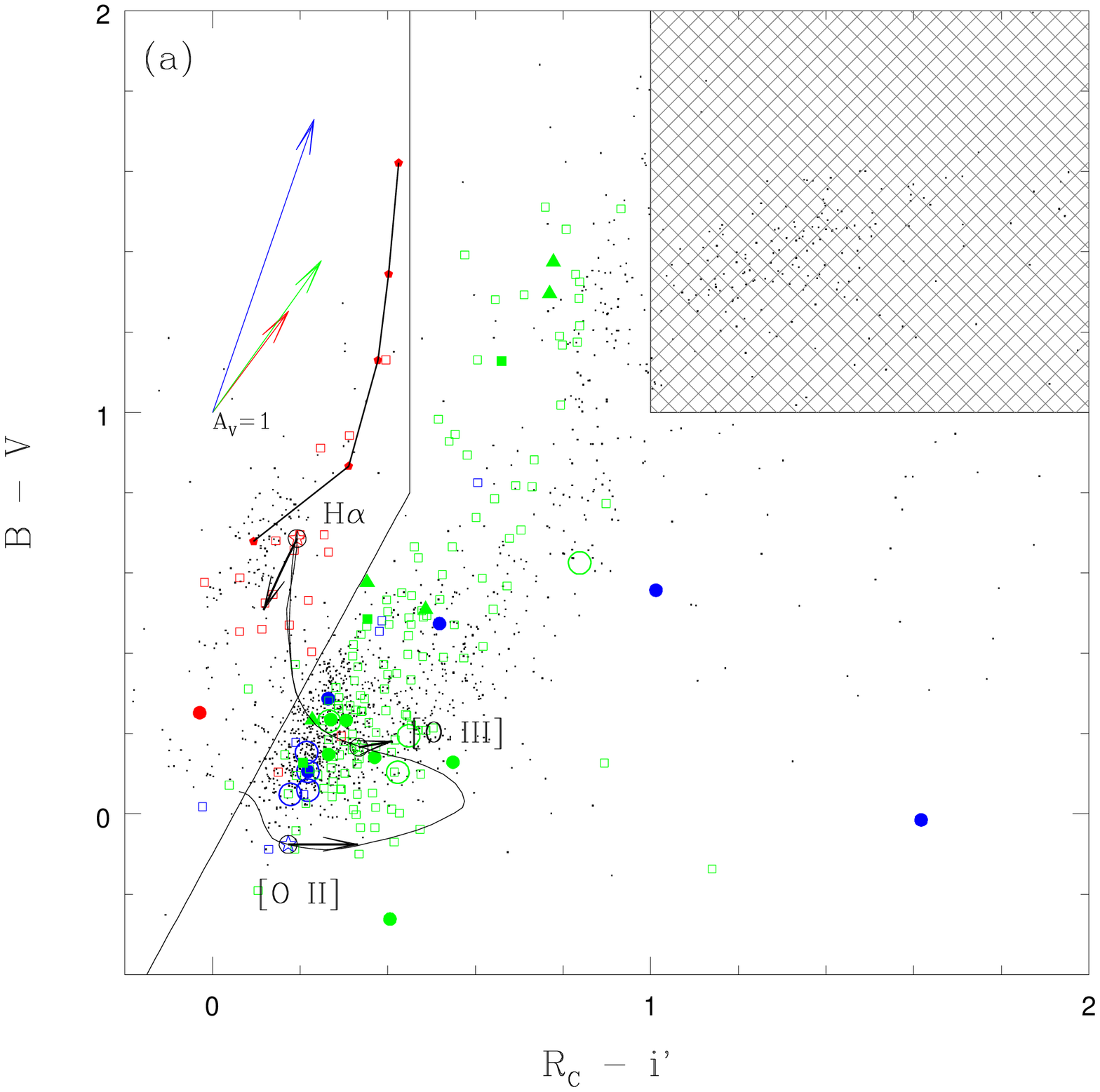}{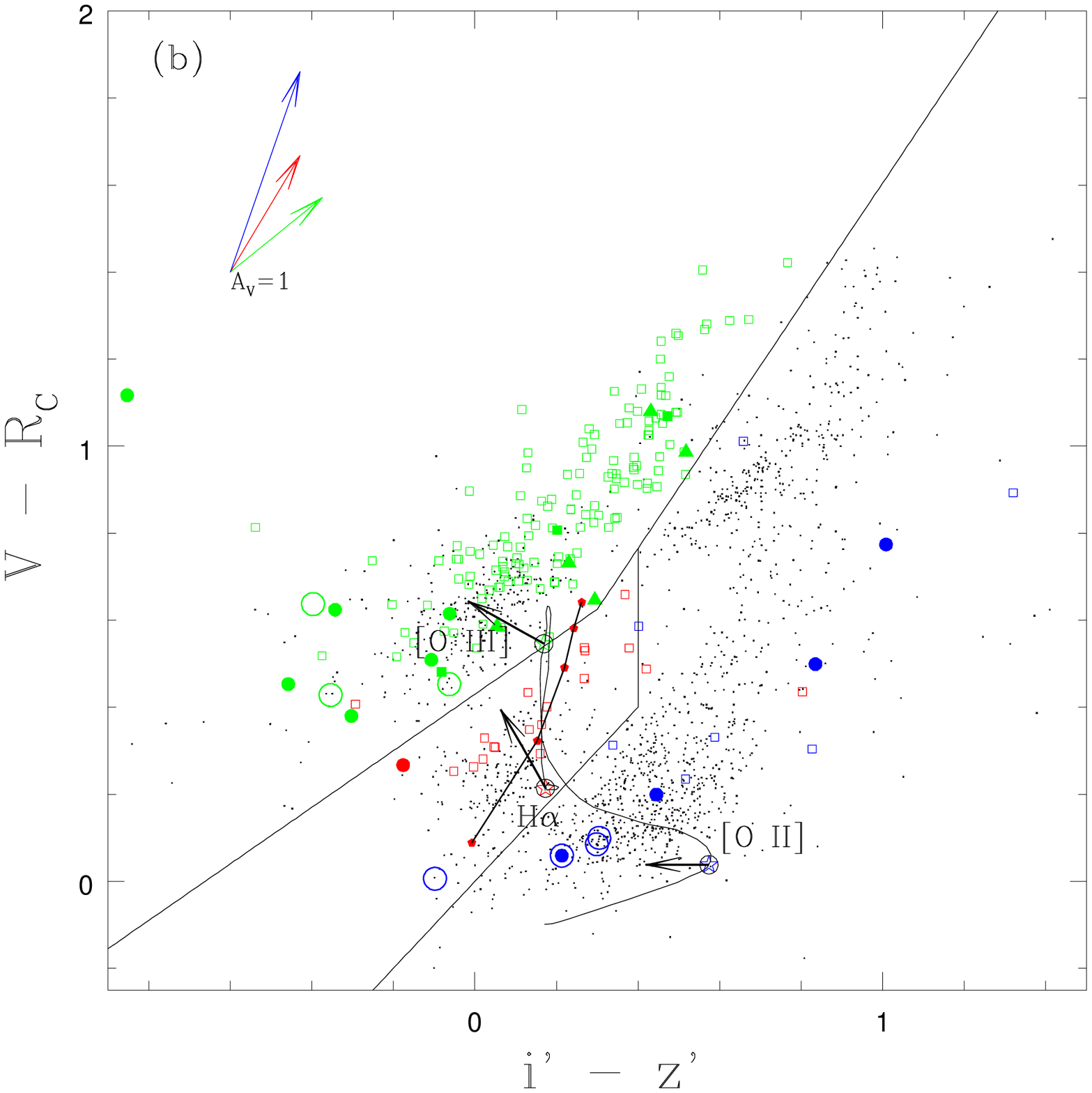}
  \caption{Two-color diagrams in $B-V$ vs. $\Rc-i\arcmin$ ({\it a}) and $V - \Rc$ vs. $i\arcmin-z\arcmin$ ({\it b}) for
    NB816 emitters. In the electronic edition, red points are \Ha, blue points are \Oii, and green points are \Oiii~and
    \Hb~emitters identified by FOCAS (open circles) or DEIMOS (filled circles). Open squares are galaxies from the Hawaii
    HDF-N, filled squares are serendipitous sources, and filled triangles are fortuitous sources. The solid lines in (a)
    are $B-V = 2.0(\Rc-i\arcmin) + 0.20$ and $\Rc-i\arcmin = 0.45$, and in (b) $V - \Rc = 0.65(i\arcmin-z\arcmin) + 0.43$,
    $V-\Rc = 1.4(i\arcmin-z\arcmin) + 0.21$, $i\arcmin-z\arcmin = 0.40$, and $V - \Rc = i\arcmin-z\arcmin$.
    A theoretical model from \citet{bc03} with constant star-formation (without dust extinction) is shown by the
    solid black curve for $z=0$ to 1.5. Along this curve, the broad-band colors at specific redshifts of 0.24, 0.64,
    and 1.20 are shown by starred symbols with a circle surrounding it. Black vectors at these points indicate the
    direction that the colors follow with different emission line strengths. The vectors in the upper left-hand corner
    correspond to 1 magnitude of V extinction using the reddenning curve of \citet{cardelli89}. Filled pentagons (red
    in the electronic edition) and thick black lines represent the colors of \citet{yip04}'s spectra from early to late
    for $z=0.24$.}
  \label{fig6}
\end{figure}

Figure~\ref{fig6}a and b show $B-V$ vs. $\Rc-i\arcmin$ and $V-\Rc$ vs. $i\arcmin-z\arcmin$ for NB816 emitters. The first
plot isolates \Ha~emitters while the second plot primarily separates \Oii~and \Oiii~emitters. \Ha~emitters are identified
by $B-V \geq 2(\Rc-i\arcmin) - 0.1$ and $\Rc-i\arcmin < 0.45$. \Oiii~emitters are selected by
$V-\Rc \geq 0.65(i\arcmin-z\arcmin) + 0.43$ and $V-\Rc \geq 1.4(i\arcmin-z\arcmin) + 0.21$ (solid black lines in
Figure~\ref{fig6}b). \Oii~emitters are classified by $V-\Rc \geq 1.4(i\arcmin-z\arcmin) + 0.21$ and
$i\arcmin - z\arcmin \geq 0.40$ or $V-\Rc \leq i\arcmin-z\arcmin$. Sources within the shaded region of Figure~\ref{fig6}a
are ``unknown'' objects as no spectral identification is available in that area. Initially, these sources were thought
to be \Oii~emitters as their colors were $V-\Rc \approx 0.9$ and $i\arcmin-z\arcmin \approx 0.6$, but this resulted in
an excess (N = 192) of sources with line luminosities above $L_{\star}$. Hypothetically, these objects may be
\Oiii~emitters, therefore, two LFs (including and excluding the unknown sources) will be presented in \S\S\S~\ref{3.3.2}.
The total number of NB816 line emitters identified as \Ha, \Oiii, and \Oii~emitters is 205, 280 (472 including
unknown NB816 emitters), and 831, respectively.

The contamination of \Oii~and \Oiii~or \Hb~line emitters into \Ha~is 1/14 (7\%) and 3/150 (2\%). There is 2/20 (10\%)
contamination from \Ha~into \Oii~and zero contamination by \Oiii~or \Hb. And for \Oiii, a contamination of 1/20 (5\%)
from \Ha~is found while \Oii~contributes zero contamination.

\subsubsection{NB921 Emitters}
\label{3.1.2}
\begin{figure}
  \plottwo{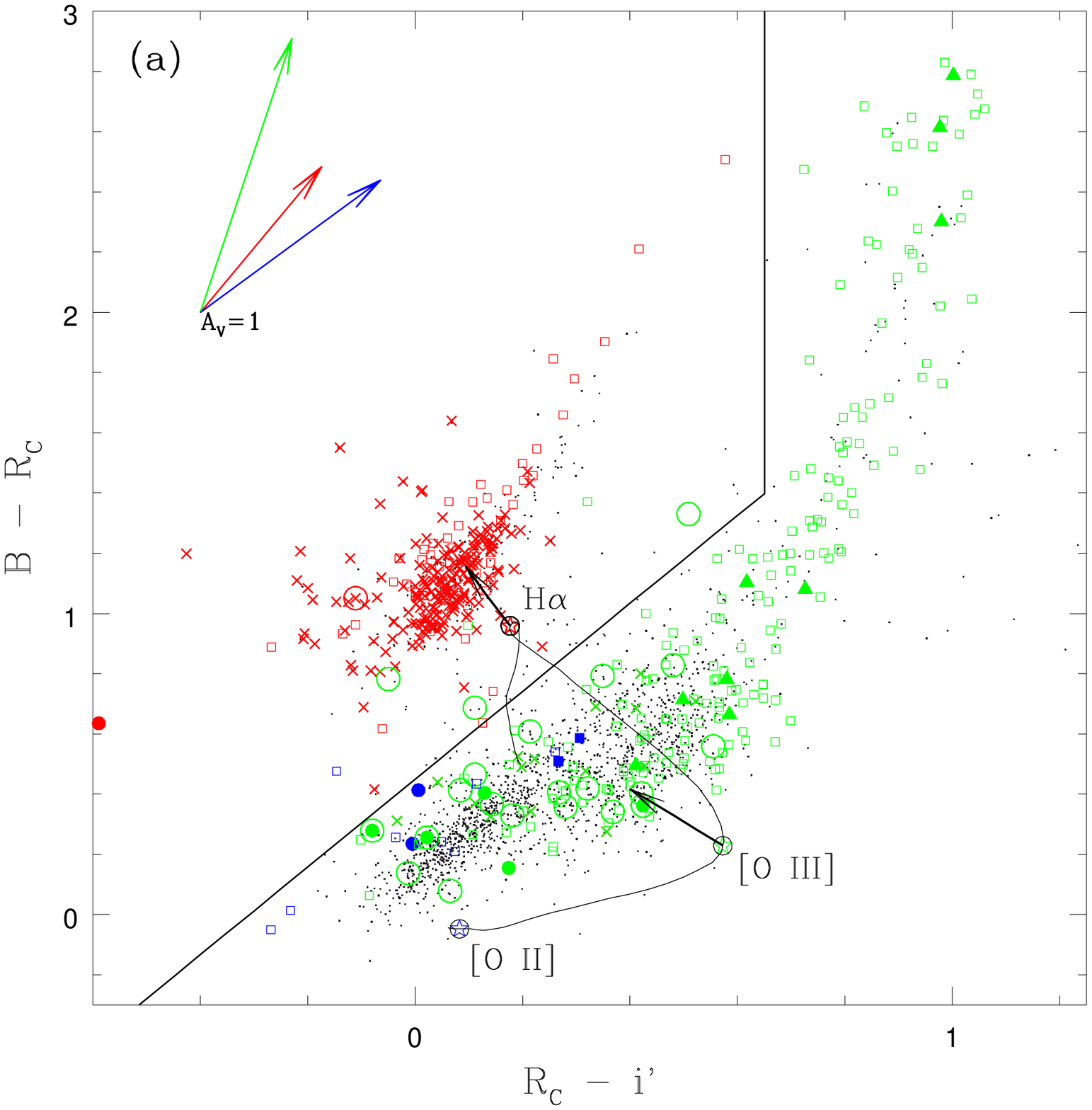}{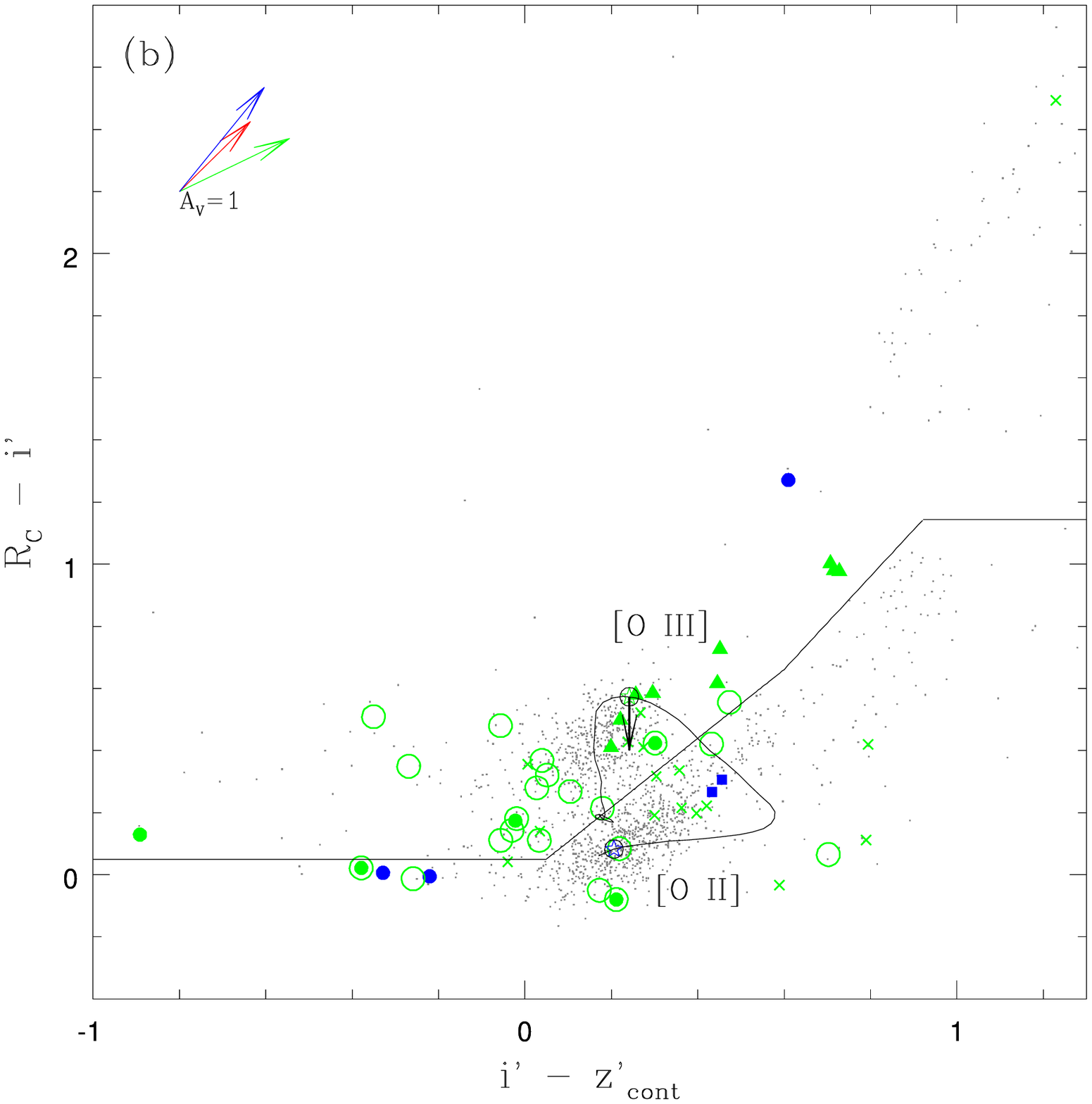}
  \caption{Two-color diagrams in $B-\Rc$ vs. $\Rc-i\arcmin$ ({\it a}) and $\Rc-i\arcmin$ vs. $i\arcmin-z\arcmin_{\rm cont}$
    ({\it b}) for NB921 emitters. In the electronic edition, red points are \Ha, blue points are \Oii, and green points are
    \Oiii~and \Hb~emitters identified by FOCAS (open circles) or DEIMOS (filled circles). Open squares are galaxies from
    the Hawaii HDF-N, filled squares are serendipitous sources, and filled triangles are fortuitous sources. NB704 and
    NB921 dual emitters are shown as crosses. The solid lines in (a) are $B-\Rc=1.46(\Rc-i\arcmin)+0.58$, and
    $\Rc-i\arcmin=0.45$, and in (b) $\Rc-i\arcmin=0.05$, $\Rc-i\arcmin=1.11(i\arcmin-z\arcmin_{\rm cont})-0.01$,
    $\Rc-i\arcmin = 1.5(i\arcmin-z\arcmin_{\rm cont})-0.24$, and $\Rc-i\arcmin=1.14$.
    A theoretical model from \citet{bc03} with constant star-formation (without dust extinction) is shown by the
    solid black curve for $z=0$ to 1.5. Along this curve, the broad-band colors at specific redshifts of 0.40, 0.84,
    and 1.45 are shown by starred symbols with a circle surrounding it. Black vectors at these points indicate the
    direction that the colors follow with different emission line strengths. The vectors in the upper left-hand corner
    correspond to 1 magnitude of V extinction using the reddenning curve of \citet{cardelli89}.}
  \label{fig7}
\end{figure}

In Figure~\ref{fig7}a, the $B-\Rc$ and $\Rc-i\arcmin$ colors for NB921 emitters are shown. Two $z$ = 0.40 \Ha~(red
circles), and 196 NB704 and NB921 emitters at $z$ = 0.40 are plotted as red crosses while 16 $z$ = 0.89 NB704
and NB921 \Hb~emitters are plotted as green crosses. The two types of NB704 and NB921 emitters are distinguished
by their similarities in broad-band colors with galaxies that have been identified spectroscopically. \Ha~NB921 emitters
are identified for having $B-\Rc \geq 1.46(\Rc-i\arcmin) + 0.58$ and $\Rc - i\arcmin \leq 0.45$. In Figure~\ref{fig7}b,
NB921 emitters that are not identified as \Ha~are plotted (as grey points) in $\Rc~-~i\arcmin$ vs.
$i\arcmin - z\arcmin_{\rm cont}$, where $z\arcmin_{\rm cont}$ accounts for a brighter measurement in $z\arcmin$ due to
a bright emission line (see Equation~\ref{eqn1} below). This correction will shift points bluer. The total number of
NB921 \Ha, \Oiii, and \Oii~emitters is 337, 655, and 899, respectively.

The amount of contamination of \Oii~and \Oiii~or \Hb~into our \Ha~selection criteria are 1/12 (8\%) and 5/209 (2\%),
respectively. And from the SDF spectroscopic sample, the contamination amount is 6/32 (19\%) of \Oiii~or \Hb~into
\Oii~and 1/5 (20\%) of \Oii~into \Oiii. The \Oii~NB921 contamination rate is higher due to small statistics. The low
contamination rates for all four NB filters show that the method of determining \Ha, \Oiii, and \Oii~is highly reliable.

\subsection{Averaged Spectral Energy Distributions}
\begin{figure}
  \plottwo{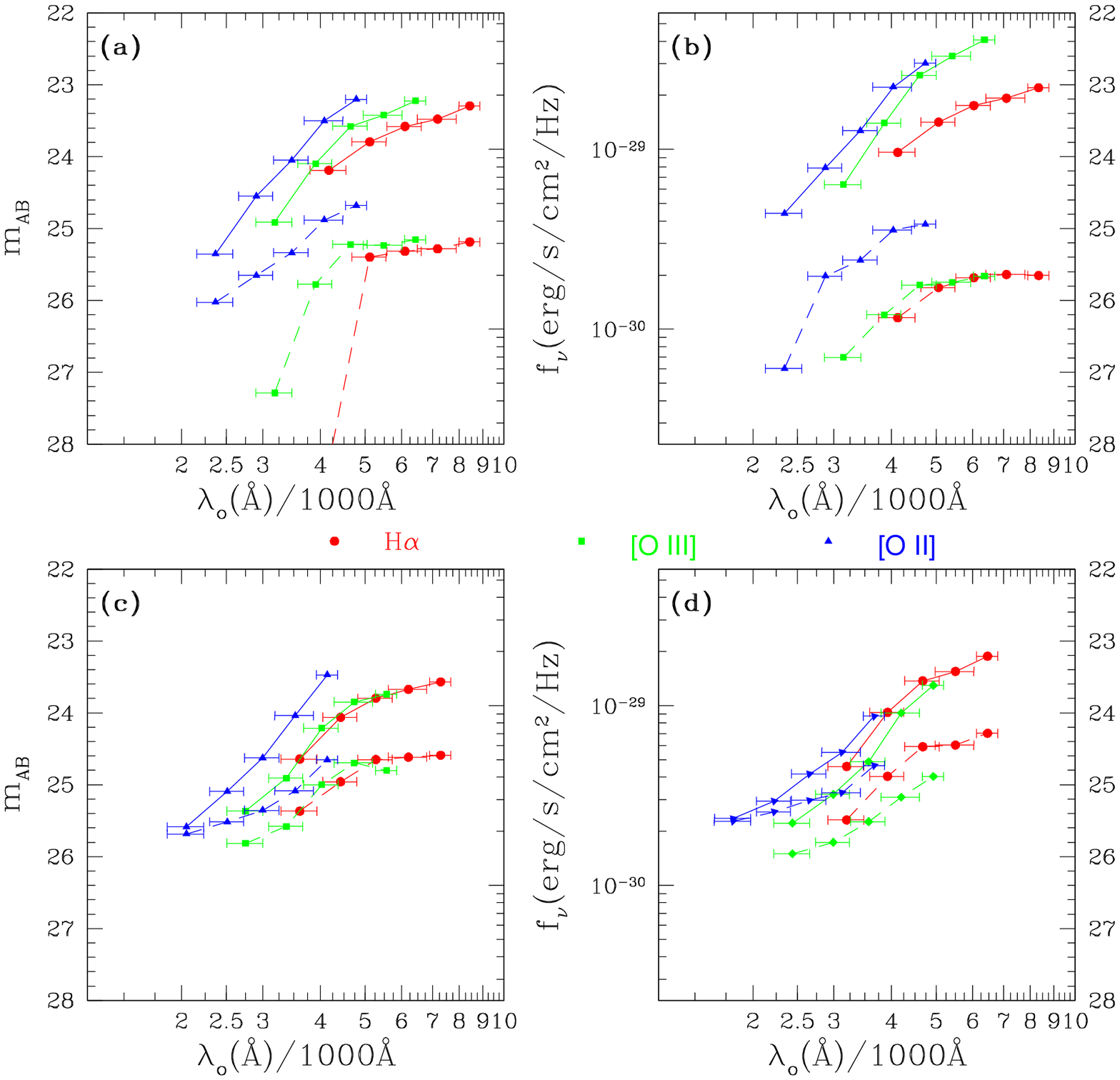}{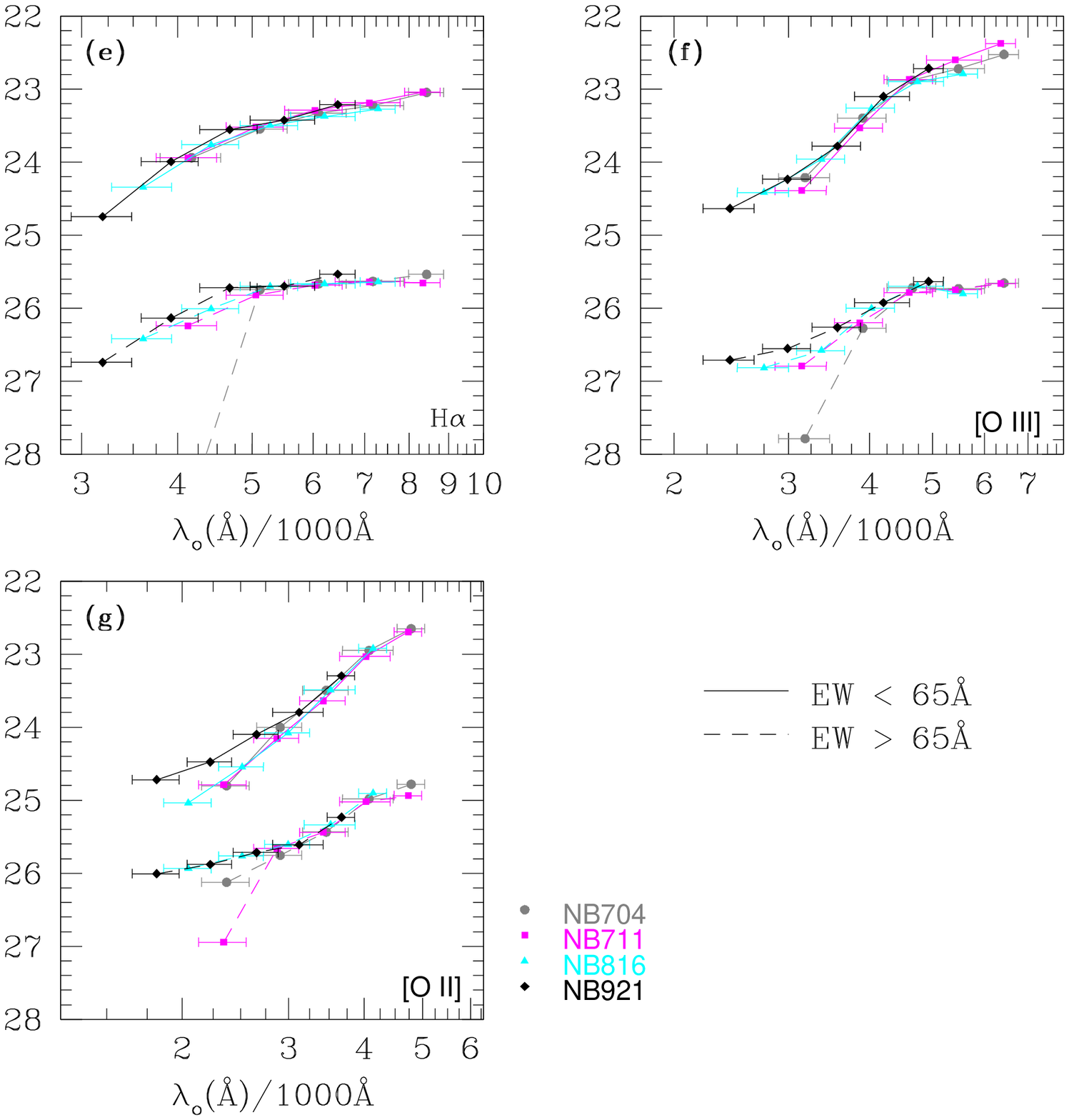}
  \caption{Spectral energy distributions for ({\it a}) NB704, ({\it b}) NB711, ({\it c}) NB816, and ({\it d}) NB921 emitters.
    The outer ordinates give the AB magnitudes and the inner ones are flux densities in ergs s$^{-1}$~cm$^{-2}$~Hz$^{-1}$.
    The rest wavelengths of the BB filters are given on the absiccas. In the electronic edition, red, green, and blue points
    correspond to \Ha~(circles), \Oiii~(squares), and \Oii~(triangles) line emitters. The solid and dashed lines correspond
    to low- and high-EWs, respectively. The division is made at observed EW of 65\AA. The SEDs for \Ha, \Oiii, and
    \Oii~emitters are compared in ({\it e}), ({\it f}), and ({\it g}), respectively, where vertical shifts are applied to
    overlap them. In (e)-(g), NB704, NB711, NB816, and NB921 emitters are given by grey circles, magenta squares, cyan
    triangles, and black diamonds, respectively in the electronic edition.}
  \label{fig8}
\end{figure}

Based on the BB color selection, averaged rest-frame optical to UV SEDs are shown in Figure~\ref{fig8}a-d for each type
of line emitters (\Ha, \Oiii, and \Oii). SEDs of high and low observed equivalent widths (EWs) are provided where
the division is made at 65\AA~(see \S\S~\ref{3.3} for a description of determinating EWs).

All high-EWs sources are bluer (flatter spectral index) compared to the low-EW sources. This is rather apparent for
the \Oii~emitters. This is not surprising as very high star-forming galaxies are expected to be blue. In addition, the
NB816 high-EW \Oiii~SED appears to peak in the $i$\arcmin~bandpass, which indicates that the \Oiii~lines may be stronger
relative to the continuum at $z=0.64$. The \Ha~SEDs show little differences among all four filters (i.e., redshifts from
$z=0.07$ to 0.4). A comparison of these SEDs with models used in photometric redshift algorithm (such as {\it hyperz}) and
a more detail analysis of these SEDs will be discussed in a future paper.

\subsection{The Luminosity Function}
\label{3.3}
\begin{figure}
  \epsscale{0.75}
  \plotone{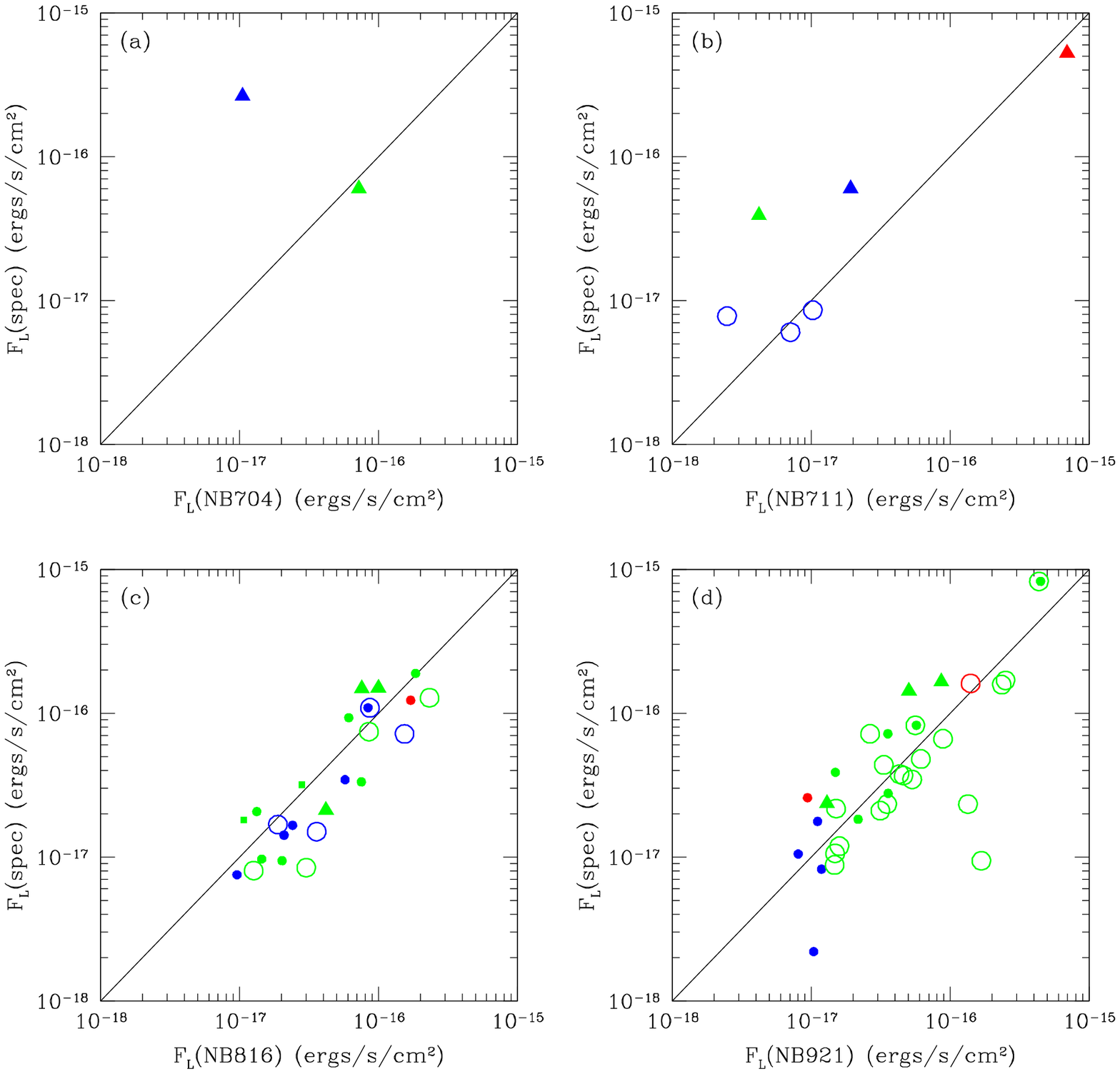}
  \caption{Comparison between spectroscopic and photometric line fluxes for ({\it a}) NB704, ({\it b}) NB711, ({\it c})
    NB816, and ({\it d}) NB921 emitters. Line fluxes are given in ergs s$^{-1}$ cm$^{-2}$. In the electronic edition, red
    points are \Ha, blue points are \Oii, and green points are \Oiii~and \Hb~emitters identified by FOCAS (open circles)
    or DEIMOS (filled circles). Filled triangles and squares are fortuitous and serendipitous sources, respectively. The
    solid lines represent one-to-one correspondence.}
  \label{fig9}
\end{figure}

The total NB flux density (in units of ergs s$^{-1}$~cm$^{-2}$~\AA$^{-1}$) can be defined as $f_{\rm NB}=f_C+F_L/\Delta$NB,
where $f_C$ is the continuum flux density (ergs s$^{-1}$~cm$^{-2}$~\AA$^{-1}$), $F_L$ is the emission line flux
(ergs s$^{-1}$~cm$^{-2}$), and $\Delta$NB is the width of the NB filter. The broad-band continuum flux density ($f_{BB}$)
is $f_{Ri}=f_C+\epsilon_1 F_L/\Delta R$ (NB704 or NB711, $\epsilon_1=0.5$), $f_{iz}=f_C+\epsilon_2 F_L/\Delta i\arcmin$
(NB816, $\epsilon_2=0.6$), and $f_{z}=f_C + \epsilon_3 F_L/\Delta z\arcmin$ (NB921, $\epsilon_3=1.0$). Here, $\epsilon_i$,
the weight of a broad-band filter to determine the broad-band continuum, is introduced to maintain generality
in the following equations. The widths are $\Delta$NB704 = 100\AA, $\Delta$NB711 = 73\AA, $\Delta$NB816 = 120\AA,
$\Delta$NB921 = 132\AA, $\Delta R$ = 1124\AA, $\Delta i\arcmin$ = 1489\AA, and $\Delta z\arcmin$ = 955\AA. Therefore the
line flux, continuum flux density, and observed equivalent widths are
\begin{eqnarray}
  \label{eqn1}
  F_{L} = &\Delta {\rm NB} \frac{f_{\rm NB} - f_{BB}}{1-\epsilon(\Delta {\rm NB}/\Delta BB)},\\
  \label{eqn2}
  f_C = &\frac{f_{BB} - \epsilon f_{\rm NB}(\Delta {\rm NB}/\Delta BB)}
  {1-\epsilon(\Delta {\rm NB}/\Delta BB)},{\rm~and}\\
  \label{eqn3}
  \rm{EW_{obs}} = &\frac{F_L}{f_C} = \Delta {\rm NB}\left[\frac{f_{\rm NB} - f_{BB}}
    {f_{BB} - \epsilon f_{\rm NB}(\Delta {\rm NB}/\Delta BB)}\right].
\end{eqnarray}
The limiting line fluxes for NB704, NB711, NB816, and NB921 are 5.3, 6.5, 6.3, and
5.7$\times$10$^{-18}$ ergs s$^{-1}$~cm$^{-2}$, respectively. \citet{fujita03} reach a limiting line flux that is twice as
bright as what is reported here for NB816 emitters. For a NB excess of 0.1 (NB704), 0.1 (NB711), 0.25 (NB816), and 0.1
mag (NB921), the observed EW is 10, 7, 33, and 15\AA, respectively. In Figure~\ref{fig9}, line fluxes derived from
photometry are compared to spectroscopic values, showing that the determination of line fluxes from photometry is accurate
over a wide range of line fluxes. The {\it observed} LFs for \Ha, \Oiii, and \Oii~are presented in
\S\S\S~\ref{3.3.1}-\ref{3.3.3} follow by an analysis of the incompleteness of the sample.

\subsubsection{\Ha~Emitters}
\label{3.3.1}
\begin{figure}
  \epsscale{0.32}
  \plotone{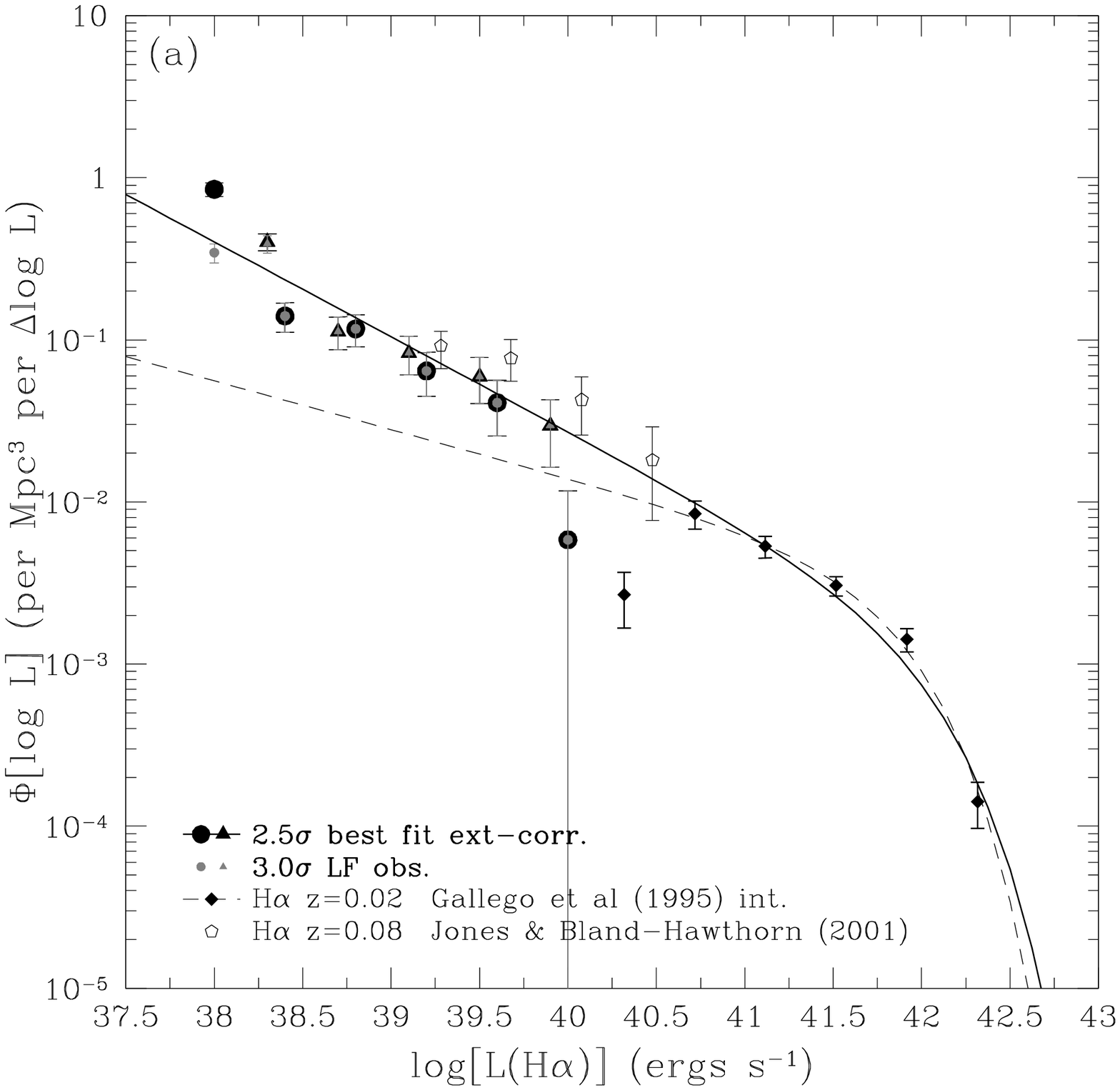}
  \plotone{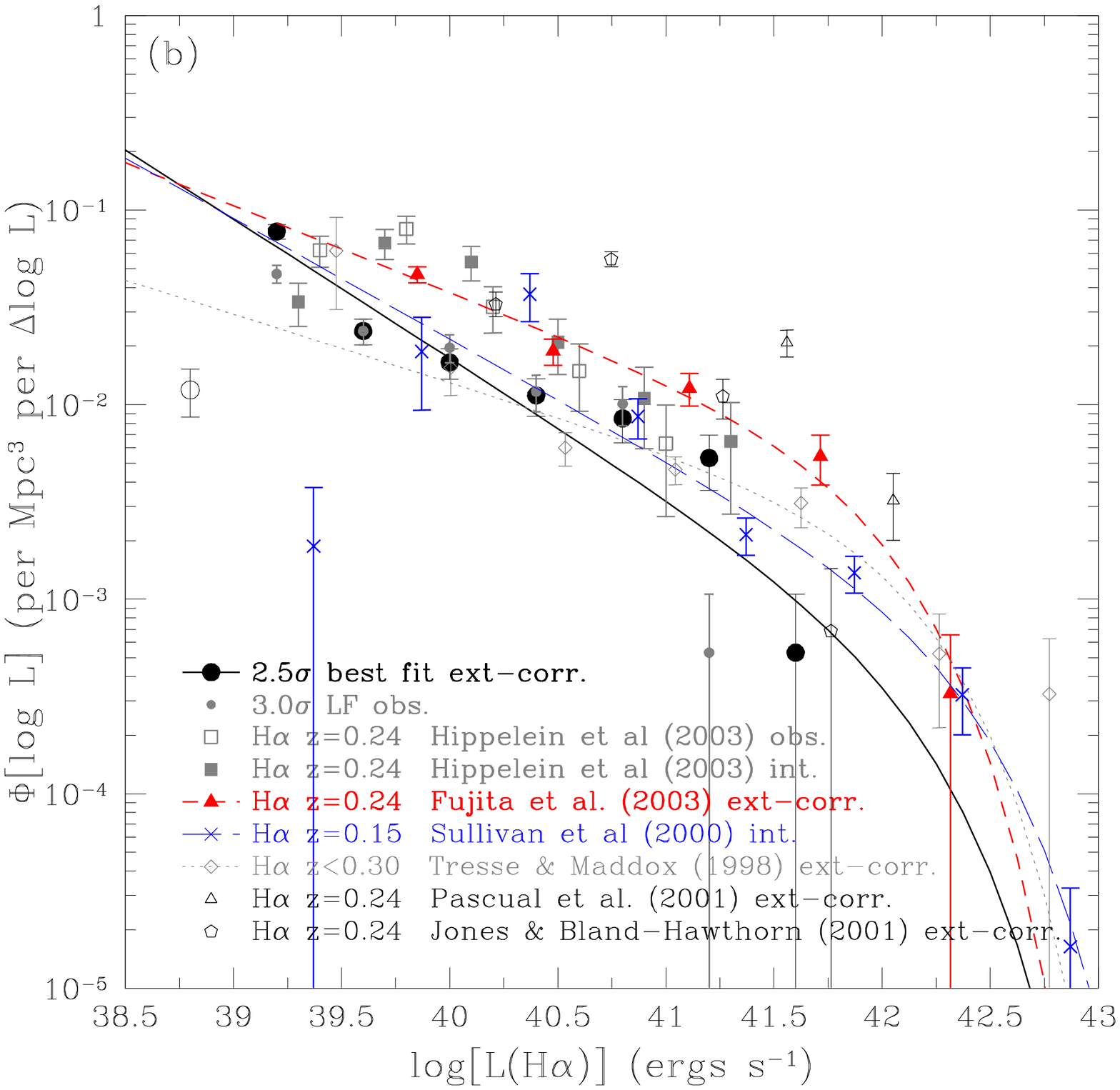}
  \plotone{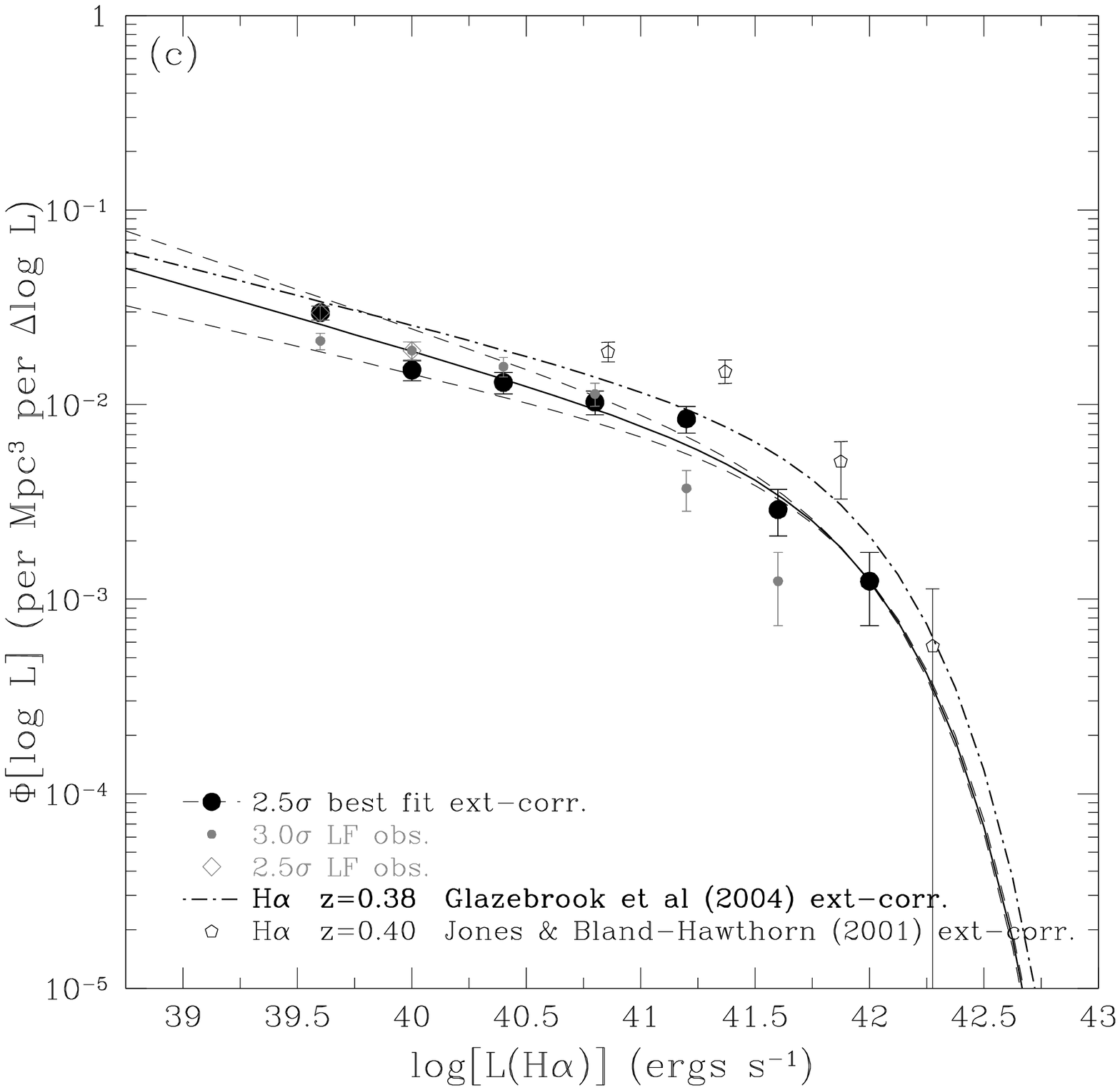}
  \caption{The luminosity function for \Ha~emitters at ({\it a}) $z=0.07-0.09$, ({\it b}) $z\approx0.24\pm0.01$, and
    ({\it c}) $z\approx0.40\pm0.01$. Ordinates are $\Phi[\log{L}]$ in units of Mpc$^{-3}~(\Delta\log{L})^{-1}$. Absiccas
    are $\log{L(\rm{H}\alpha)}$ in ergs s$^{-1}$. The LFs of NB704, NB816, and NB921 are plotted as filled circles,
    and NB711 emitters are plotted as filled triangles in (a), where black and grey filled circles are for the $2.5\sigma$
    extinction-corrected and $3\sigma$ observed sample, respectively. For NB921, the $\pm$1$\sigma$ uncertainties in
    $\alpha$ are shown by the two thin short-dashed black lines, and the 2.5$\sigma$ observed points are shown as open grey
    diamonds to illustrate the effect of incompleteness at the faint end. Open circles at the faint luminosity end
    are $2.5\sigma$ points excluded from the best fit given by the thick solid black lines. The luminosity functions from
    other studies are overlayed. \citet{jones01} at $z=0.08$, 0.24, and 0.40 are shown as open pentagons, \citet{gallego95}
    as short-dashed black lines with open (observed) and filled (intrinsic) diamonds, $z=0.24$ observed \citet{hippelein03}
    points are shown as open grey squares, and $z=0.2\pm0.1$ \citet{TM98} as filled grey diamonds and dotted line. In
    addition, \citet{fujita03} $z=0.24$ is shown as a dashed line and filled triangles (colored red in the electronic edition),
    \citet{sullivan00} as a long-dashed line (colored blue in the electronic edition) with crosses, a dot - short-dashed
    black line for \citet{glazebrook04}, and \citet{pascual05} as open triangles. All values have been converted to
    the common cosmology. Extinction-corrected values follow Equation~\ref{ext} with the exception of \citet{gallego95},
    \citet{TM98}, \citet{sullivan00}, and \citet{hippelein03} where their extinction-corrected LF are used.}
  \label{fig10}
\end{figure}

Since the NB filters include the [\textsc{N ii}] doublet with the \Ha~emission lines, these line flux measurements must be
corrected. It is assumed that $\kappa$, the flux ratio of \Ha~and the [\textsc{N ii}] doublet (\Ha/[\textsc{N ii}]), is
4.66. This is an average flux ratio from 17 DEIMOS spectra between $z=0.08$ and 0.34. \citet{TM98}, \citet{yan99},
\citet{iwamuro00}, and \citet{fujita03} used a flux ratio of 2.3, which is reported by \citet{kennicutt92} and
\citet{gallego97}. In addition, the non-square shape of the NB filters must be accounted for, so a statistical correction
of 28\% is applied for all filters. Therefore the observed luminosity is
$L_{\rm obs}({\rm H}\alpha)=4\pi d_L^2 F_L\frac{1}{1+1/\kappa}\times1.28$. With these corrections, the luminosity function
is constructed by
\begin{eqnarray}
  \Phi(\log{L_i}) = \frac{1}{\Delta\log{L}}\sum_j\frac{1}{V_i}\nonumber \\
  \textrm{with~} |\log{L_j} - \log{L_i}| < \case{1}{2}\Delta\log{L}.
\end{eqnarray}
The number of \Ha~line emitting galaxies per Mpc$^3$ per $\Delta\log{L({\rm H}\alpha)}$ is plotted in Figure~\ref{fig10}
for (a) NB704 and NB711, (b) NB816, and (c) NB921 as small filled grey circles. The logarithmic bin size for \Ha~is
$\Delta\log{L({\rm H}\alpha)} = 0.4$. The comoving volume for each galaxy is corrected for the shape of the filter being
triangular, which can be as high as 27\% of the total accessible volume for the faintest galaxies. The LFs are fitted to
a Schechter profile \citep{schechter76}:
\begin{equation}
  \Phi(L)dL =
  \phi_{\star}\left(\frac{L}{L_{\star}}\right)^{\alpha}\exp{\left(-\frac{L}{L_{\star}}\right)}\frac{dL}{L_{\star}},
\end{equation}
where $\Phi(\log{L})d\log{L} = \Phi(L)dL$. The LF can be integrated to obtain the luminosity density
$\mathcal{L} = \int_0^{\infty} dL L \Phi(L) = \phi_{\star}L_{\star}\Gamma(\alpha+2)$ in ergs s$^{-1}$ Mpc$^{-3}$.

\subsubsection{\textsc{O iii} Emitters}
\label{3.3.2}
Although the strongest \Oiii~line is located at 5007\AA, the measured total line flux includes the 4959\AA~line for
redshifts of 0.411-0.417, 0.430-0.431, 0.631-0.639, and 0.841-0.850. Assuming that $\frac{F_{5007}}{F_{4959}}$ = 3,
the corrected \Oiii~$\lambda$5007 luminosity for redshifts when both lines are in the filter is
$L_{\rm obs}(\textsc{O iii}) = 4\pi d_L^2 F_L\frac{3}{4}\times1.28$.
From the SDF spectroscopic sample (excluding fortuitous and serendipitous sources) of \Oiii~emitters, it is statistically
estimated that 4/10 (NB816) and 2/22 (NB921) include both \Oiii~lines. If each NB filter is divided into 5\AA~bins, and
the redshift distribution is uniform across the filter, then the fraction of detected line emitters where both lines
are present is 6/24 (NB704), 1/15 (NB711), 8/24 (NB816), and 8/26 (NB921) must have their luminosity reduced by 25\%.
However, the non-square shape of the filters will lower the percentage for objects detected in these redshift intervals.
Accounting for the filters' shape, the corrections are 21.2\% (NB704), 3.3\% (NB711), 23.3\% (NB816), and 22.0\% (NB921).
It should be noted that these corrections should affect lower luminosity objects, but due to the degeneracy of line fluxes
(truly faint versus off-center from $\bar z$), these corrections are applied regardless of their line flux.

\begin{figure}[htp]
  \epsscale{1.17}
  \plottwo{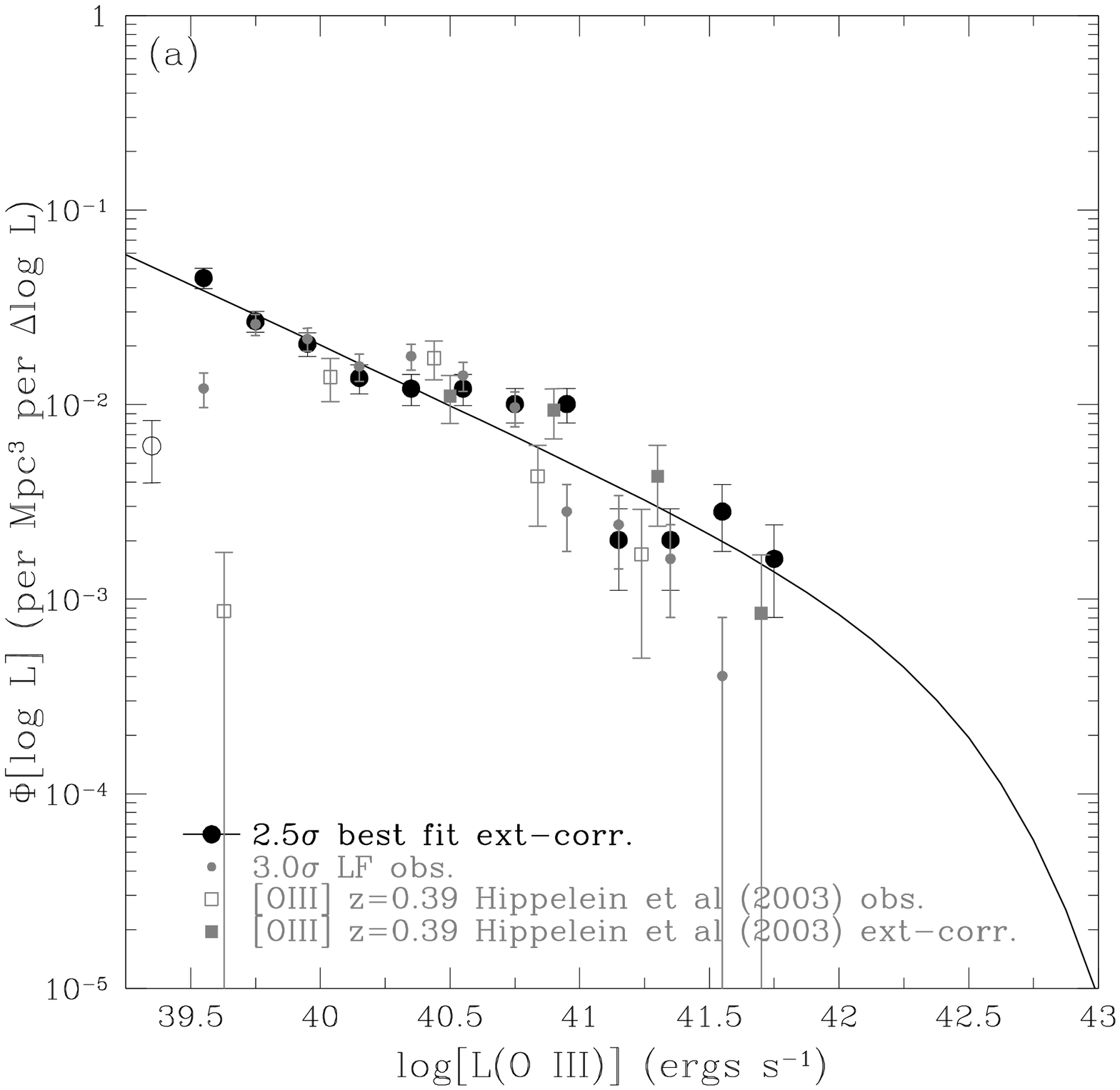}{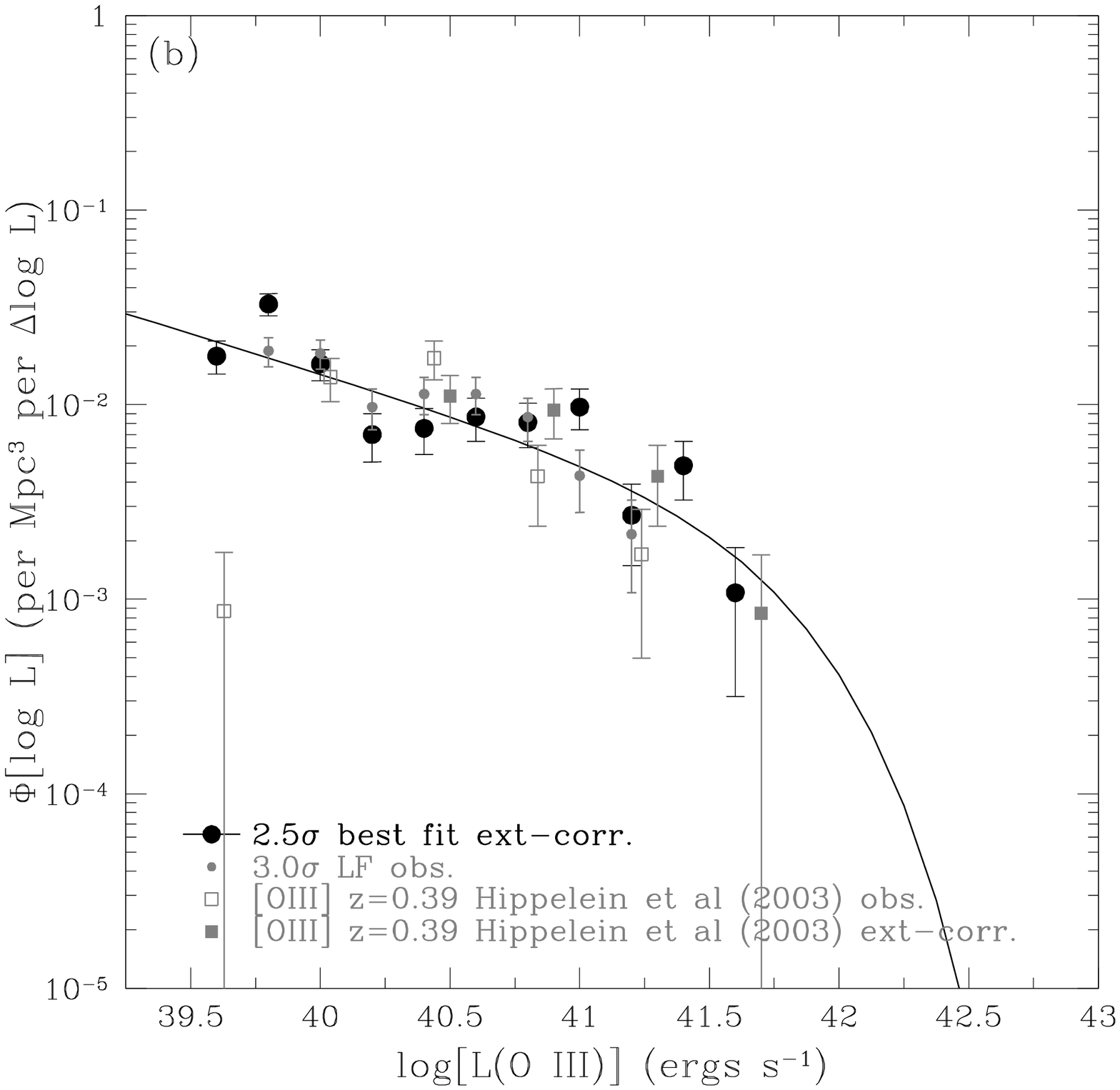}\\
  \plottwo{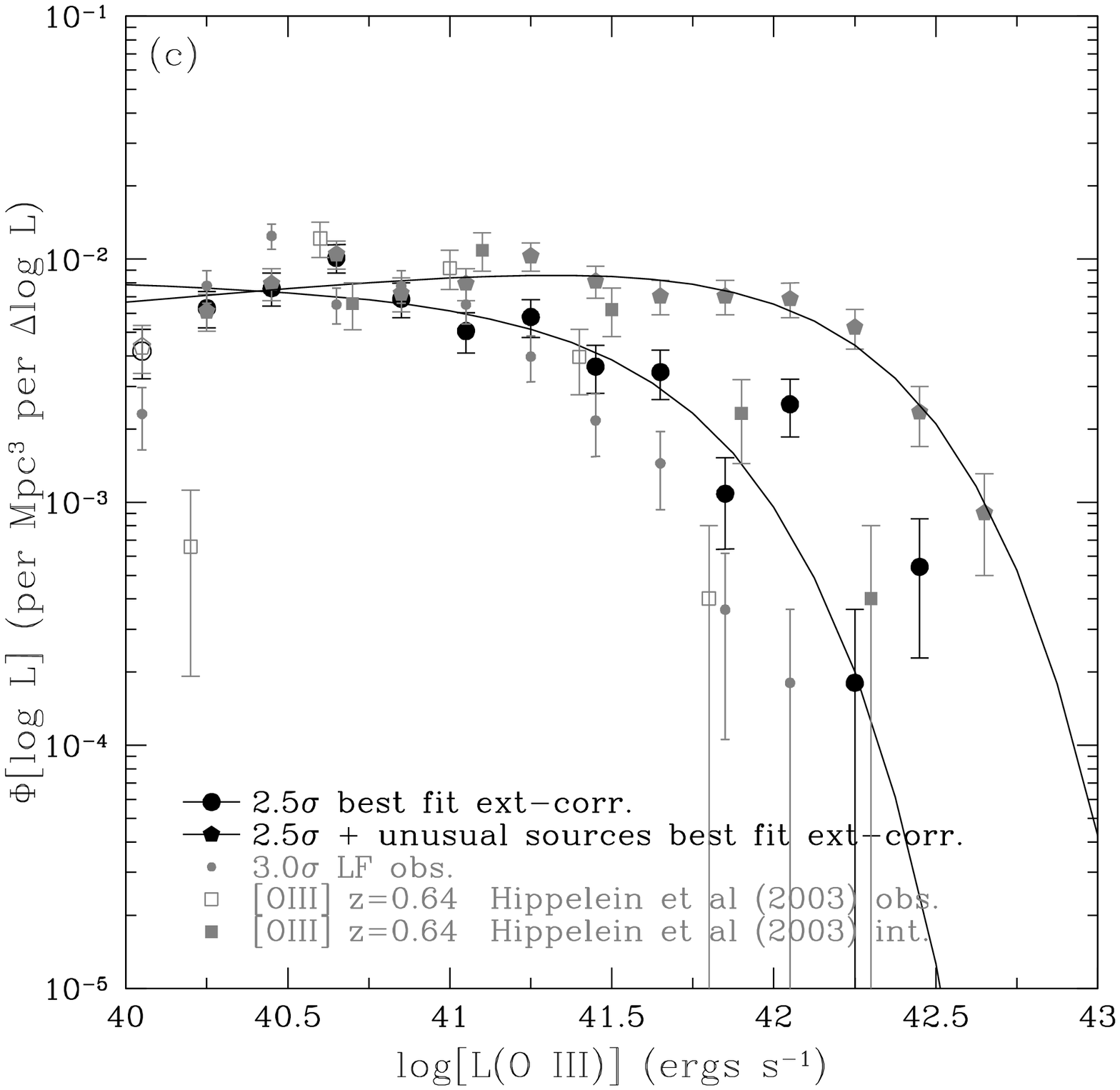}{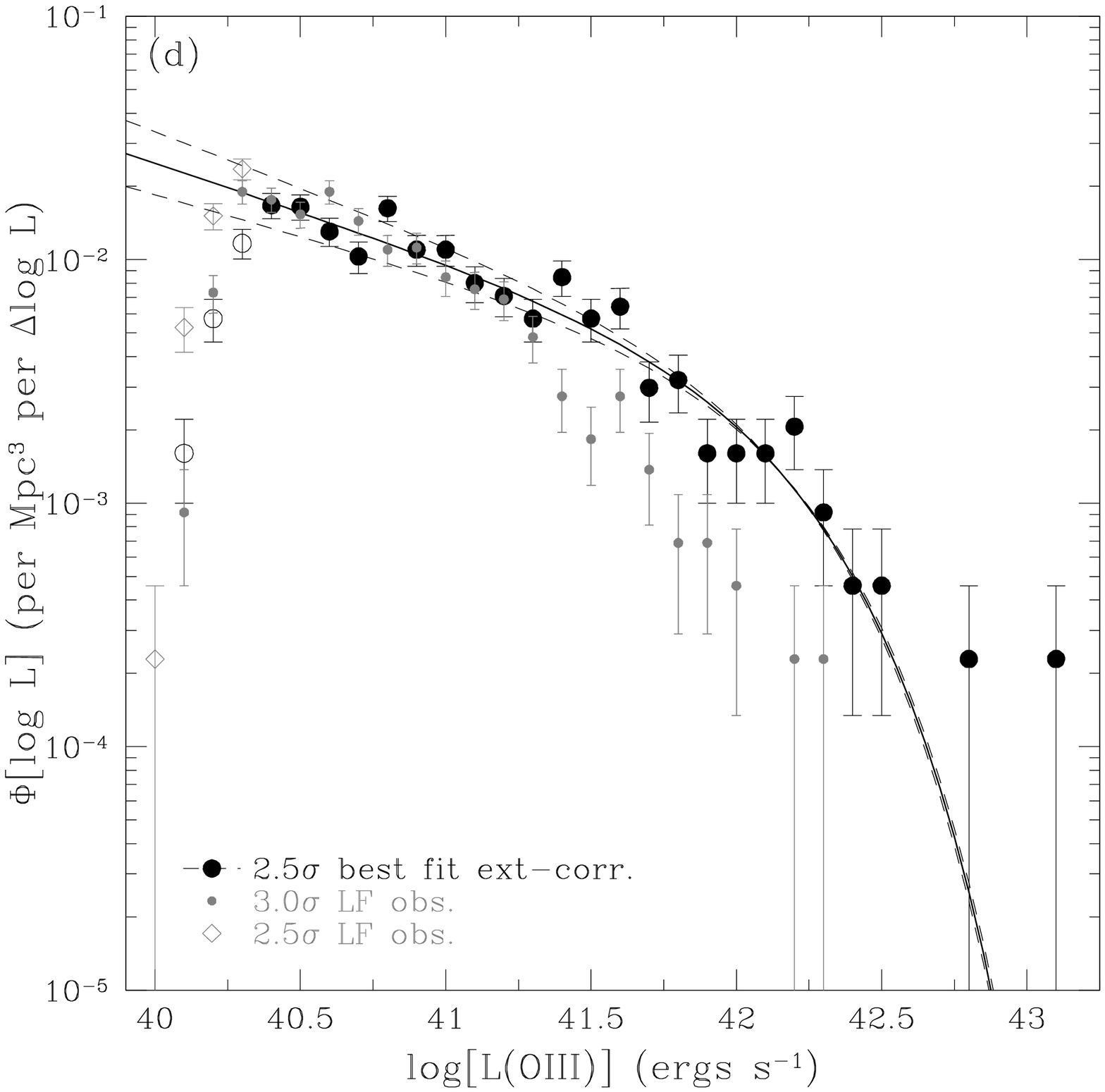}
  \caption{The luminosity function for \Oiii~emitters at (a) $z\approx0.41\pm0.01$, (b) $z\approx0.42\pm0.01$, (c)
    $z\approx0.63\pm0.01$, and (d) $z\approx0.84\pm0.01$. Ordinates are $\Phi[\log{L}]$ in units of
    Mpc$^{-3}~(\Delta\log{L})^{-1}$. Absiccas are $\log{L(\textsc{O iii})}$ in ergs s$^{-1}$. The LFs of NB704,
    NB711, NB816, and NB921 are plotted as filled circles for the $2.5\sigma$ extinction-corrected (black) and $3\sigma$
    observed (grey) sample. For NB921, the $\pm$1$\sigma$ uncertainties in $\alpha$ are shown by the two thin short-dashed
    black lines, and the 2.5$\sigma$ observed points are shown as open grey diamonds to illustrate the effect
    of incompleteness at the faint end.
    Open circles at the faint luminosity end are $2.5\sigma$ points excluded from the best fit given by the thick solid
    black lines. For NB816, the filled grey pentagons and solid line are the LF including the 192 unknown sources. The
    luminosity functions of \citet{hippelein03} at $z=0.39-0.41$ and 0.63 - 0.65 are shown as open (observed) and filled
    (extinction-corrected) grey squares.}
  \label{fig11}
\end{figure}

The luminosity functions for \Oiii~are shown in Figure~\ref{fig11} for NB704 (a), NB711 (b), NB816 (c), and NB921 (d) as
small filled grey circles. The LF are binned into $\Delta\log{(\textsc{O iii})} = 0.2$ except for the NB921 LF where 0.1
is used. The Schechter parameters for the \Oiii~NB816 LF including the unknown sources are $\log{L_{\star}}=41.69\pm0.08$,
$\log{\phi_{\star}}=-2.16\pm0.07$, and $\alpha=-0.87\pm0.08$.

\subsubsection{\textsc{O ii} Emitters}
\label{3.3.3}
\begin{figure}[htp]
  \epsscale{1.17}
  \plottwo{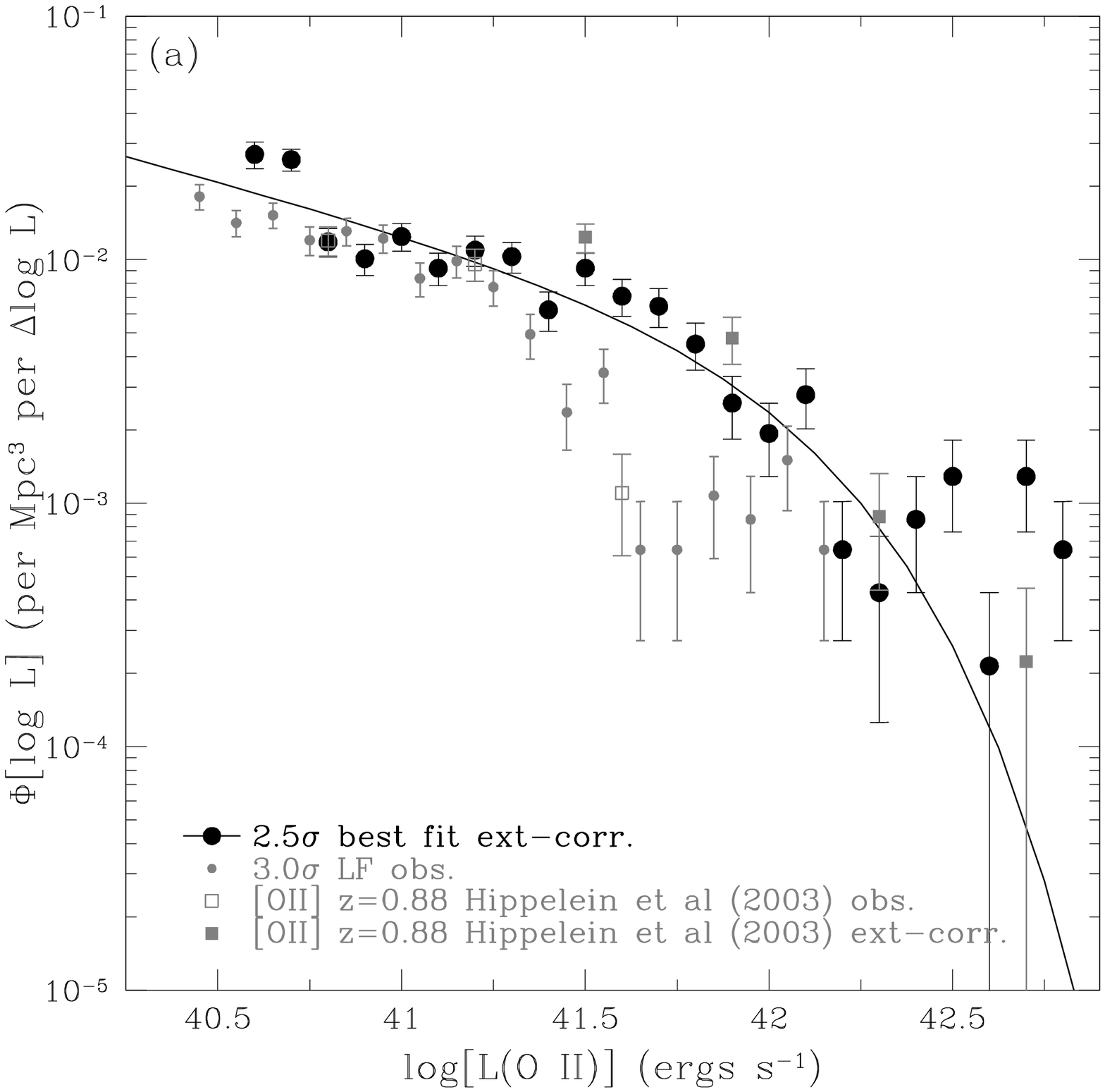}{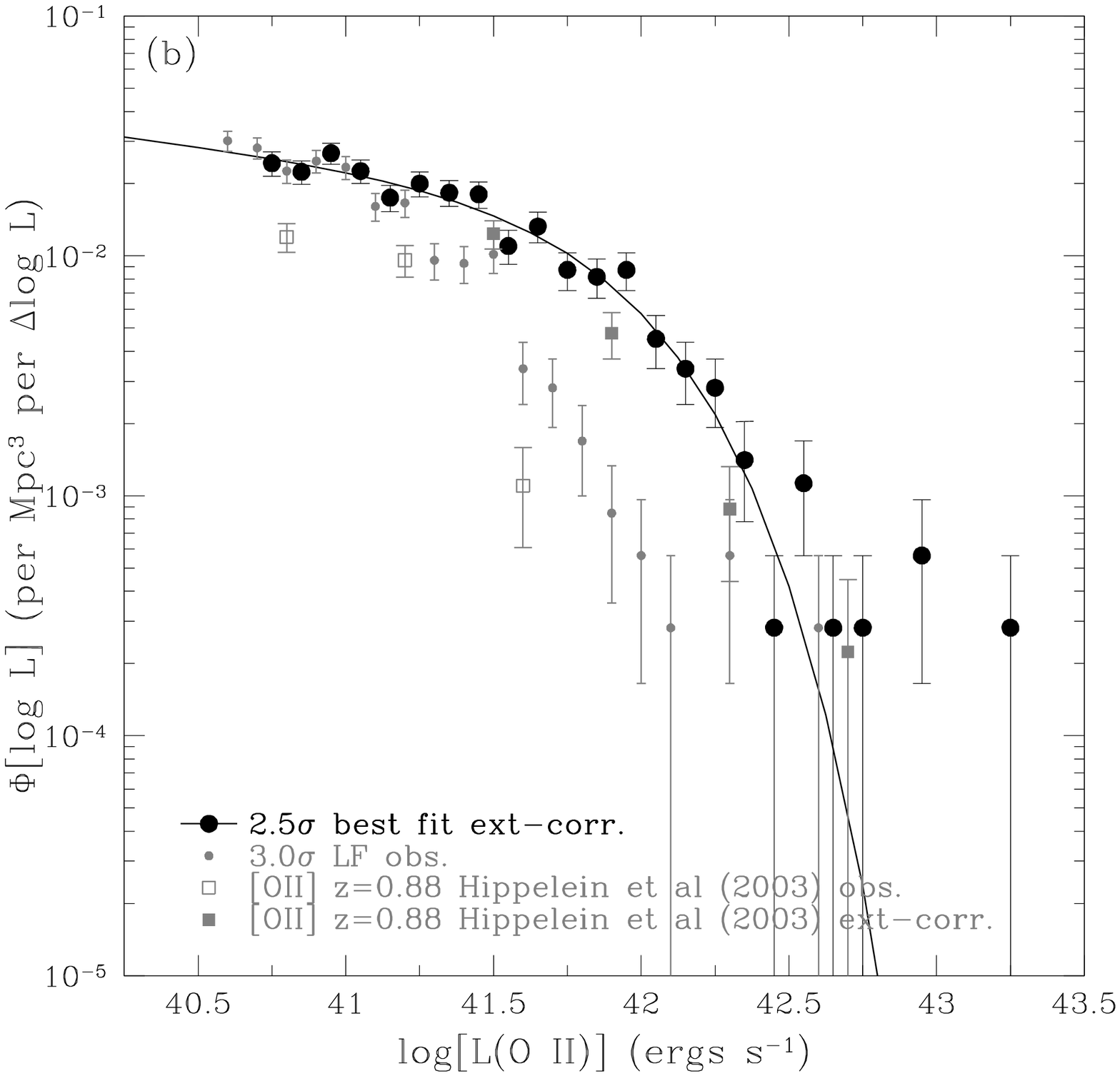}\\
  \plottwo{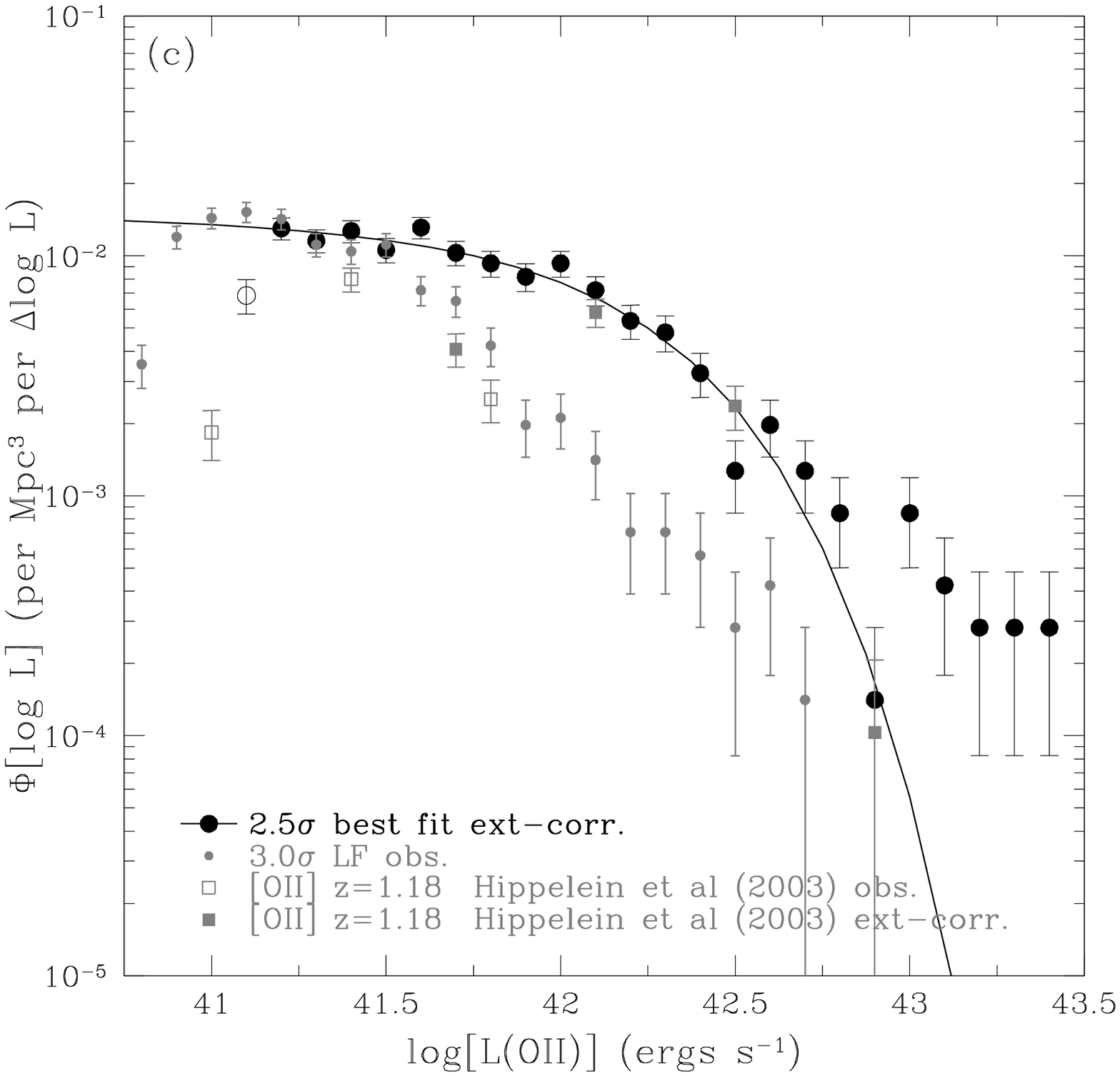}{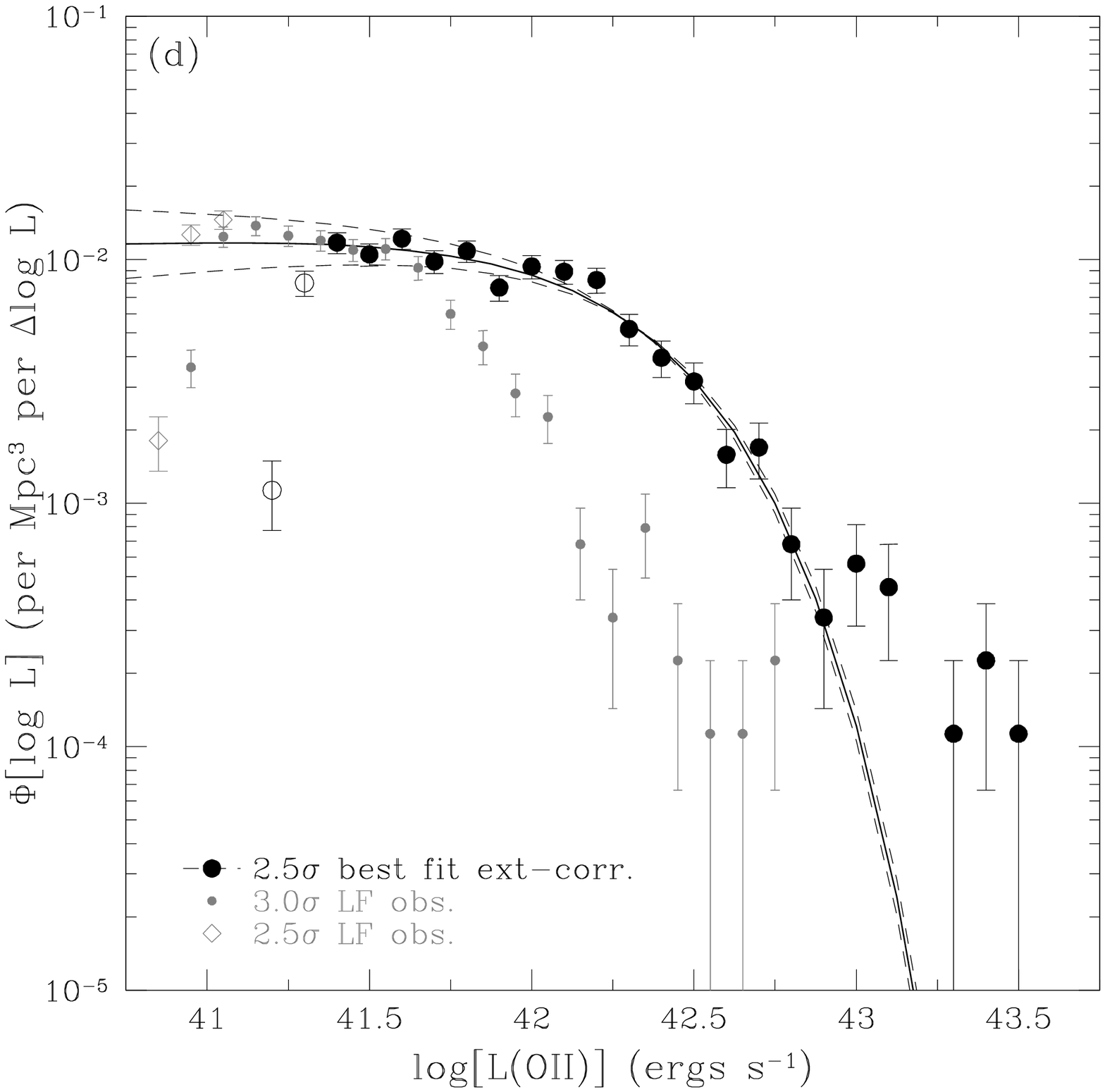}
  \caption{The luminosity function for \Oii~emitters at (a) $z\approx0.89\pm0.01$, (b) $z\approx0.91\pm0.01$, (c)
    $z\approx1.19\pm0.02$, and (d) $z\approx1.47\pm0.02$. Ordinates are $\Phi[\log{L}]$ in units of
    Mpc$^{-3}~(\Delta\log{L})^{-1}$. Absiccas are $\log{L(\textsc{O ii})}$ in ergs s$^{-1}$. The LFs for NB704, NB711, NB816,
    and NB921 are plotted as filled circles for the $2.5\sigma$ extinction-corrected (black) and $3\sigma$ observed (grey)
    sample. Open circles at the faint end are $2.5\sigma$ points excluded from the best fit given by the thick solid black
    lines. For NB921, the $\pm$1$\sigma$ uncertainties in $\alpha$ are shown by the two thin short-dashed black lines,
    and the 2.5$\sigma$ observed points are shown as open grey diamonds to illustrate the effect of incompleteness at the
    faint end. The luminosity functions of \citet{hippelein03} at $z=0.87-0.89$ and 1.18 - 1.21 are shown as open
    (observed) and filled (extinction-corrected) grey squares.}
  \label{fig12}
\end{figure}

Fortunately there are no strong lines near \Oii~that fall within the NB filters, so the NB filter measures only the
\Oii~line and underlying continuum: $L_{\rm obs}(\textsc{O ii}) = 4\pi d_L^2 F_L\times1.28$. The luminosity functions for
\Oii~are shown in Figure~\ref{fig12} as small filled grey circles. A bin size of $\Delta\log{L(\textsc{O ii})} = 0.1$ is
used.

\subsection{Completeness of the Luminosity Functions}
\label{3.4}
One major question when constructing these luminosity functions is the completeness at the faint luminosity end.
A common technique used to determine completeness is to distribute artificial sources on the images and then
see what fraction of those are detected with SExtractor. \citet{kashik06} determined that the completeness for the
SDF NB921 image used in this paper is about 50\% at 26.0 mag. Since the depth of the other images is comparable to the
NB921 or half a magnitude shallower, the completeness of these filters are given by adjusting the NB magnitude of
50\% completeness based on the limiting magnitude of the images. This scaling implies that the magnitude for 50\%
completeness is 26.0 for NB704, 25.5 for NB711, and 26.1 for NB816. Using the number of NB emitters within a
$\Delta$NB = 0.25 mag and the completeness curve as a function of magnitude \citep[available at][]{kashik04}, the
number of NB emitters missed due to incompleteness is 109 (NB704), 137 (NB711), 160 (NB816), and 214 (NB921).

As a consistency check, the amount of incompleteness can also be determined by loosening the 3$\sigma$ BB-NB excess
selection criteria to a depth of 2.5$\sigma$. The larger 2.5$\sigma$ limits are shown as dashed red lines in
Figure~\ref{fig1}a-d. This threshold results in detecting an additional 203, 188, 196, and 226 NB emitters for NB704,
NB711, NB816, and NB921, respectively. Thus the 2.5$\sigma$ method and the first method of artificial adding sources to
the images yield similar results. For example, the NB816 and NB921 values are higher than method 1 by 22\% and 6\%,
respectively. By extending the sample to 2.5$\sigma$, additional sources with an emission in the NB filter will be detected,
but there will also be spurious detections. The predicted number of additional spurious detections based on Gaussian
statistics is about 1\%, which is rather optimistic. Even with 10\% false detections, for example, this would mean that
20 sources are spurious (for each NB filter), but the remaining $\sim180$ sources are real NB emitters. Only spectra of
these faint emission lines can determine how many spurious sources exist at 2.5$\sigma$.

After re-identifying the NB emitters down to 2.5$\sigma$, the broad-band color classification described in
\S\S~\ref{3.1} are applied and the luminosity functions are then recalculated for each line emitter in all four NB filters.
For the completeness-corrected Schechter fits provided in Table~\ref{table4}, this 2.5$\sigma$ method is adopted as it
shows the effects of incompleteness on the faint end slope. For the lowest luminosity bins (typically one to three,
see Figures~\ref{fig10}-\ref{fig12}), the completeness of the \Ha~$3\sigma$ sample as determined from this method
is 52\% (NB704, $\log{L}$=38.0), 67\% (NB711, $\log{L}$=37.9 and 38.3), 59\% (NB816, $\log{L}$=39.2), and 72\% (NB921,
$\log{L}$=39.60). For \Oiii, the completeness is 36\% (NB704, $\log{L}$=39.55), 58\% (NB711, $\log{L}$=39.8), 66\% (NB816,
$\log{L}$=40.05 and 40.25), and 62\% (NB921, $\log{L}$=40.1, 40.2, and 40.3). Finally, the \Oii~completeness is 68\%
(NB704, $\log{L}$=40.45), 51\% (NB711, $\log{L}$=40.5), 54\% (NB816, $\log{L}$=40.8 and 40.9), and 55\% (NB921, $\log{L}$=40.95
and 41.05). These fits must be corrected for reddenning to derive the star formation rate density.

\subsection{Correcting for Dust Extinction}
UV and optical measurements of the SFR are subject to significant dust obscuration. The amount of extinction can be estimated
by comparing the observed and intrinsic Balmer decrements ($F_{\rm H\alpha}/F_{\rm H\beta}$), but this should be done for
each individual galaxy. Generally, spectroscopy of all sources is not feasible, so to mitigate this problem, some studies
apply a constant extinction throughout. For example, \citet{fujita03} and \citet{pascual05} adopted one magnitude of
extinction for \Ha~as determined by \citet{kennicutt92}. However, it has been apparent that more active star-forming
galaxies suffer greater obscuration. In particular, \citet{jansen01} and \citet{aragon03} reveal that the Balmer-decrement
derived color excess $E(B-V)$ depends on the $M_B$ magnitude, and \citet{hopkins01} find a dependence of the color excess
on the FIR luminosity. So far, observed SFRs have been reported.

To correct for extinction, three methods are considered. First, a standard $A_{H\alpha}$ = 1 is applied. The second method
utilizes available DEIMOS spectra. From 16 sources with redshift between 0.29 and 0.40, the average Balmer decrement after
correcting for stellar absorption\footnotemark[11] is $4.38\pm1.86$. The color excess can be determined from
\begin{equation}
  E(B-V) = \frac{\log{(R_{\rm obs}/R_{\rm int})}}{0.4[k(\rm{H}\beta) - k(\rm{H}\alpha)]},
\end{equation}
where $R_{\rm obs}$ and $R_{\rm int}$ are the observed and intrinsic Balmer decrements.
\footnotetext[11]{Following \citet{kennicutt92}, 5\AA~of stellar absorption is assumed for all objects.}
The latter is 2.86 for Case B recombination \citep{osterbrock89}. $k(\lambda) = A(\lambda)/E(B-V)$ is the reddenning curve
of \citet{cardelli89} with $k({\rm H}\beta)=3.61$ and $k(\rm{H}\alpha)=2.54$. This corresponds to a color excess of
$E(B-V)=0.43\pm0.43$ or $A_{H\alpha}=1.44\pm1.43$, which is reasonable compared to other studies that obtain
$A_{H\alpha}=0.5-1.8$ \citep[and references therein]{kennicutt98}. The last method is the SFR-dependent correction of
\citet{hopkins01}:
\begin{eqnarray}
  \nonumber
  \log{SFR_{\rm obs}(H\alpha)}&=\log{SFR_{\rm int}}-2.360\\
  &\times\log{\left[\frac{0.797\log{(SFR_{\rm int})}+3.786}{2.86}\right]},
  \label{ext}
\end{eqnarray}
where differences are due to cosmological corrections and using a Cardelli reddenning curve rather than the Calzetti
reddenning curve (Hopkins, priv. comm). For \Oii~and \Oiii~emitters, Equation~\ref{ext} can be used so long as their
luminosities are converted to \Ha~prior to properly correcting for reddenning. The conversions are given in \S\S~\ref{3.6}.
The Schechter fits with the extinction correction of Equation~\ref{ext} are provided in Table~\ref{table4} and are
shown in Figure~\ref{fig10}-\ref{fig12} as filled black circles. For the \Ha~NB711 LF, filled triangles are used to
distinguish from the \Ha~NB704 emitters. For \Oiii~measured in NB816, the extinction and completeness-corrected
Schechter parameters including the unknown sources are $\log{L_{\star}}=42.23\pm0.07$, $\log{\phi_{\star}}=-2.27\pm0.06$,
and $\alpha=-0.88\pm0.05$. The first two methods of extinction corrections are also reported in this table.

\subsection{The Star Formation Rate Density}
\label{3.6}
The following conversions of luminosity to star-formation rate density $\dot\rho_{\rm SFR}$ (in
$M_{\sun}$ yr$^{-1}$ Mpc$^{-3}$) are assumed:
\begin{eqnarray}
  \label{eqn8}
  \dot\rho_{\rm SFR}({\rm H}\alpha)  = &7.9\times10^{-42}\mathcal{L}({\rm H}\alpha),\phm{ (\pm 0.4)~and}\\
  \label{eqn9}
  \dot\rho_{\rm SFR}(\textsc{O ii})  = &(1.4 \pm 0.4)\times10^{-41}\mathcal{L}(\textsc{Oii}),\rm{~and}\\
  \label{eqn10}
  \dot\rho_{\rm SFR}(\textsc{O iii}) = &(7.6 \pm 3.7)\times10^{-42}\mathcal{L}(\textsc{Oiii}),\phm{~and}
\end{eqnarray}
where the \Ha~and \Oii~conversions are from \citet{kennicutt98}. The \Ha~SFR conversion assumes a Salpeter initial mass
function (IMF) with masses between 0.1 and 100 M$_{\sun}$. The \Oii~conversion is from local emission line studies with
an \Oii/\Ha~=~0.57.

\begin{figure}
  \epsscale{0.65}
  \plotone{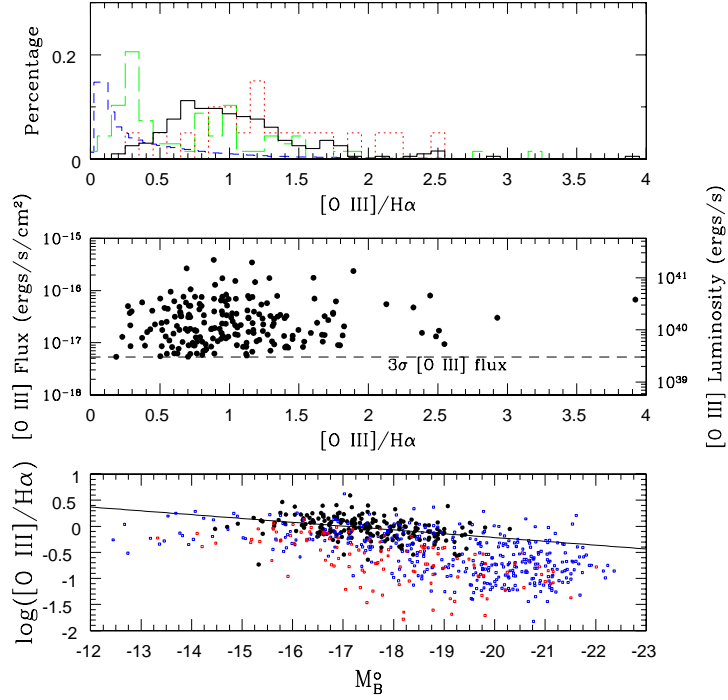}
  \caption{\Oiii-\Ha~flux ratio for NB704+921 emitters. The {\it top} diagram shows a solid black histogram for 196
    $z\approx0.40$ NB704+921 line emitters. The dotted and long-dashed histograms are from \citet{hippelein03} for
    20 $z=0.40$ and 68 $z=0.25$ objects, respectively (colored red and green in the electronic edition). The SDSS DR2
    histogram is shown as a short-dashed histogram (colored blue in the electronic edition). The \textit{middle} figure
    shows the \Oiii~flux and luminosity as a function of the ratio. The \textit{bottom} figure shows the logarithm of
    the ratio as a function of $M_B^o$. In the electronic edition, the red and blue open squares are nearby star-forming
    galaxies from \citet{jansen00} and \citet{moustakas06}, respectively. The best fit to the black points is given
    in Equation~\ref{eqn11}.}
  \label{fig13}
\end{figure}

The conversion factor for \Oiii~is obtained from 196 NB704+NB921 emitters. Line fluxes for the two filters were obtained
using Equation~\ref{eqn1}. A histogram plot of the \Oiii-to-\Ha~ratio is shown in Figure~\ref{fig13}. This is compared
with \citet{hippelein03}'s $z=0.24$ and 0.64 \Oiii~emitters and the SDSS DR2 sample\footnotemark[12] \citep{brinchmann04}.
\footnotetext[12]{The emission line catalogue can be found at\\\url{http://www.mpa-garching.mpg.de/SDSS/DR2/Data/emission\_lines.html}.}
These \Oiii~NB emitters have a larger \Oiii/\Ha~ratio compared with SDSS. This may be caused by a difference in the
metallicity content between the two samples, and the selection requirement that both NB filters (NB704 and NB921) have
an excess.
The average and standard deviation for \Oiii~to \Ha+[\textsc{N ii}] ratio are 0.86 and 0.42. Correcting the NB921 line flux
by the previous assumption that \Ha/[\textsc{N ii}] = 4.66, an \Oiii/\Ha~flux ratio of $1.05 \pm 0.51$ is used. This is
similar to \citet{teplitz00}, who fix this ratio to unity, and it is known to vary between 0.5 and 2 from \citet{KKP99}.
In addition, the logarithm of the \Oiii/\Ha~ratio as a function of $M_B^{\rm o}$ is shown in Figure~\ref{fig13}. The best
fit is
\begin{equation}
  \label{eqn11}
  \log{\left[\frac{\textsc{O iii}}{H\alpha}\right]} = (0.073\pm0.013)M_B^{\rm o} + 1.274\pm0.233,
\end{equation}
which has less scatter and a higher \Oiii-\Ha~ratio compared to nearby star-forming galaxies of \citet{jansen00} and
\citet{moustakas06}. Although the \Oiii-\Ha~ratio is an average that allows for the determination of the SFR, a large
dispersion in the ratio is found. A more appropriate calculation is to use the full histogram (see Figure~\ref{fig13})
of \Oiii-\Ha~ratios known from 196 NB704+921 emitters. To use the histogram, a random integer between 1 and 196 is
assigned for each \Oiii~emitter in all four NB filters. Each integer has a corresponding \Oiii/\Ha~value given from the
sample of NB704+921 emitters. Then this ratio is used to convert from an \Oiii~LF to a \Ha~LF for a SFR density. This
method accounts for objects with low \Oiii/\Ha, which will produce some luminous \Ha~emission and hence an increase in
$L_{\star}$. Star formation rate densities for \Oiii~are reported using this method rather than integrating the \Oiii~LF.

For the remaining 16 NB704+921 emitters which are believed to be \Oii+\Hb~emitters based on BB colors, the \Oii/\Hb~ratio
is $0.51\pm0.32$. This ratio is for seven emitters, as seven sources with line fluxes near the 3$\sigma$ limit and two
sources with \Oii/\Hb~ratio of 5.2 and 8.7 are excluded.

\begin{figure}
  \epsscale{1.17}
  \plottwo{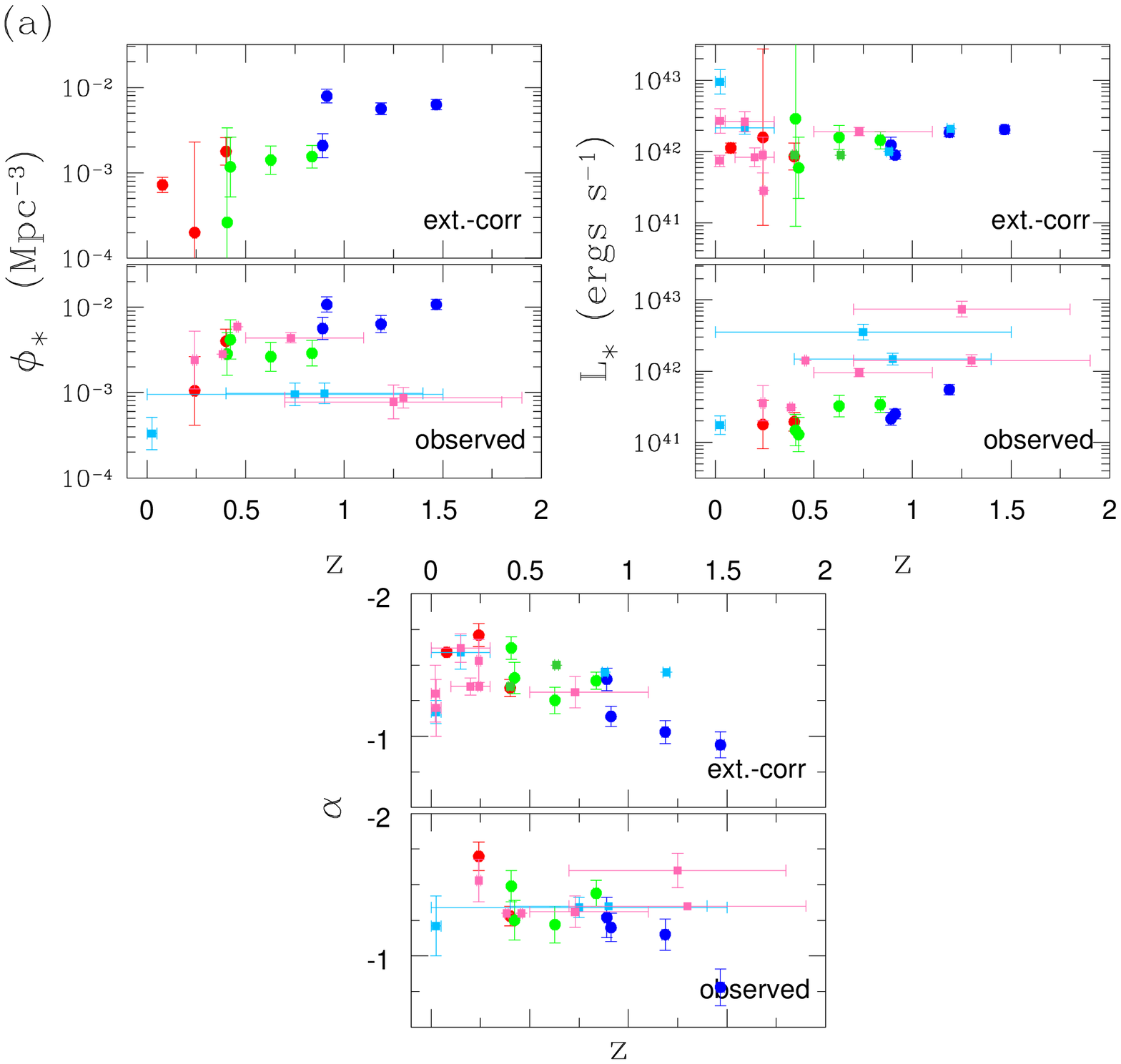}{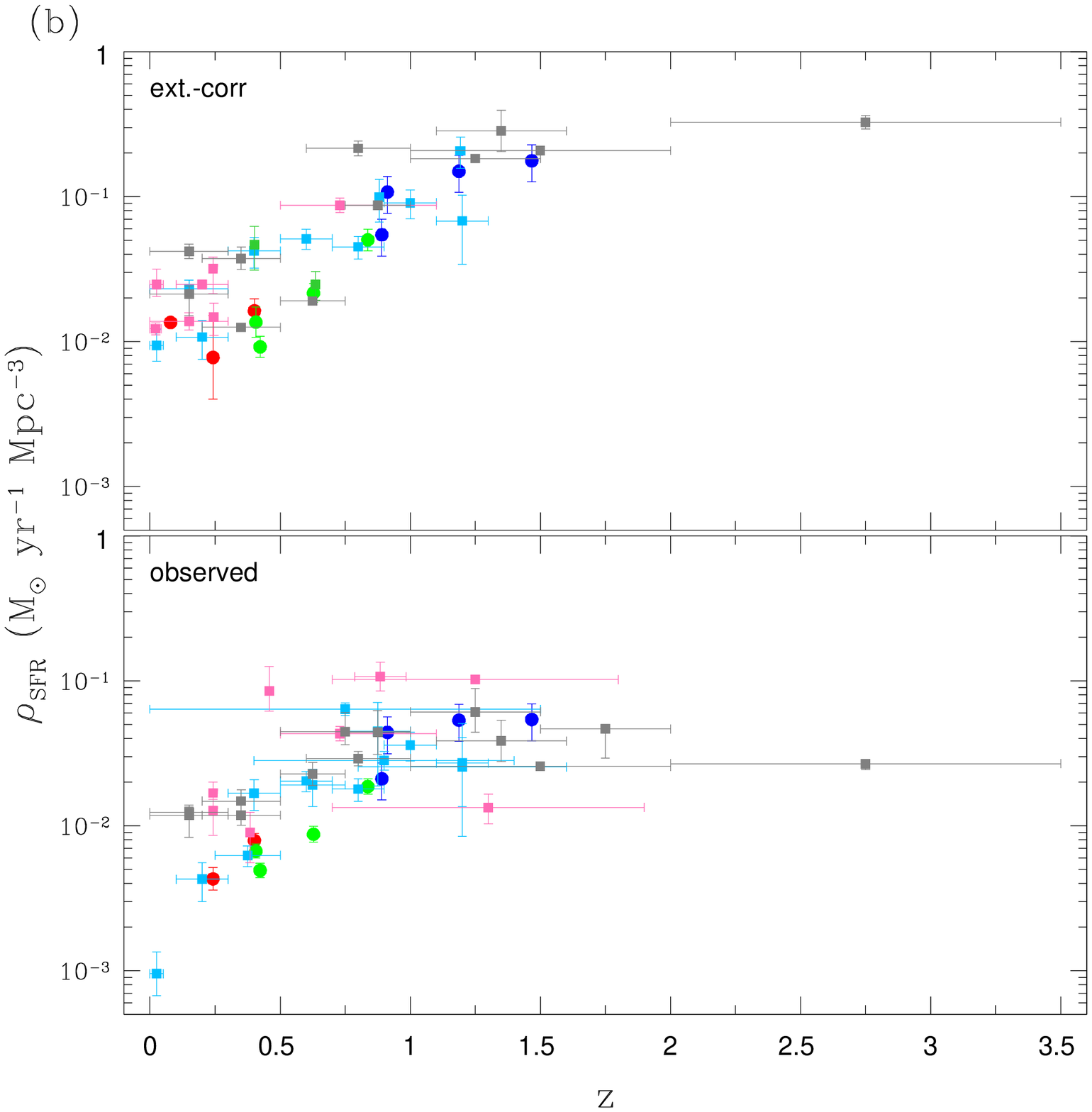}
  \caption{Schechter LF parameters and SFR density as a function of redshift. NB emitters are shown as circles and squares
    represent previous emission line surveys: \Oii~(light blue in the electronic edition) values are from \citet{hammer97},
    \citet{hogg98}, \citet{sullivan00}, \citet{gallego02}, \citet{hicks02}, and \citet{teplitz03}. \Ha~(pink in the
    electronic edition) points are from \citet{gallego95}, \citet{TM98}, \citet{glazebrook99}, \citet{yan99},
    \citet{hopkins00}, \citet{moorwood00}, \citet{sullivan00}, \citet{tresse02}, \citet{fujita03}, \citet{perez03},
    \citet{glazebrook04}, and \citet{pascual05}. \Oii, \Oiii~(dark green in electronic edition), and \Ha~data from
    \citet{hippelein03} are also included. Grey points are UV measurements from \citet{cowie99}, \citet{sullivan00},
    \citet{mass01}, and \citet{wilson02}.}
  \label{fig14}
\end{figure}

In Table~\ref{table4} the observed best fit Schechter parameters and the measured $\dot\rho_{\rm SFR}$ for each redshift are
summarized. The uncertainties in the Schechter parameters are determined using the non-linear least squares curve fitting
package, MPFIT. These Schechter parameters are plotted as circles in Figure~\ref{fig14}a, and the inferred SFR densities are
plotted as circles in Figure~\ref{fig14}b. Also plotted in Figure~\ref{fig14}a as squares are measurements from
\citet[$z<0.045$ \Ha]{gallego95}, \citet[$z<1.5$ \Oii]{hogg98}, \citet[$z=1.3\pm0.6$ \Ha]{yan99},
\citet[$z=1.25\pm0.55$ \Ha]{hopkins00}, \citet[$z<0.05$ \Oii]{gallego02}, \citet[$0.5\leq z\leq1.1$ \Ha]{tresse02}, and
\citet[$z=0.242\pm0.01$ \Ha]{fujita03}. Extinction-corrected Schechter parameters (shown as filled squares) from
\citet{gallego95}, \citet[$z=0.2\pm0.1$ \Ha]{TM98}, \citet[$z<0.3$ \Ha~and \Oii]{sullivan00}, \citet{tresse02},
\citet{fujita03}, and \citet[$z<0.05$ \Ha]{perez03} are provided in the upper panels. Also plotted are data points from
\citet{hippelein03} for $z=0.25\pm0.01$ \Ha, $z=0.40\pm0.01$ and $0.64\pm0.01$ \Oiii, and $0.88\pm0.01$ and $1.19\pm0.02$
\Oii. All measurements have been converted from the published cosmology to the cosmology chosen in this paper.

For Figure~\ref{fig14}b, SFR densities derived via integration of the luminosity function of previous surveys are included.
In many cases, a LF is not available, and so these SFR densities are derived from a binned luminosity density.
Additionally, studies that only report the luminosity density or SFR density are included. The \Ha~$\dot\rho_{\rm SFR}$
measurements are from \citet{glazebrook99} at $z=0.89\pm0.1$, \citet{glazebrook04} for $z=0.38$ and 0.46, and
\citet{pascual05} at $z=0.24\pm0.01$. \Oii~measurements from \citet{hammer97} at $z$ = 0.25-0.50, 0.50-0.75, and 0.75-1.0, \citet{hogg98}
at $z=0.2$, 0.4, 0.6, 0.8, 1.0, and 1.2, and \citet{hicks02} at $z=1.2\pm0.4$. Table~\ref{table5} summarizes the Schechter
profiles and SFR densities for studies plotted in Figure~\ref{fig14}a-b. UV-determined SFR densities from \citet{lilly96},
\citet{connolly97}, \citet{treyer98}, \citet{cowie99}, \citet{sullivan00}, \citet{mass01}, and \citet{wilson02} are shown
as grey squares in Figure~\ref{fig14}b. The conversion from the UV continuum between 1500-2800\AA~to a star-formation
rate is SFR$_{\rm UV}$(M$_{\sun}$ yr$^{-1}$) = $1.4\times10^{-28}L_{\nu}$(ergs s$^{-1}$ Hz$^{-1}$)
\citep{kennicutt98,hopkins04}. Again a Salpeter IMF with masses between 0.1 and 100$M_{\sun}$ is assumed.

\section{COMPARISON WITH OTHER STUDIES}
\label{4}
\subsection{\Ha~Emitters}
\label{4.1}
The LF reported here for $z\approx0.07$ extends an order of magnitude fainter than \citet{jones01} and about two orders
compared to \citet{gallego95}, giving a better constraint on the faint end slope. The NB704 and NB711 \Ha~LFs indicate
that $\alpha$ is steeper by about 0.3 compared to \citet{gallego95}; it is also steeper than that of \citet{treyer05}
  and \citet{wyder05}. This effect makes little difference in the SFR density; however, it reveals that there are more low
luminosity star-forming galaxies than previously predicted for $z\lesssim0.1$. At similar luminosities, the NB704 and
NB711 number densities agree with those of \citet{jones01}.

For $z\approx0.24$, while many emission-line studies are available, there is still significant scatter in the resulting
\Ha~LFs. The \Ha~NB816 LF is somewhat consistent with \citet{TM98} and \citet{sullivan00} particularly the latter with a
steep $\alpha$: $<-1.5$. However, the LFs of \citet{jones01}, \citet{fujita03}, \citet{hippelein03}, and \citet{pascual05}
have a higher number density by a factor of two or more. This is probably the result of cosmic variance as an estimate
of the relative cosmic variance is significant: following \citet{somerville04}, the bias $b$ is about 0.7 for a comoving
number density of 0.05 Mpc$^{-3}$ and $\sigma_{\rm DM} \approx 0.9$ for a volume of $4.7\times10^3$ Mpc$^{3}$, therefore,
$\sigma_v = b\sigma_{\rm DM} \approx 0.6$.

\begin{figure}
  \epsscale{0.45}
  \plotone{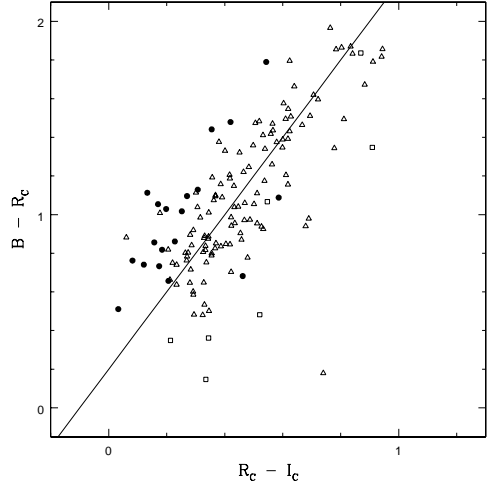}
  \caption{$B-\Rc$ and $\Rc-I_{\rm C}$ colors for Hawaii HDF-N galaxies with NB816 redshifts. Filled circles are
    \Ha~($z\approx0.24$), open triangles are \Oiii~or \Hb~($z\approx0.63$), and \Oii~are shown as open squares. The
    solid line is the selection criterion of \citet{fujita03}: $B-\Rc = 2(\Rc-I_{\rm C}) + 0.20$.}
  \label{fig15}
\end{figure}

The \Ha~NB816 LF of \citet{fujita03} has twice as many line emitters per logarithmic bin than the NB816 emitters in this
paper. But the $B-\Rc$ and $\Rc-I_{\rm C}$ colors were examined for Hawaii HDF-N sources with NB816 redshifts, and a
significant (about 50\%) amount of contamination from \Oiii~into their \Ha~selection criterion was found, which will
certainly reduce their number densities. This can be seen in Figure~\ref{fig15}, and indicates that using population
synthesis models [as \citet{fujita03} have done] is not enough; spectroscopic identification is required to obtain a
sample with low contamination. It should be pointed out that the selection criterion of \citet{fujita03} does distinguish
\Oii~from \Ha. At $z=0.40$, \citet{jones01} report a higher density by about a factor of two. The estimate of the LF that
\citet{glazebrook04} made with a small number of \Ha~emitters is consistent with the \Ha~NB921 LF.

The extinction-corrected \Ha~SFR densities ($z<0.5$) appear consistent with the \Ha~surveys of \citet{gallego95},
\citet{sullivan00}, and \citet{hippelein03}, \Oii~measurements from \citet{hogg98} and \citet{gallego02}, and a UV
measurement from \citet{cowie99}. However, \Ha~measurements from \citet{fujita03} and \citet{perez03}, \Oiii~from
\citet{hippelein03}, UV from \citet{wilson02}, and \Oii~from \citet{sullivan00} are twice as high. Assuming that
each of these surveys do not have any systematic differences, cosmic variance can certainly explain a difference by a
factor of two.

\subsection{\textsc{O iii} and \textsc{O ii} Emitters}
The Fabry-Perot interferometry technique employed by \citet{hippelein03} also selects similar redshift intervals with
the exception of the NB921 emitters ($z=0.84$ and 1.46). Figure~\ref{fig11}a-c and \ref{fig12}a-c reveal good agreement
between the observed LFs. The extinction-corrected LFs cannot be compared, as different reddenning assumptions are used.
The observed \Oii~luminosity densities (or SFR densities since the same \Oii/\Ha~ratio is used) for $z\approx0.9$
to 1.5 agree with those of \citet{hogg98}.

\section{DISCUSSION}
\label{5}
\subsection{Differences between NB704 and NB711 LFs}
The LFs for \Ha~at $z=0.074$ and 0.086 show little differences. This could be coincidental given that cosmic variance
is expected to be large in a small comoving volume. For \Oiii, the lower redshift LF shows a steeper faint end slope and
has 351 versus 209 emitters. This can be the result of differences in the NB filter as the NB711 has a comoving volume
that is smaller by 25\%. For \Oii, the density of NB711 emitters below a luminosity of $10^{41.5}$ ergs s$^{-1}$ is
twice as high compared to NB704 emitters, and there is 818 versus 673 emitters. The difference in number of line emitters
cannot be explained by the smaller NB711 bandpass as more objects are seen in the NB711 filter. However, the difference
can be attributed to cosmic variance.

\subsection{Evolution of the LF and SFR Densities}
In Figure~\ref{fig14}a, a steep evolution in the number density $\phi_{\star}$ is found while the luminosity $L_{\star}$
has a milder evolution. The comoving density and luminosity can be fitted with $\phi_{\star} \propto (1+z)^P$ and
$L_{\star} \propto (1+z)^Q$ with $P =3.85\pm0.95$ and $Q =0.47\pm0.58$. This is contrary to \citet{hopkins04} who reports
$P=0.15\pm0.60$ and $Q=2.70\pm0.60$. In addition, the faint end slope $\alpha$ appears to flatten out at higher redshifts,
evolving from -1.6 to -1.0 as $z$ goes from 0.08 to 1.47 with $\alpha \propto (1.94\pm0.37)\log{(1+z)}$. The flattening at
higher redshifts resembles the effects of incompleteness at the faint end.
However, even with the completeness correction described in \S\S~\ref{3.4}, $\alpha$ still evolves toward a flatter
slope at higher redshifts. Moreover, at the highest redshifts, $\alpha$ is well determined with several bins in the LF. One
other possible concern is that \Oiii~and \Oii~emitters were improperly identified, and resulted in a contamination at the
faint end of the \Ha~LF. The contamination reported from spectra indicated that less than 10\% would be mis-identified, so
this could not account completely for the steep faint end slope at low redshift. \citet{gabasch04,gabasch06} also see an
evolution (although mild) in $\alpha$ from the blue and red band LFs of 5500 galaxies in the FORS Deep Field.
However, \citet{arnouts05} and \citet{wyder05} find the converse with ultraviolet continuum measurements.

The integrated \Oii~SFR densities at $z\sim1$ are found to be 10 times higher than at $z<0.5$ from the \Ha~and \Oiii~NB
emitters. While this is consistent with some studies, other studies report a SFR density that is twice as high for similar
redshifts. There is agreement between the extinction-corrected \Oii~NB emitters with measurements above $z$ of 0.75 from
UV continuum and \Oii~emission line surveys.

The \Oiii~SFR densities appear half as large as the \Ha~and \Oii~measurements. This is the result of not detecting
sources with low \Oiii/\Ha~ratios, as these sources will be detected in the NB921 filter, but will be too faint for the
NB704 filter. Figure~\ref{fig13} shows that local studies have found sources with \Oiii/\Ha~ratios as small as 0.03.
Even if a small fraction of \Oiii~NB emitters have low \Oiii/\Ha~ratios, this can affect the LF enough to raise the SFR by
an additional factor of two.

\subsection{Future Work}
\label{5.3}
The method described in this paper can be applied to other large fields. For example, deep NB816 imaging
with Suprime-Cam of SSA22 \citep{hu04} and around SDSSp J104433.04-012502.2 \citep{fujita03,ajiki06} were intended to
identify LAEs. The low-$z$ NB emitter sample can be obtained from these fields at three additional redshifts
of 0.24, 0.64, and 1.18. For SDSSp J104433.04, the previous work of \citet{fujita03} for $z\approx0.24$
\Ha~emitters can be improved by reducing contamination (see \S\S~\ref{4.1}). Also, imaging at the remaining three NB
filters will significantly expand the sample with nine more redshift intervals. The results of these individual fields
can be compared to the results presented for the SDF, and the effect of cosmic variance will be reduced when all NB
emitters are combined together. Moreover, NB imaging of deep spectroscopic galaxy surveys (e.g., DEEP2) will ($i$)
further improve these fields as this method provides the redshift of several hundred objects in the fields, ($ii$)
provide existing spectra of NB emitters to further examine the NB technique, and ($iii$) NB921 imaging of known DEEP2
galaxies with the redshifts of $z=0.84$ and 1.47 would provide additional spectroscopic points to be overlayed on
Figure~\ref{fig7}b as the $z\arcmin_{\rm cont}$ is unknown without NB921 photometry.

\section{CONCLUSION}
\label{6}
Using four NB and five BB filters, one to two thousand NB line-emitters (for each filter) are photometrically identified.
Considering the strongest emission lines (\Ha, \Oii, and \Oiii), broad-band colors are used to distinguish them into
twelve redshifts intervals (some of which overlap). With a large sample of NB emitters, an averaged rest-frame optical to
UV SED is obtained for each redshift. The \Ha~SEDs show little differences for all four redshifts. Generally, high-EW
emitters appear bluer relative to the low-EW objects.

The luminosity functions are generated for eleven redshift windows between $z = 0.07$ and 1.47. These luminosity
functions are integrated to obtain a luminosity density, and converted to a measured star-formation rate density after
correcting for extinction. The lowest redshifts covered by the NB704 and NB711 filters indicate a steep faint end slope.
These NB emitters show that the SFR at $z\approx1$ is ten times higher than $z\sim0$ with a steep decline to $z\approx0.4$.
Moreover, the \Oii~SFR is consistent with UV and other \Oii~measurements. Below $z$ of 0.5, $\dot\rho_{\rm SFR}$
measurements from \Ha~and \Oiii~emitters are consistent with several studies; however, there appears to be a discrepancy
in $\dot\rho_{\rm SFR}$ by a factor of two or more from other studies. Cosmic variance may be imposed to explain the
discrepancy.

\acknowledgements 
The data presented herein were obtained at the W.M. Keck Observatory, which is operated as a scientific partnership
among the California Institute of Technology, the University of California and the National Aeronautics and Space
Administration. The Observatory was made possible by the generous financial support of the W.M. Keck Foundation.
The analysis pipeline used to reduce the DEIMOS data was developed at UC Berkeley with support from NSF grant
AST-0071048. This research was supported, in part, by NASA grant 1275777. C. Ly is an Eugene V. Cota-Robles fellow. This
research was also supported by a scientific research grant (15204012) from the Ministry of Education, Science, Culture,
and Sports of Japan. This research has made use of the NASA/IPAC Extragalactic Database, which is operated by the
Jet Propulsion Laboratory, Caltech, under contract with the National Aeronautics and Space Administration, and NASA
Astrophysics Data System. C. Ly would like to thank A. Hopkins for helpful discussions. In addition, we thank the Subaru
and Keck staff for their invaluable help, and Drew Phillips for his successful DEIMOS mask construction program
\textsf{dsim}. We also appreciate the comments of an anonymous referee that improved this paper.

\clearpage
\LongTables
\newcommand{\B}{\tablenotemark{b}}
\newcommand{\C}{\tablenotemark{c}}
\begin{landscape}
\begin{deluxetable}{lclrrrrrrrrrcccc}
  \tabletypesize{\footnotesize}
  \tablewidth{0pc}
  \tablecaption{Photometric Properties of Spectroscopically Confirmed Narrow-Band Line-Emitting Galaxies}
  \tablehead{
  \colhead{NB ID\tablenotemark{a}}& \colhead{Name}&\colhead{redshift}&\multicolumn{9}{c}{Optical AB magnitude}&\multicolumn{4}{c}{Line fluxes}\\
      \colhead{}& & &\colhead{$B$}&\colhead{$V$}&\colhead{$R_{\rm C}$}&\colhead{$i$\arcmin}&\colhead{$z$\arcmin} &\colhead{NB704}&\colhead{NB711}&\colhead{NB816}&\colhead{NB921}&\colhead{\Oii}&\colhead{H$\beta$}&\colhead{\Oiii}&\colhead{\Ha}\\
      \colhead{(1)}&\colhead{(2)}&\colhead{(3)}&\colhead{(4)}&\colhead{(5)}&\colhead{(6)}&\colhead{(7)}&\colhead{(8)}&\colhead{(9)}&\colhead{(10)}&\colhead{(11)}&\colhead{(12)}&\colhead{(13)}&\colhead{(14)}&\colhead{(15)}&\colhead{(16)}}
    \startdata
  \multicolumn{16}{c}{FOCAS NB816 emitters}\\
  28247  & SDFJ132411.7+271531 & 1.1807   &   26.38&   26.33\phm{)}&   26.33\phm{)}&   26.15\phm{)}&   26.25\phm{)}&   26.09\phm{)}&   26.98\phm{)}&   24.76\phm{)}&   26.44 & 16.8 &\ldots&\ldots&\ldots\\
  38133  & SDFJ132356.3+271726 & 1.1798   &   23.97&   23.82\phm{)}&   23.72\phm{)}&   23.50\phm{)}&   23.20\phm{)}&   23.76\phm{)}&   23.78\phm{)}&   22.47\phm{)}&   23.30 & 71.7 &\ldots&\ldots&\ldots\\
  42561  & SDFJ132403.4+271817 & 0.6303   &   24.08&   23.84\phm{)}&   23.39\phm{)}&   23.13\phm{)}&   23.19\phm{)}&   23.38\phm{)}&   23.40\phm{)}&   22.15\phm{)}&   23.27 & 61.6 & 29.7 &  128 &\ldots\\
  76702  & SDFJ132405.8+272537 & 1.1847\B &   24.58&   24.47\phm{)}&   24.41\phm{)}&   24.19\phm{)}&   23.98\phm{)}&   24.46\phm{)}&   24.60\phm{)}&   23.16\phm{)}&   24.11 &  109 &\ldots&\ldots&\ldots\\
  78892  & SDFJ132415.3+272559 & 0.6150   &   25.68&   25.49\phm{)}&   24.85\phm{)}&   24.40\phm{)}&   24.80\phm{)}&   24.67\phm{)}&   25.02\phm{)}&   23.42\phm{)}&   24.87 &  5.7 & 11.3 & 74.4 &\ldots\\
  96705  & SDFJ132425.6+272947 & 0.6319   &   28.54&   27.91\phm{)}&   27.86\phm{)}&   27.03\phm{)}&   28.60\phm{)}&  (27.38)      &  (26.52)      &   25.35\phm{)}&   27.50 &\ldots&\ldots&  5.6 &\ldots\\
  99588  & SDFJ132357.8+273030 & 0.6359   &   26.95&   26.84\phm{)}&   26.41\phm{)}&   25.99\phm{)}&   26.34\phm{)}&   26.40\phm{)}&   26.29\phm{)}&   24.57\phm{)}&   26.59 &  2.9 &\ldots&  6.6 &\ldots\\
  168136 & SDFJ132403.0+274435 & 1.1783   &   25.48&   25.42\phm{)}&   25.34\phm{)}&   25.12\phm{)}&   24.82\phm{)}&   25.28\phm{)}&   25.82\phm{)}&   24.11\phm{)}&   24.82 & 15.0 &\ldots&\ldots&\ldots\\\tableline
  \multicolumn{16}{c}{FOCAS NB921 emitters}\\
  41910  & SDFJ132438.4+271612 & 0.8390   &   27.32&   26.84\phm{)}&   26.54\phm{)}&   26.59\phm{)}&   25.38\phm{)}&   26.57\phm{)}&   26.70\phm{)}&   26.93\phm{)}&   23.67 &\ldots&  8.0 & 34.6 &\ldots\\
  46399  & SDFJ132416.7+271655 & 0.8308\B &   26.54&   26.42\phm{)}&   26.26\phm{)}&   26.34\phm{)}&   25.36\phm{)}&   26.17\phm{)}&   26.49\phm{)}&   26.85\phm{)}&   23.82 &  7.9 & 16.2 & 82.3 &\ldots\\
  54902  & SDFJ132413.7+271825 & 0.8378   &   25.67&   25.54\phm{)}&   25.33\phm{)}&   24.96\phm{)}&   24.61\phm{)}&   25.09\phm{)}&   25.07\phm{)}&   24.88\phm{)}&   23.58 &  9.4 &  5.9 & 37.7 &\ldots\\
  58816  & SDFJ132404.7+271912 & 0.8369   &   26.95&   26.89\phm{)}&   26.74\phm{)}&   26.31\phm{)}&   25.60\phm{)}& \ldots\phm{))}& \ldots\phm{))}&   26.75\phm{)}&   24.04 &\ldots&  4.0 & 23.3 &\ldots\\
  62897  & SDFJ132409.9+272009 & 0.8322   &   24.46&   24.25\phm{)}&   23.86\phm{)}&   23.64\phm{)}&   23.14\phm{)}&   23.77\phm{)}&   23.56\phm{)}&   23.52\phm{)}&   22.11 & 88.1 & 55.0 &  185 &\ldots\\
  63322  & SDFJ132354.9+272016 & 0.3991   &   24.56&   24.03\phm{)}&   23.51\phm{)}&   23.62\phm{)}&   23.53\phm{)}&   22.36\phm{)}&   23.74\phm{)}&   23.76\phm{)}&   22.44 &\ldots& 30.8 &  161 & 31.6 \\
  87190  & SDFJ132358.2+272539 & 0.8391   &   25.21&   25.15\phm{)}&   24.85\phm{)}&   24.70\phm{)}&   24.05\phm{)}&   24.93\phm{)}&   24.44\phm{)}&   24.73\phm{)}&   22.56 & 23.8 & 13.0 &  101 &\ldots\\
  92017  & SDFJ132404.8+272645 & 0.8334\B &   26.86&   26.73\phm{)}&   26.50\phm{)}&   26.08\phm{)}&   25.36\phm{)}&   26.28\phm{)}&   25.99\phm{)}&   26.11\phm{)}&   24.18 &  5.1 & 14.0 & 71.9 &\ldots\\
  95258  & SDFJ132511.9+272731 & 0.8387\B &   24.73&   24.67\phm{)}&   24.48\phm{)}&   24.46\phm{)}&   23.49\phm{)}&   24.53\phm{)}&   23.98\phm{)}&   24.67\phm{)}&   21.66 & 47.7 &  131 &  824 &\ldots\\
  96981  & SDFJ132423.1+272752 & 0.8287   &   26.04&   25.94\phm{)}&   25.64\phm{)}&   25.37\phm{)}&   24.82\phm{)}&   25.44\phm{)}&   25.23\phm{)}&   25.32\phm{)}&   23.61 &  9.8 &\ldots& 36.8 &\ldots\\
  97394  & SDFJ132408.8+272757 & 0.8318   &   27.17&   26.93\phm{)}&   26.48\phm{)}&   26.37\phm{)}&   25.70\phm{)}&   26.43\phm{)}&   26.42\phm{)}&   26.36\phm{)}&   24.19 &  4.5 &\ldots& 21.0 &\ldots\\
  99909  & SDFJ132414.4+272833 & 0.8286   &   27.34&   27.20\phm{)}&   26.54\phm{)}&   26.20\phm{)}&   26.01\phm{)}&   26.37\phm{)}&   26.03\phm{)}&   26.30\phm{)}&   24.76 &  2.0 &\ldots& 11.9 &\ldots\\
  108717 & SDFJ132411.3+273042 & 0.8343   &   27.60&   27.30\phm{)}&   26.77\phm{)}&   26.29\phm{)}&   25.96\phm{)}&   26.32\phm{)}&   26.34\phm{)}&   26.22\phm{)}&   24.81 &\ldots&\ldots&  8.8 &\ldots\\
  109516 & SDFJ132406.3+273057 & 0.8391   &   26.06&   25.71\phm{)}&   25.59\phm{)}&   25.48\phm{)}&   24.83\phm{)}&   25.52\phm{)}&   25.31\phm{)}&   25.44\phm{)}&   23.41 &  1.2 &  8.9 & 48.0 &\ldots\\
  109948 & SDFJ132352.9+273106 & 0.8399   &   24.45&   24.33\phm{)}&   24.09\phm{)}&   23.81\phm{)}&   23.34\phm{)}&   24.03\phm{)}&   23.68\phm{)}&   23.77\phm{)}&   22.13 & 28.8 & 22.5 &  169 &\ldots\\
  111896 & SDFJ132403.3+273131 & 0.8391   &   23.93&   23.83\phm{)}&   23.60\phm{)}&   23.42\phm{)}&   23.02\phm{)}&   23.62\phm{)}&   23.29\phm{)}&   23.42\phm{)}&   21.83 & 58.7 & 28.1 &  158 &\ldots\\
  114783 & SDFJ132351.0+273205 & 0.8358   &   27.24&   27.02\phm{)}&   26.82\phm{)}&   26.50\phm{)}&   26.04\phm{)}&   26.69\phm{)}&   26.38\phm{)}&   26.18\phm{)}&   24.86 &\ldots&\ldots& 10.6 &\ldots\\
  116106 & SDFJ132508.1+273223 & 0.8492   &   27.62&   27.56\phm{)}&   27.54\phm{)}&   27.47\phm{)}&   26.33\phm{)}&$>$27.86\phm{)}&  (26.88)      &   27.84\phm{)}&   25.11 &\ldots&  3.0 & 16.4 &\ldots\\
  123068 & SDFJ132502.7+273403 & 0.8387   &   26.34&   26.34\phm{)}&   26.20\phm{)}&   26.21\phm{)}&   25.19\phm{)}&   26.77\phm{)}&   26.02\phm{)}&   26.08\phm{)}&   23.38 &\ldots& 10.2 & 66.4 &\ldots\\
  154431 & SDFJ132407.2+274056 & 0.8276   &   26.23&   26.01\phm{)}&   25.82\phm{)}&   25.73\phm{)}&   25.12\phm{)}&   25.71\phm{)}&   25.88\phm{)}&   26.00\phm{)}&   23.97 &  6.6 & 12.1 & 43.7 &\ldots\\\tableline
  \multicolumn{16}{c}{DEIMOS NB816 emitters}\\
  29275  & SDFJ132517.9+271546 & 1.1818   &   25.84&   25.37\phm{)}&   24.87\phm{)}&   24.35\phm{)}&   23.52\phm{)}&   24.76\phm{)}&   24.88\phm{)}&   23.33\phm{)}&   23.65 & 34.5 &\ldots&\ldots&\ldots\\
  31925  & SDFJ132515.6+271611 & 1.1813   &   26.37&   26.08\phm{)}&   25.88\phm{)}&   25.62\phm{)}&   25.17\phm{)}&   26.25\phm{)}&   25.79\phm{)}&   24.59\phm{)}&   25.38 & 17.3 &\ldots&\ldots&\ldots\\
  34775  & SDFJ132508.1+271648 & 1.1830   &   27.90&   27.34\phm{)}&   26.57\phm{)}&   25.56\phm{)}&   24.55\phm{)}&   26.42\phm{)}&   26.07\phm{)}&   24.38\phm{)}&   24.50 & 16.6 &\ldots&\ldots&\ldots\\
  56181  & SDFJ132525.9+272112 & 0.6197   &   27.52&   27.39\phm{)}&   26.77\phm{)}&   26.22\phm{)}&   26.56\phm{)}&  (27.21)      &  (26.91)      &   25.20\phm{)}&   26.82 &\ldots&\ldots&  9.7 &\ldots\\
  59788  & SDFJ132510.2+272153 & 0.6283   &   28.05&  (28.31)      &   27.20\phm{)}&   26.79\phm{)}&  (27.64)      &   26.64\phm{)}&  (26.59)      &   25.27\phm{)}&   26.94 &\ldots&\ldots& 20.7 &\ldots\\
  68251  & SDFJ132340.7+272346 & 0.6347   &   25.81&   25.57\phm{)}&   25.06\phm{)}&   24.76\phm{)}&   24.87\phm{)}&   25.19\phm{)}&   24.71\phm{)}&   23.71\phm{)}&   24.88 & 19.5 & 13.8 & 71.0 &\ldots\\
  70071  & SDFJ132434.9+272410 & 0.6229   &   24.55&   24.40\phm{)}&   24.02\phm{)}&   23.76\phm{)}&   24.06\phm{)}&   23.86\phm{)}&   24.03\phm{)}&   22.54\phm{)}&   24.20 & 32.0 & 12.9 &  189 &\ldots\\
  110439 & SDFJ132524.7+273244 & 0.6293   &   25.95&   25.81\phm{)}&   25.35\phm{)}&   24.98\phm{)}&   25.44\phm{)}&   25.24\phm{)}&   25.23\phm{)}&   23.52\phm{)}&   25.41 &\ldots& 12.9 & 33.3 &\ldots\\
  122518 & SDFJ132513.5+273518 & 1.1735   &$>$29.64&$>$28.93\phm{)}&$>$28.99\phm{)}&  (28.33)      &$>$27.81\phm{)}&\ldots\phm{))} &\ldots\phm{))} &   25.95\phm{)}&$>$27.73 & 7.5  &\ldots&\ldots&\ldots\\
  136295 & SDFJ132505.5+273810 & 0.2438   &   24.21&   23.96\phm{)}&   23.69\phm{)}&   23.72\phm{)}&   23.90\phm{)}&   24.04\phm{)}&   24.03\phm{)}&   22.41\phm{)}&   24.05 &\ldots&\ldots& 129  & 123\\
  165225 & SDFJ132453.4+274358 & 0.6373   &   26.93&   26.69\phm{)}&   26.08\phm{)}&   25.81\phm{)}&   25.87\phm{)}&   26.28\phm{)}&   25.94\phm{)}&   24.80\phm{)}&   25.88 &\ldots&\ldots&  9.4 &\ldots\\\tableline
  \multicolumn{16}{c}{DEIMOS NB921 emitters}\\
  31248  & SDFJ132459.8+271423 & 1.4733   &   27.11&   26.80\phm{)}&   26.70\phm{)}&   26.69\phm{)}&   26.65\phm{)}&  (27.22)      &$>$27.18\phm{)}&   26.40\phm{)}&   25.55 & 17.7 &\ldots&\ldots&\ldots\\
  69400  & SDFJ132428.7+272136 & 0.3986   &   28.02&   27.47\phm{)}&   27.39\phm{)}&  (27.98)      &  (27.69)      &   26.57\phm{)}&  (26.97)      &  (27.68)      &   25.69 &\ldots&  6.7 & 1.1  & 25.8 \\
  71165  & SDFJ132422.3+272202 & 1.4692   &   26.77&   26.56\phm{)}&   26.53\phm{)}&   26.54\phm{)}&   26.40\phm{)}&   26.85\phm{)}&  (26.59)      &   26.57\phm{)}&   25.31 &  8.2 &\ldots&\ldots&\ldots\\
  78567  & SDFJ132444.1+272344 & 0.8358   &   26.48&   26.31\phm{)}&   26.33\phm{)}&   26.15\phm{)}&   25.62\phm{)}&   26.25\phm{)}&   26.02\phm{)}&   26.36\phm{)}&   24.26 &\ldots&  7.0 & 27.7 &\ldots\\
  84040  & SDFJ132509.0+272455 & 0.8482   &   27.96&   27.79\phm{)}&   27.56\phm{)}&   27.43\phm{)}&  (27.12)      &$>$27.86\phm{)}&   27.39\phm{)}&  (27.53)      &   25.34 &\ldots&\ldots& 38.8 &\ldots\\
  89013  & SDFJ132353.4+272602 & 0.8316   &$>$29.64&$>$28.93\phm{)}&$>$28.99\phm{)}&  (28.29)      &   26.78\phm{)}&\ldots\phm{))} &\ldots\phm{))} &$>$27.18\phm{)}&   24.94 &\ldots&  3.0 & 18.3 &\ldots\\
  128889 & SDFJ132520.5+273520 & 1.4771   &$>$29.64&$>$28.93\phm{)}&  (28.95)      &   27.67\phm{)}&   26.67\phm{)}&\ldots\phm{))} &\ldots\phm{))} &  (27.18)      &   25.51 &  2.2 &\ldots&\ldots&\ldots\\
  134603 & SDFJ132507.4+273638 & 1.4513   &$>$29.64&$>$28.93\phm{)}&$>$28.99\phm{)}&$>$28.62\phm{)}&  (27.36)      &\ldots\phm{))} &\ldots\phm{))} &$>$27.18\phm{)}&   25.89 & 10.5 &\ldots&\ldots&\ldots\\\tableline
  \multicolumn{16}{c}{Serendipitous Sources\tablenotemark{d}}\\
  59317  & SDFJ132510.3+272151 & 0.6750   &   25.22&   25.09\phm{)}&   24.61\phm{)}&   24.41\phm{)}&   24.49\phm{)}&   24.56\phm{)}&   24.57\phm{)}&   24.15\phm{)}&   24.67 & 24.1 & 18.1 & 57.4 &\ldots\\
  67280  & SDFJ132515.2+272340 & 0.6300   &   24.43&   23.94\phm{)}&   23.14\phm{)}&   22.78\phm{)}&   22.58\phm{)}&   23.03\phm{)}&   23.04\phm{)}&   22.55\phm{)}&   22.66 & 11.4 &\ldots&\ldots&\ldots\\
  104363 & SDFJ132505.8+273135 & 0.6382   &   25.59&   24.46\phm{)}&   23.39\phm{)}&   22.73\phm{)}&   22.26\phm{)}&   23.10\phm{)}&   23.14\phm{)}&   22.40\phm{)}&   22.28 &\ldots&  9.1 & 31.8 &\ldots\\
  41681  & SDFJ132415.8+271611 & 1.4920   &   25.44&   25.29\phm{)}&   24.93\phm{)}&   24.67\phm{)}&   24.24\phm{)}&   24.90\phm{)}&   24.78\phm{)}&   24.44\phm{)}&   24.28 & 24.5 &\ldots&\ldots&\ldots\\
  132483 & SDFJ132507.8+273608 & 1.4328   &   24.53&   24.31\phm{)}&   23.95\phm{)}&   23.64\phm{)}&   23.19\phm{)}&   23.80\phm{)}&   23.83\phm{)}&   23.48\phm{)}&   23.20 &  8.9 &\ldots&\ldots&\ldots\\\tableline
  \multicolumn{16}{c}{Fortuitous Sources\tablenotemark{e}}\\
  13370  & SDFJ132517.3+271325 & 0.4306   &   22.45&   21.65\phm{)}&   21.10\phm{)}&   20.89\phm{)}&   20.58\phm{)}&   20.97\phm{)}&   20.91\phm{)}&   20.65\phm{)}&   20.83 &\ldots& 59.9 & 24.2 &\ldots\\
  18344  & SDFJ132456.2+271400 & 0.8224   &   24.29&   23.71\phm{)}&   23.18\phm{)}&   22.57\phm{)}&   22.14\phm{)}&   22.80\phm{)}&   22.84\phm{)}&   22.27\phm{)}&   22.29 & 38.5 & 34.8 & 77.2 &\ldots\\
  30036  & SDFJ132455.4+271610 & 0.4570   &   24.02&   23.33\phm{)}&   22.83\phm{)}&   22.66\phm{)}&   22.47\phm{)}&   22.76\phm{)}&   22.78\phm{)}&   22.52\phm{)}&   22.67 &\ldots& 10.1 & 17.1 &\ldots\\
  39015  & SDFJ132520.5+271738 & 0.8242   &   26.09&   24.39\phm{)}&   23.30\phm{)}&   22.30\phm{)}&   21.61\phm{)}&   23.12\phm{)}&   22.90\phm{)}&   21.82\phm{)}&   21.76 &\ldots&\ldots&\ldots&\ldots\\
  40607  & SDFJ132524.6+271800 & 0.8306   &   25.88&   24.31\phm{)}&   23.26\phm{)}&   22.29\phm{)}&   21.58\phm{)}&   23.06\phm{)}&   22.93\phm{)}&   21.82\phm{)}&   21.73 &\ldots&\ldots&\ldots&\ldots\\
  57481  & SDFJ132452.2+272104 & 0.4498   &   22.62&   21.99\phm{)}&   21.57\phm{)}&   21.48\phm{)}&   21.31\phm{)}&   21.48\phm{)}&   21.53\phm{)}&   21.37\phm{)}&   21.45 &\ldots& 58.9 & 86.3 &\ldots\\
  58410  & SDFJ132520.4+272109 & 0.6375   &   23.69&   23.46\phm{)}&   22.87\phm{)}&   22.64\phm{)}&   22.59\phm{)}&   22.84\phm{)}&   22.81\phm{)}&   22.15\phm{)}&   22.72 & 92.0 & 38.0 &  110 &\ldots\\
  59053  & SDFJ132517.8+272119 & 0.4658   &   23.88&   22.88\phm{)}&   22.09\phm{)}&   21.69\phm{)}&   21.22\phm{)}&   21.95\phm{)}&   21.88\phm{)}&   21.41\phm{)}&   21.36 &\ldots& 39.1 &\ldots&\ldots\\
  60042  & SDFJ132420.9+272126 & 0.6360   &   23.79&   23.22\phm{)}&   22.49\phm{)}&   22.13\phm{)}&   21.90\phm{)}&   22.33\phm{)}&   22.32\phm{)}&   21.81\phm{)}&   21.94 &  151 &  113 &  110 &\ldots\\
  69992  & SDFJ132416.2+272315 & 0.8352   &   23.49&   23.15\phm{)}&   22.77\phm{)}&   22.28\phm{)}&   22.03\phm{)}&   22.52\phm{)}&   22.45\phm{)}&   22.10\phm{)}&   21.86 &  171 &  103 &  143 &\ldots\\
  80579  & SDFJ132414.7+272506 & 0.8978   &   23.63&   23.37\phm{)}&   23.13\phm{)}&   22.72\phm{)}&   22.51\phm{)}&   22.45\phm{)}&   22.27\phm{)}&   22.64\phm{)}&   22.46 &  265 &  166 &\ldots&\ldots\\
  93969  & SDFJ132507.1+272735 & 0.4676   &   24.64&   23.41\phm{)}&   22.30\phm{)}&   21.75\phm{)}&   21.31\phm{)}&   21.99\phm{)}&   22.01\phm{)}&   21.48\phm{)}&   21.39 &\ldots&\ldots&\ldots&\ldots\\
  104779 & SDFJ132521.1+272932 & 0.8984   &   23.91&   23.53\phm{)}&   23.24\phm{)}&   22.66\phm{)}&   22.37\phm{)}&   22.85\phm{)}&   22.63\phm{)}&   22.45\phm{)}&   22.41 & 60.3 &\ldots&\ldots&\ldots\\
  104825 & SDFJ132523.0+272937 & 0.8983   &   24.69&   23.33\phm{)}&   22.39\phm{)}&   21.41\phm{)}&   20.70\phm{)}&   22.01\phm{)}&   21.93\phm{)}&   21.05\phm{)}&   20.76 & 24.9 &\ldots&\ldots&\ldots\\
  106829 & SDFJ132520.6+272949 & 0.8988   &   24.02&   23.43\phm{)}&   22.94\phm{)}&   22.22\phm{)}&   21.77\phm{)}&   22.66\phm{)}&   22.53\phm{)}&   21.94\phm{)}&   21.81 & 36.4 &\ldots&\ldots&\ldots\\
  120415 & SDFJ132523.6+273229 & 0.6236   &   24.66&   23.36\phm{)}&   22.29\phm{)}&   21.52\phm{)}&   21.09\phm{)}&   21.96\phm{)}&   21.85\phm{)}&   21.26\phm{)}&   21.12 &\ldots&-18.1 &\ldots&\ldots\\
  134198 & SDFJ132511.1+273539 & 0.0842   &   21.18&   20.69\phm{)}&   20.47\phm{)}&   20.29\phm{)}&   20.18\phm{)}&   20.43\phm{)}&   19.96\phm{)}&   20.17\phm{)}&   20.29 &\ldots&\ldots&\ldots&  422 \\
  139473 & SDFJ132523.6+273549 & 0.8487   &   24.21&   23.80\phm{)}&   23.43\phm{)}&   22.85\phm{)}&   22.58\phm{)}&   23.23\phm{)}&   23.13\phm{)}&   22.64\phm{)}&   22.50 & 41.2 & 13.0 & 19.1 &\ldots\\
  169168 & SDFJ132500.7+274109 & 0.6871   &   23.72&   23.21\phm{)}&   22.57\phm{)}&   22.08\phm{)}&   21.79\phm{)}&   22.31\phm{)}&   22.26\phm{)}&   21.84\phm{)}&   21.85 & 23.6 & 21.2 &  6.2 &\ldots\\
  27743\C& SDFJ132410.8+271554 & 0.6316   &   24.46&   23.09\phm{)}&   22.10\phm{)}&   21.32\phm{)}&   20.81\phm{)}& \ldots\phm{))}& \ldots\phm{))}&   21.03\phm{)}&   20.87 &\ldots&\ldots&\ldots&\ldots\\\tableline
  \multicolumn{16}{c}{FOCAS NB711 Emitters}\\
  165413 & SDFJ132422.0+274016 & 0.9034   &   28.44&   27.50\phm{)}&   27.11\phm{)}&   26.60\phm{)}&   26.89\phm{)}&   26.20\phm{)}&   25.22\phm{)}&   26.66\phm{)}&   26.57 &  8.6 &\ldots&\ldots&\ldots\\
  176956 & SDFJ132417.5+274221 & 0.9106   &   27.94&   27.61\phm{)}&   26.90\phm{)}&   26.21\phm{)}&   25.80\phm{)}&   27.05\phm{)}&   25.43\phm{)}&   26.21\phm{)}&   26.02 &  6.0 &\ldots&\ldots&\ldots\\
  183380 & SDFJ132411.0+274331 & 0.9000   &   26.66&   26.55\phm{)}&   26.25\phm{)}&   25.83\phm{)}&   25.42\phm{)}&   25.77\phm{)}&   25.68\phm{)}&   25.12\phm{)}&   25.59 &  7.8 &\ldots&\ldots&\ldots\\
  \enddata
  \label{table1}
  \tablenotetext{a}{The NB catalog number corresponds to the NB filter that the line emission is within. $R_{\rm C}$-band ID's are provided for FOCAS NB711 emitters and fortuitous sources.}
  \tablenotetext{b}{These FOCAS objects were also observed with DEIMOS. The reported redshift is from the DEIMOS observation.}
  \tablenotetext{c}{This ID is for the $i$\arcmin-band catalog.}
  \tablenotetext{d}{Serendipitous sources are secondary sources detected within the DEIMOS long slits and have the appropriate NB redshift.}
  \tablenotetext{e}{Fortuitous sources are lower priority targets for the DEIMOS observations with the appropriate NB redshift.}
  \tablecomments{Properties of NB line-emitting galaxies. Col. (1) provides the NB catalog ID, Col. (2) lists the SDF J2000 ID, redshifts are provided in Col. (3), and photometric
  information is given in Col. (4) - (12). Photometric values in parentheses are between 1 and 2$\sigma$. The 1$\sigma$ magnitude is provided as a lower limit for sources below
  1$\sigma$. \Oii, \Hb, \Oiii, and \Ha~lines fluxes in units of $10^{-18}$ ergs s$^{-1}$ cm$^{-2}$ are provided in Col. (13) - (16).}
  \end{deluxetable}
  \clearpage
\end{landscape}
\clearpage
\begin{deluxetable}{clccccc}
  \tablewidth{0pc}
  \tablecaption{Summary of All 75 Spectra.}
  \tablehead{
    \colhead{Redshift range} & \colhead{Type} & \colhead{FOCAS} & \colhead{DEIMOS} & \colhead{`S'} & \colhead{`F'} & \colhead{Total}\\
    \colhead{(1)} & \colhead{(2)} & \colhead{(3)} & \colhead{(4)} & \colhead{(5)} & \colhead{(6)} & \colhead{(7)}}
  \startdata
  0.080 - 0.091 & \Ha~711          &  0 & 0 & 0 & 1 &  1\\
  0.233 - 0.251 & \Ha~816          &  0 & 1 & 0 & 0 &  1\\
  0.391 - 0.431 & \Ha~921/\Oiii~704&  1 & 1 & 0 & 0 &  2\\
  0.416 - 0.444 & \Oiii~711        &  0 & 0 & 0 & 1 &  1\\
  0.616 - 0.656 & \Oiii~816        &  4 & 6 & 2 & 4 & 16\\
  0.823 - 0.868 & \Oiii~921        & 19 & 3 & 0 & 5 & 27\\
  0.439 - 0.460 & \Hb~704          &  0 & 0 & 0 & 2 &  2\\
  0.458 - 0.473 & \Hb~711          &  0 & 0 & 0 & 2 &  2\\
  0.664 - 0.689 & \Hb~816          &  0 & 0 & 1 & 1 &  2\\
  0.877 - 0.905 & \Hb~921/\Oii~704 &  0 & 0 & 0 & 4 &  4\\
  0.902 - 0.922 & \Oii~711         &  3 & 0 & 0 & 0 &  3\\
  1.171 - 1.203 & \Oii~816         &  4 & 4 & 0 & 0 &  8\\
  1.450 - 1.485 & \Oii~921         &  0 & 4 & 2 & 0 &  6
  \enddata
  \label{table2}
  \tablecomments{Summary of different line emitters with spectroscopic confirmation. Col. (1) lists the redshift range, Col. (2)
    gives the emission line and the NB filter corresponding to the redshift, and Col. (3)-(6) list the
    number of sources that are FOCAS, DEIMOS, serendipitous (`S'), and fortuitous (`F'), respectively.
    The total number of sources for each redshift range is given in Col. (7).}
\end{deluxetable}
\begin{deluxetable}{lcccc}
  \tablewidth{0pc}
  \tablecaption{Redshift Range, Comoving Volume, and Luminosity Distance for Possible Emission
    Lines Detected in Narrow-bands}
  \tablehead{
    & \multicolumn{4}{c}{Redshift range $z_1\leq z\leq z_2$}\\
    \colhead{Line}&\colhead{NB704}&\colhead{NB711}&\colhead{NB816}&\colhead{NB921}\\
    \colhead{(1)} &\colhead{(2)}  &\colhead{(3)}  &\colhead{(4)}  &\colhead{(5)}}
  \startdata
  \Lya                            & 4.753-4.836 & 4.830-4.890 & 5.653-5.752 & 6.508-6.617\\
  \Oii~$\lambda$3727              & 0.877-0.904 & 0.902-0.922 & 1.171-1.203 & 1.450-1.485\\
  H$\beta$                        & 0.439-0.460 & 0.458-0.473 & 0.664-0.689 & 0.878-0.905\\
  \Oiii~$\lambda\lambda$4959, 5007& 0.397-0.417 & 0.416-0.430 & 0.616-0.640 & 0.823-0.850\\
  \Ha                             & 0.066-0.081 & 0.080-0.091 & 0.233-0.251 & 0.391-0.411\\\tableline
  & \multicolumn{4}{c}{Comoving volume in 10$^3$ $h_{70}^{-3}$ Mpc$^3$}\\
  \Lya                            &      198.57 &      142.70 &      214.13 &      214.52\\
  \Oii~$\lambda$3727              &\phm{1}46.59 &\phm{1}35.46 &\phm{1}70.88 &\phm{1}88.48\\
  H$\beta$                        &\phm{1}15.15 &\phm{1}11.45 &\phm{1}31.65 &\phm{1}46.64\\
  \Oiii~$\lambda\lambda$4959, 5007&\phm{1}12.40 &\phm{11}9.26 &\phm{1}27.66 &\phm{1}43.65\\
  \Ha                             &\phm{11}0.43 &\phm{11}0.42 &\phm{11}4.71 &\phm{1}12.11\\\tableline
  & \multicolumn{4}{c}{Luminosity distance in $h_{70}^{-1}$ Mpc}\\
  \Lya                            &      44407 &      45124 &      54406 &      64057\\
  \Oii~$\lambda$3727              &\phm{4}5726 &\phm{4}5897 &\phm{5}8167 &      10618\\
  H$\beta$                        &\phm{4}2494 &\phm{4}2604 &\phm{5}4086 &\phm{6}5736\\
  \Oiii~$\lambda\lambda$4959, 5007&\phm{4}2219 &\phm{4}2322 &\phm{5}3729 &\phm{6}5302\\
  \Ha                             &\phm{44}333 &\phm{45}391 &\phm{5}1213 &\phm{6}2180
\enddata
\label{table3}
\tablecomments{The redshift range, comoving volume, and luminosity distance for the strongest line
  emitters [Col. (1)] in four narrow bandpasses [Col. (2)-(5)] as given by their FWHM.}
\end{deluxetable}
\clearpage
\newcommand{\A}{\tablenotemark{a}}
\begin{landscape}
\begin{deluxetable}{cccccccccccccc}
  \tabletypesize{\scriptsize}
    \tablewidth{0pc}
    \tablecaption{Schechter Fits and Inferred SFR Densities}
    \tablehead{
    & &\multicolumn{4}{c}{Observed fit (completeness-corrected)}&\multicolumn{3}{c}{$\log{\dot\rho_{\rm SFR}}$} & \multicolumn{4}{c}{Extinction-corrected fit} &\colhead{$\log{\dot\rho_{\rm SFR}}$}\\
     \cline{3-6} \cline{10-13}
     \colhead{$z$}&\colhead{N}&\colhead{$\log{\phi_{\star}}$}&\colhead{$\log{L_{\star}}$}&\colhead{$\alpha$}&\colhead{$\log{\mathcal{L}}$}&
     \colhead{A= 0}& \colhead{A = 1.0} & \colhead{A = 1.44}&
    \colhead{$\log{\phi_{\star}}$}&\colhead{$\log{L_{\star}}$}&\colhead{$\alpha$}&\colhead{$\log{\mathcal{L}}$}&\colhead{Eq.~\ref{ext}}\\
    \colhead{(1)} &\colhead{(2)}  &\colhead{(3)}  &\colhead{(4)}  &\colhead{(5)} &\colhead{(6)}&\colhead{(7)}&\colhead{(8)}  &\colhead{(9)}  &\colhead{(10)} &\colhead{(11)}&\colhead{(12)} &\colhead{(13)} &\colhead{(14)}}
  \startdata
  \multicolumn{14}{c}{\Ha~emitters}\\
  0.07, 0.09 & 171, 147 &\ldots&\ldots&\ldots&\ldots&\ldots&\ldots&\ldots&-3.14$\pm$0.09&42.05$\pm$0.07& -1.59$\pm$0.02& 39.23$\pm$0.03&-1.87\\
  0.24 & 259 & -2.98$\pm$0.40& 41.25$\pm$0.34& -1.70$\pm$0.10& 38.74$\pm$0.08 & -2.37  & -1.97 & -1.79 & -3.70$\pm$1.06& 42.20$\pm$1.24& -1.71$\pm$0.08& 38.99$\pm$0.29& -2.11\\
  0.40 & 391 & -2.40$\pm$0.14& 41.29$\pm$0.13& -1.28$\pm$0.07& 39.00$\pm$0.05 & -2.10  & -1.70 & -1.53 & -2.75$\pm$0.16& 41.93$\pm$0.19& -1.34$\pm$0.06& 39.31$\pm$0.08& -1.79\\\tableline
  \multicolumn{14}{c}{\Oiii~emitters}\\
  0.41 & 351 & -2.55$\pm$0.25& 41.17$\pm$0.22& -1.49$\pm$0.11& 38.85$\pm$0.06 & -2.17\A& -1.77 & -1.59 & -3.58$\pm$1.11& 42.46$\pm$1.51& -1.62$\pm$0.08& 39.25$\pm$0.48& -1.87\A\\
  0.42 & 209 & -2.38$\pm$0.22& 41.11$\pm$0.24& -1.25$\pm$0.13& 38.81$\pm$0.09 & -2.31\A& -1.91 & -1.73 & -2.93$\pm$0.35& 41.77$\pm$0.43& -1.41$\pm$0.11& 39.02$\pm$0.17& -2.03\A\\
  0.62 & 293 & -2.58$\pm$0.17& 41.51$\pm$0.15& -1.22$\pm$0.13& 39.00$\pm$0.05 & -2.06\A& -1.66 & -1.48 & -2.51$\pm$0.11& 41.70$\pm$0.10& -1.03$\pm$0.09& 39.20$\pm$0.05& -1.66\A\\
  0.83 & 662 & -2.54$\pm$0.15& 41.53$\pm$0.11& -1.44$\pm$0.09& 39.19$\pm$0.03 & -1.73\A& -1.33 & -1.15 & -2.81$\pm$0.13& 42.16$\pm$0.12& -1.39$\pm$0.06& 39.51$\pm$0.04& -1.30\A\\\tableline
  \multicolumn{14}{c}{\Oii~emitters}\\
  0.89 & 673 & -2.25$\pm$0.13& 41.33$\pm$0.09& -1.27$\pm$0.14& 39.18$\pm$0.03 & -1.68  & -1.28 & -1.10 & -2.68$\pm$0.14& 42.09$\pm$0.11& -1.40$\pm$0.08& 39.59$\pm$0.03& -1.26\\
  0.91 & 818 & -1.97$\pm$0.09& 41.40$\pm$0.07& -1.20$\pm$0.10& 39.50$\pm$0.02 & -1.36  & -0.96 & -0.78 & -2.10$\pm$0.08& 41.95$\pm$0.06& -1.14$\pm$0.07& 39.89$\pm$0.02& -0.97\\
  1.18 & 894 & -2.20$\pm$0.10& 41.74$\pm$0.07& -1.15$\pm$0.11& 39.58$\pm$0.02 & -1.27  & -0.87 & -0.69 & -2.25$\pm$0.07& 42.27$\pm$0.06& -1.03$\pm$0.08& 40.03$\pm$0.02& -0.82\\
  1.47 & 951 & -1.97$\pm$0.06& 41.60$\pm$0.05& -0.78$\pm$0.13& 39.59$\pm$0.02 & -1.27  & -0.87 & -0.69 & -2.20$\pm$0.06& 42.31$\pm$0.05& -0.94$\pm$0.09& 40.10$\pm$0.02& -0.75
  \enddata
  \label{table4}
  \tablenotetext{a}{The \Oiii~$\log{\dot\rho_{\rm SFR}}$ measurements do not use Equation~\ref{eqn8}, but follows the random integer approach described in \S\S~\ref{3.6}.}
  \tablecomments{A summary of the Schechter parameters. Col. (1)-(2) list the redshift and the size of the 2.5$\sigma$ sample. Schechter variables
    $\phi_{\star}$ (Mpc$^{-3}$), $L_{\star}$ (ergs s$^{-1}$), and the faint end slope $\alpha$ are listed in Col. (3)-(5) uncorrected for extinction,
    and Col. (10)-(12) with the extinction correction of Equation~\ref{ext}. The integrated luminosity density $\mathcal{L}$ (ergs s$^{-1}$ Mpc$^{-3}$)
    and the SFR density ($M_{\sun}$ yr$^{-1}$ Mpc$^{-3}$) are given in Col. (6)-(7) and (13)-(14) uncorrected and corrected for extinction. The SFR
    densities assuming $A_{H\alpha}=1.0$ and 1.44 are given in Col. (8)-(9).}
    \end{deluxetable}
  \clearpage
\end{landscape}

\clearpage
\begin{landscape}
  \begin{deluxetable}{lcccccccccccc}
    \tabletypesize{\scriptsize}
    \tablewidth{0pc}
    \tablecaption{Compilation of Emission-line SFR Density Measurements}
    \tablehead{
      & & & & &\multicolumn{3}{c}{Observed}& & \multicolumn{3}{c}{Extinction-corrected} & \\
      \cline{6-8} \cline{10-12}
      \colhead{Reference}&\colhead{Estimator}&\colhead{Redshift}&\colhead{$C_{L_{\star}}$}&\colhead{$C_{\phi_{\star}}$}&\colhead{$\log{L_{\star}}$}&\colhead{$\log{\phi_{\star}}$}&\colhead{$\alpha$}&\colhead{$\dot\rho_{\rm SFR,obs}$\tablenotemark{a}}&\colhead{$\log{L_{\star}}$}&\colhead{$\log{\phi_{\star}}$}&\colhead{$\alpha$}&\colhead{$\dot\rho_{\rm SFR,int}$\tablenotemark{a}}\\
    \colhead{(1)} &\colhead{(2)}  &\colhead{(3)}  &\colhead{(4)}  &\colhead{(5)} &\colhead{(6)} &\colhead{(7)} &\colhead{(8)}&\colhead{(9)}&\colhead{(10)}&\colhead{(11)}&\colhead{(12)}&\colhead{(13)} }
    \startdata
    \citet{hammer97}    & \Oii & 0.375$\pm$0.125	 & 0.6880 & 1.5103 &\ldots        &\ldots        &\ldots        &-2.20$^{+0.07}_{-0.08}$&\ldots        &\ldots        &\ldots        &\ldots	               \\[0.10cm]
                        &      & 0.625$\pm$0.125	 & 0.7801 & 1.2168 &\ldots        &\ldots        &\ldots        &-1.72$^{+0.11}_{-0.15}$&\ldots        &\ldots        &\ldots        &\ldots	               \\[0.10cm]
                        &      & 0.875$\pm$0.125	 & 0.8538 & 1.0445 &\ldots        &\ldots        &\ldots        &-1.35$^{+0.20}_{-0.38}$&\ldots        &\ldots        &\ldots        &\ldots	               \\[0.05cm]
    \citet{hogg98}      & \Oii & 0.750$\pm$0.750	 & 2.5650 & 0.2147 &42.55$\pm$0.11&-3.02$\pm$0.13&-1.34$\pm$0.07&-1.20$\pm$0.04         &\ldots        &\ldots        &\ldots        &\ldots	               \\[0.05cm]
            	        &      & 0.200$\pm$0.100	 & 2.2799 & 0.2689 &\ldots        &\ldots        &\ldots        &-2.37$^{+0.11}_{-0.16}$&\ldots        &\ldots        &\ldots        &\ldots	               \\[0.10cm]
          	        &      & 0.400$\pm$0.100	 & 2.4363 & 0.2384 &\ldots        &\ldots        &\ldots        &-1.77$^{+0.09}_{-0.12}$&\ldots        &\ldots        &\ldots        &\ldots	               \\[0.10cm]
          	        &      & 0.600$\pm$0.100	 & 2.5280 & 0.2212 &\ldots        &\ldots        &\ldots        &-1.69$^{+0.06}_{-0.08}$&\ldots        &\ldots        &\ldots        &\ldots	               \\[0.10cm]
          	        &      & 0.800$\pm$0.100	 & 2.5725 & 0.2126 &\ldots        &\ldots        &\ldots        &-1.75$^{+0.07}_{-0.08}$&\ldots        &\ldots        &\ldots        &\ldots	               \\[0.10cm]
          	        &      & 1.000$\pm$0.100	 & 2.5839 & 0.2093 &\ldots        &\ldots        &\ldots        &-1.44$^{+0.09}_{-0.11}$&\ldots        &\ldots        &\ldots        &\ldots	               \\[0.10cm]
          	        &      & 1.200$\pm$0.100	 & 2.5731 & 0.2094 &\ldots        &\ldots        &\ldots        &-1.57$^{+0.18}_{-0.30}$&\ldots        &\ldots        &\ldots        &\ldots	               \\[0.05cm]
    \citet{gallego02}   & \Oii & 0.025$\pm$0.025	 & 2.0940 & 0.3180 &41.24$\pm$0.13&-3.48$\pm$0.19&-1.21$\pm$0.21&-3.02$\pm$0.15         &42.98$\pm$0.17&-4.21$\pm$0.16&-1.17$\pm$0.08&-2.03$\pm$0.110          \\[0.05cm]
    \citet{hicks02}     & \Oii & 1.200$\pm$0.400	 & 0.9279 & 0.9151 &\ldots        &\ldots        &\ldots        &-1.59$^{+0.30}_{-0.48}$&\ldots        &\ldots        &\ldots        &\ldots	               \\[0.05cm]
    \citet{sullivan00}  & \Oii & 0.150$\pm$0.150	 & 2.3483 & 0.2351 &\ldots        &\ldots        &\ldots        &\ldots                 &42.33$\pm$0.09&-3.45$\pm$0.18&-1.59$\pm$0.12&-1.64$\pm$0.06           \\[0.05cm]
    \citet{teplitz03}   & \Oii & 0.900$\pm$0.500	 & 1.0000 & 1.0000 &42.15$\pm$0.08&-3.06$\pm$0.12&-1.35         &-1.55$\pm$0.06         &\ldots        &\ldots        &\ldots        &\ldots	               \\[0.05cm]
    \citet{hippelein03} & \Oii & 0.881$\pm$0.014	 & 1.0000 & 1.0000 &\ldots        &\ldots        &\ldots        &\ldots                 &42.00         &-2.36         &-1.45         &-1.00$^{+0.12}_{-0.17}$  \\[0.10cm]
                        &      & 1.193$\pm$0.018	 & 1.0000 & 1.0000 &\ldots        &\ldots        &\ldots        &\ldots                 &42.32         &-2.36         &-1.45         &-0.68$^{+0.09}_{-0.12}$  \\[0.10cm]
    \citet{gallego95}   & \Ha  & 0.022$\pm$0.022	 & 0.5219 & 2.5663 &\ldots        &\ldots        &\ldots        &\ldots                 &41.87$\pm$0.08&-2.79$\pm$0.20&-1.30$\pm$0.20&-1.91$^{+0.04}_{-0.04}$  \\[0.10cm]
    \citet{TM98}	& \Ha  & 0.200$\pm$0.100	 & 0.6110 & 1.8585 &\ldots        &\ldots        &\ldots        &\ldots                 &41.92$\pm$0.13&-2.56$\pm$0.09&-1.35$\pm$0.06&-1.61$^{+0.03}_{-0.03}$  \\[0.05cm]
    \citet{glazebrook99}& \Ha  & 0.885$\pm$0.099	 & 0.8564 & 1.0394 &\ldots        &\ldots        &\ldots        &-0.97$\pm$0.10         &\ldots        &\ldots        &\ldots        &\ldots	               \\[0.05cm]
    \citet{yan99}	& \Ha  & 1.300$\pm$0.600	 & 0.9468 & 0.8905 &42.82         &-2.82         &-1.35         &-0.96$^{+0.09}_{-0.11}$&\ldots        &\ldots        &\ldots        &\ldots	               \\[0.10cm]
    \citet{hopkins00}   & \Ha  & 1.250$\pm$0.550	 & 2.1095 & 0.2675 &42.87$\pm$0.11&-3.11$\pm$0.20&-1.60$\pm$0.12&-1.00		        &\ldots        &\ldots        &\ldots        &\ldots	               \\
    \citet{moorwood00}  & \Ha  & 2.200$\pm$0.050	 & 1.0645 & 0.7456 &\ldots        &\ldots        &\ldots        &\ldots                 &\ldots        &\ldots        &\ldots        &\ldots	               \\
    \citet{sullivan00}  & \Ha  & 0.150$\pm$0.150	 & 2.3483 & 0.2351 &\ldots        &\ldots        &\ldots        &\ldots                 &42.42$\pm$0.14&-3.55$\pm$0.20&-1.62$\pm$0.10&-1.86$\pm$0.06           \\[0.10cm]
    \citet{pascual05}   & \Ha  & 0.242$\pm$0.014	 & 0.6305 & 1.8103 &\ldots        &\ldots        &\ldots        &-1.77$^{+0.08}_{-0.09}$&\ldots        &\ldots        &\ldots        &\ldots	               \\[0.10cm]
    \citet{tresse02}    & \Ha  & 0.730$^{+0.37}_{-0.23}$ & 0.8131 & 1.0811 &41.98$\pm$0.06&-2.36$\pm$0.06&-1.31$\pm$0.11&-1.37$\pm$0.050        &42.28$\pm$0.06&-2.36$\pm$0.06&-1.31$\pm$0.11&-1.06$\pm$0.05           \\[0.10cm]
    \citet{fujita03}    & \Ha  & 0.242$\pm$0.009	 & 1.0000 & 1.0000 &41.55$\pm$0.25&-2.62$\pm$0.34&-1.53$\pm$0.15&-1.90$^{+0.08}_{-0.17}$&41.95$\pm$0.25&-2.62$\pm$0.34&-1.53$\pm$0.15&-1.50$^{+0.08}_{-0.17}$  \\[0.10cm]
    \citet{hippelein03} & \Ha  & 0.245$\pm$0.007	 & 1.0000 & 1.0000 &\ldots        &\ldots        &\ldots        &\ldots                 &41.45         &-2.32         &-1.35         &-1.83$^{+0.10}_{-0.13}$  \\[0.10cm]
    \citet{perez03}     & \Ha  & 0.025$\pm$0.025	 & 1.0000 & 1.0000 &\ldots        &\ldots        &\ldots        &\ldots		        &42.43$\pm$0.17&-3.00$\pm$0.20&-1.20$\pm$0.20&-1.61$^{+0.11}_{-0.08}$  \\[0.10cm]
    \citet{glazebrook04}& \Ha  & 0.384$\pm$0.006	 & 1.0000 & 1.0000 &41.49         &-2.55         &-1.30         &-2.05$^{+0.14}_{-0.21}$&\ldots        &\ldots        &\ldots        &\ldots	               \\[0.10cm]
                        &      & 0.458$\pm$0.006	 & 1.0000 & 1.0000 &42.15         &-2.23         &-1.30         &-1.07$^{+0.17}_{-0.14}$&\ldots        &\ldots        &\ldots        &\ldots	               \\[0.10cm]
    \citet{hippelein03} & \Oiii& 0.401$\pm$0.011	 & 1.0000 & 1.0000 &\ldots        &\ldots        &\ldots        &\ldots                 &41.95         &-2.32         &-1.35         &-1.33$^{+0.13}_{-0.18}$  \\[0.10cm]
                        &      & 0.636$\pm$0.010	 & 1.0000 & 1.0000 &\ldots        &\ldots        &\ldots        &\ldots                 &41.95         &-2.70         &-1.50         &-1.61$^{+0.09}_{-0.11}$  \\[0.05cm]
    \citet{lilly96}     &2800\AA&0.350$\pm$0.150         & 0.6777 & 1.5367 &\ldots        &\ldots        &\ldots        &-1.93$\pm$0.07         &\ldots        &\ldots        &\ldots        &\ldots                   \\[0.05cm]
                        &       &0.625$\pm$0.125	 & 0.7801 & 1.2168 &\ldots        &\ldots        &\ldots        &-1.64$\pm$0.08	        &\ldots        &\ldots        &\ldots        &\ldots	               \\[0.05cm]
                        &       &0.875$\pm$0.125	 & 0.8538 & 1.0445 &\ldots        &\ldots        &\ldots        &-1.36$\pm$0.15  	&\ldots        &\ldots        &\ldots        &\ldots	               \\[0.10cm]
    \citet{connolly97}  &2800\AA&0.750$\pm$0.250	 & 0.8190 & 1.1141 &\ldots        &\ldots        &\ldots        &-1.35$^{+0.13}_{-0.09}$&\ldots        &\ldots        &\ldots        &\ldots	               \\[0.10cm]
                        &       &1.250$\pm$0.250	 & 0.9376 & 0.8991 &\ldots        &\ldots        &\ldots        &-1.21$^{+0.16}_{-0.14}$&\ldots        &\ldots        &\ldots        &\ldots	               \\[0.10cm]
                        &       &1.750$\pm$0.250	 & 1.0154 & 0.7977 &\ldots        &\ldots        &\ldots        &-1.33$^{+0.03}_{-0.20}$&\ldots        &\ldots        &\ldots        &\ldots	               \\[0.10cm]
    \citet{treyer98}    &2000\AA&0.150$\pm$0.150	 & 2.3483 & 0.2351 &\ldots        &\ldots        &\ldots        &-1.93$^{+0.06}_{-0.15}$&\ldots        &\ldots        &\ldots        &-1.67$^{+0.06}_{-0.15}$  \\[0.10cm]
    \citet{cowie99}	&2000\AA&0.350$\pm$0.150	 & 1.1453 & 0.6995 &\ldots        &\ldots        &\ldots        &\ldots		        &\ldots        &\ldots        &\ldots        &-1.90$^{+0.004}_{-0.003}$\\[0.10cm]
                        &	&0.625$\pm$0.125	 & 1.3184 & 0.5539 &\ldots        &\ldots        &\ldots        &\ldots 		&\ldots        &\ldots        &\ldots        &-1.72$^{+0.01}_{-0.004}$ \\[0.10cm]
                        &	&0.875$\pm$0.125	 & 1.4429 & 0.4754 &\ldots        &\ldots        &\ldots        &\ldots		        &\ldots        &\ldots        &\ldots        &-1.06$^{+0.01}_{-0.003}$ \\[0.10cm]
                        &	&1.250$\pm$0.250	 & 1.5845 & 0.4092 &\ldots        &\ldots        &\ldots        &\ldots		        &\ldots        &\ldots        &\ldots        &-0.74$^{+0.01}_{-0.006}$ \\[0.05cm]
    \citet{sullivan00}  &2000\AA&0.150$\pm$0.150	 & 2.3483 & 0.2351 &\ldots        &\ldots        &\ldots        &-1.91$\pm$0.05	        &\ldots        &\ldots        &\ldots        &-1.38$\pm$0.05           \\[0.05cm]
    \citet{mass01}      &1500\AA&1.500$\pm$0.500	 & 0.9803 & 0.8440 &\ldots        &\ldots        &\ldots        &-1.59$^{+0.02}_{-0.02}$&\ldots        &\ldots        &\ldots        &-0.68$^{+0.03}_{-0.03}$  \\[0.10cm]
                        &       &2.750$\pm$0.750         & 1.1080 & 0.7118 &\ldots        &\ldots        &\ldots        &-1.57$^{+0.02}_{-0.04}$&\ldots        &\ldots        &\ldots        &-0.49$^{+0.04}_{-0.05}$  \\[0.05cm]
    \citet{wilson02}	&2500\AA&0.350$\pm$0.150         & 2.0408 & 0.3430 &\ldots        &\ldots        &\ldots        &-1.83$\pm$0.08         &\ldots        &\ldots        &\ldots        &-1.43$\pm$0.08           \\[0.05cm]
                    	&       &0.800$\pm$0.200         & 2.0408 & 0.3430 &\ldots        &\ldots        &\ldots        &-1.54$\pm$0.05         &\ldots        &\ldots        &\ldots        &-0.67$\pm$0.05           \\[0.05cm]
                    	&       &1.350$\pm$0.250         & 2.0408 & 0.3430 &\ldots        &\ldots        &\ldots        &-1.42$\pm$0.14         &\ldots        &\ldots        &\ldots        &-0.555$\pm$0.14          \\
    \enddata
    \label{table5}
    \tablenotetext{a}{The luminosity density is obtained from the available binned data instead of integrating over all luminosities if no LF parameters are provided.}
    \tablecomments{Compilation of Schechter fits and SFR densities from emission-line techniques and UV continuum measurements. References are listed in Col. (1) with the SFR estimator
      reported in Col. (2), and the redshift range in Col. (3). Col. (4) and (5) provide factors to convert $L_{\star}$ (ergs s$^{-1}$) and $\phi_{\star}$ (Mpc$^{-3}$) to the common
  cosmology. Schechter parameters ($\log{L_{\star}}$, $\log{\phi_{\star}}$, and $\alpha$) when available are given in Col. (6)-(8) for observed measurements, and Col. (10)-(12) for
  extinction-corrected measurements. Observed and extinction-corrected $\dot\rho_{\rm SFR}$ (in $M_{\sun}$ yr$^{-1}$ Mpc$^{-3}$) are given in Col. (9) and (13), respectively. 
}
  \end{deluxetable}
  \clearpage
\end{landscape}
\clearpage

\clearpage

\begin{thebibliography}{}
\bibitem[Arag{\'o}n-Salamanca et al.(2003)]{aragon03}
  Arag{\'o}n-Salamanca, A., Alonso-Herrero, A., Gallego, J., Garc{\'{\i}}a-Dab{\'o}, C.~E., P{\'e}rez-Gonz{\'a}lez, P.~G.,
  Zamorano, J., \& Gil de Paz, A.\ 2003, ASP Conf.~Ser.~297: Star Formation Through Time, 297, 191 

\bibitem[Ajiki et al.(2003)]{ajiki03}
  Ajiki, M., et al.\ 2003, \aj, 126, 2091 
 
\bibitem[Ajiki et al.(2006)]{ajiki06}
  Ajiki, M., et al.\ 2006, \pasj, 58, 113

\bibitem[Arnouts et al.(2005)]{arnouts05}
  Arnouts, S., et al.\ 2005, \apjl, 619, L43

\bibitem[Bertin \& Arnouts(1996)]{bertin96}
  Bertin, E., \& Arnouts, S.\ 1996, \aaps, 117, 393 
 
\bibitem[Brinchmann et al.(2004)]{brinchmann04}
  Brinchmann, J., Charlot, S., White, S.~D.~M., Tremonti, C., Kauffmann, G., Heckman, T., \& 
  Brinkmann, J.\ 2004, \mnras, 351, 1151

\bibitem[Bruzual \& Charlot(2003)]{bc03}
  Bruzual, G., \& Charlot, S.\ 2003, \mnras, 344, 1000 

\bibitem[Capak et al.(2004)]{capak04}
  Capak, P., et al.\ 2004, \aj, 127, 180 

\bibitem[Cardelli et al.(1989)]{cardelli89}
  Cardelli, J.~A., Clayton, G.~C., \& Mathis, J.~S.\ 1989, \apj, 345, 245 
 
\bibitem[Connolly et al.(1997)]{connolly97}
  Connolly, A.~J., Szalay, A.~S., Dickinson, M., Subbarao, M.~U., \& Brunner, R.~J.\ 1997, 
  \apjl, 486, L11 
 
\bibitem[Cowie et al.(1999)]{cowie99}
  Cowie, L.~L., Songaila, A., \& Barger, A.~J.\ 1999, \aj, 118, 603 

\bibitem[Cowie et al.(2004)]{cowie04}
  Cowie, L.~L., Barger, A.~J., Hu, E.~M., Capak, P., \& Songaila, A.\ 2004, \aj, 127, 3137 
 
\bibitem[Doherty et al.(2006)]{doherty06}
  Doherty, M., Bunker, A., Sharp, R., Dalton, G., Parry, I., \& Lewis, I.\ 2006, \mnras, 370, 331

\bibitem[Drozdovsky et al.(2005)]{drozdovsky05}
  Drozdovsky, I., Yan, L., Chen, H.-W., Stern, D., Kennicutt, R.~J., Spinrad, H., \& Dawson, S.\
  2005, \aj, 130, 1324 
 
\bibitem[Faber et al.(2003)]{faber03}
  Faber, S.~M., et al.\ 2003, \procspie, 4841, 1657 
 
\bibitem[Fujita et al.(2003)]{fujita03}
  Fujita, S.~S., et al.\ 2003, \apjl, 586, L115 
 
\bibitem[Gabasch et al.(2004)]{gabasch04}
  Gabasch, A., et al.\ 2004, \aap, 421, 41

\bibitem[Gabasch et al.(2006)]{gabasch06}
  Gabasch, A., et al.\ 2006, \aap, 448, 101

\bibitem[Gallego et al.(1995)]{gallego95}
  Gallego, J., Zamorano, J., Aragon-Salamanca, A., \& Rego, M.\ 1995, \apjl, 455, L1 
 
\bibitem[Gallego et al.(1997)]{gallego97}
  Gallego, J., Zamorano, J., Rego, M., \& Vitores, A.~G.\ 1997, \apj, 475, 502 
 
\bibitem[Gallego et al.(2002)]{gallego02}
  Gallego, J., Garc{\'{\i}}a-Dab{\'o}, C.~E., Zamorano, J., Arag{\'o}n-Salamanca, A., \& Rego, M.\ 2002, \apjl, 570, L1 
 
\bibitem[Glazebrook et al.(1999)]{glazebrook99}
  Glazebrook, K., Blake, C., Economou, F., Lilly, S., \& Colless, M.\ 1999, \mnras, 306, 843 
 
\bibitem[Glazebrook et al.(2004)]{glazebrook04}
  Glazebrook, K., Tober, J., Thomson, S., Bland-Hawthorn, J., \& Abraham, R.\ 2004, \aj, 128, 2652 
 
\bibitem[Hammer et al.(1997)]{hammer97}
  Hammer, F., et al.\ 1997, \apj, 481, 49 
 
\bibitem[Hicks et al.(2002)]{hicks02}
  Hicks, E.~K.~S., Malkan, M.~A., Teplitz, H.~I., McCarthy, P.~J., \& Yan, L.\ 2002, \apj, 581, 205 

\bibitem[Hippelein et al.(2003)]{hippelein03}
  Hippelein, H., et al.\ 2003, \aap, 402, 65 
 
\bibitem[Hogg et al.(1998)]{hogg98}
  Hogg, D.~W., Cohen, J.~G., Blandford, R., \& Pahre, M.~A.\ 1998, \apj, 504, 622 
 
\bibitem[Hopkins et al.(2000)]{hopkins00}
  Hopkins, A.~M., Connolly, A.~J., \& Szalay, A.~S.\ 2000, \aj, 120, 2843 

\bibitem[Hopkins et al.(2001)]{hopkins01}
  Hopkins, A.~M., Connolly, A.~J., Haarsma, D.~B., \& Cram, L.~E.\ 2001, \aj, 122, 288 
 
\bibitem[Hopkins(2004)]{hopkins04}
  Hopkins, A.~M.\ 2004, \apj, 615, 209 
 
\bibitem[Hu et al.(2002)]{hu02}
  Hu, E.~M., Cowie, L.~L., McMahon, R.~G., Capak, P., Iwamuro, F., Kneib, J.-P., Maihara, T., \& 
  Motohara, K.\ 2002, \apjl, 568, L75 
 
\bibitem[Hu et al.(2004)]{hu04}
  Hu, E.~M., Cowie, L.~L., Capak, P., McMahon, R.~G., Hayashino, T., \& Komiyama, Y.\ 2004, \aj, 127, 563

\bibitem[Iwamuro et al.(2000)]{iwamuro00}
  Iwamuro, F., et al.\ 2000, \pasj, 52, 73 
 
\bibitem[Iye et al.(2004)]{iye04}
  Iye, M., et al.\ 2004, \pasj, 56, 381 

\bibitem[Jansen et al.(2000)]{jansen00}
  Jansen, R.~A., Fabricant, D., Franx, M., \& Caldwell, N.\ 2000, \apjs, 126, 331 
 
\bibitem[Jansen et al.(2001)]{jansen01}
  Jansen, R.~A., Franx, M., \& Fabricant, D.\ 2001, \apj, 551, 825 
 
\bibitem[Jones \& Bland-Hawthorn(2001)]{jones01}
  Jones, D.~H., \& Bland-Hawthorn, J.\ 2001, \apj, 550, 593 
 
\bibitem[Kaifu(1998)]{kaifu98}
  Kaifu, N.\ 1998, \procspie, 3352, 14 
 
\bibitem[Kashikawa et al.(2002)]{kashik02}
  Kashikawa, N., et al.\ 2002, \pasj, 54, 819 
 
\bibitem[Kashikawa et al.(2004)]{kashik04}
  Kashikawa, N., et  al.\ 2004, \pasj, 56, 1011

\bibitem[Kashikawa et al.(2006)]{kashik06}
  Kashikawa, N., et  al.\ 2006, \apj, 648, 7

\bibitem[Kennicutt(1983)]{kennicutt83}
  Kennicutt, R.~C.\ 1983, \apj, 272, 54 

\bibitem[Kennicutt(1992)]{kennicutt92}
  Kennicutt, R.~C.\ 1992, \apj, 388, 310 

\bibitem[Kennicutt(1998)]{kennicutt98}
  Kennicutt, R.~C.\ 1998, \araa, 36, 189 
 
\bibitem[Kewley, Gellar, \& Jansen(2004)]{kewley04}
  Kewley, L.~J., Geller, M.~J., \& Jansen, R.~A.\ 2004, \aj, 127, 2002 

\bibitem[Kobulnicky et al.(1999)]{KKP99}
  Kobulnicky, H.~A., Kennicutt, R.~C., \& Pizagno, J.~L.\ 1999, \apj, 514, 544 

\bibitem[Kodaira et al.(2003)]{kodaira03}
  Kodaira, K., et al.\ 2003, \pasj, 55, L17 

\bibitem[Kodama et al.(2004)]{kodama04}
  Kodama, T., Balogh, M.~L., Smail, I., Bower, R.~G., \& Nakata, F.\ 2004, \mnras, 354, 1103 
 
\bibitem[Lilly et al.(1996)]{lilly96}
  Lilly, S.~J., Le Fevre, O., Hammer, F., \& Crampton, D.\ 1996, \apjl, 460, L1 
 
\bibitem[Malkan et al.(1995)]{MTM95}
  Malkan, M., Teplitz, H., \& McLean, I.\ 1995, \apjl, 448, L5 

\bibitem[Massarotti et al.(2001)]{mass01}
  Massarotti, M., Iovino, A., \& Buzzoni, A.\ 2001, \apjl, 559, L105 
 
\bibitem[Meisenheimer \& Wolf(2002)]{combo02}
  Meisenheimer, K., \& Wolf, C.\ 2002, Astronomy and Geophysics, 43, 15

\bibitem[McCarthy et al.(1999)]{M99}
  McCarthy, P.~J., et al.\ 1999, \apj, 520, 548 
 
\bibitem[Moorwood et al.(2000)]{moorwood00}
  Moorwood, A.~F.~M., van der Werf, P.~P., Cuby, J.~G., \& Oliva, E.\ 2000, \aap, 362, 9 

\bibitem[Moustakas \& Kennicutt(2006)]{moustakas06}
  Moustakas, J., \& Kennicutt, R.~C., Jr.\ 2006, \apjs, 164, 81
 
\bibitem[Oke(1990)]{oke90}
  Oke, J.~B.\ 1990, \aj, 99, 1621 
 
\bibitem[Oke et al.(1995)]{oke95}
  Oke, J.~B., et al.\ 1995, \pasp, 107, 375 
 
\bibitem[Osterbrock(1989)]{osterbrock89}
  Osterbrock, D.~E.\ 1989, Astrophysics of Gaseous Nebulae and Active Galactic Nuclei (Mill Valley, CA: University
  Science Books)
 
\bibitem[Ouchi et al.(2003)]{ouchi03}
  Ouchi, M., et al.\ 2003, \apj, 582, 60 

\bibitem[Pascual et al.(2005)]{pascual05}
  Pascual, S., Villar, V., Gallego, J., Zamorano, J., Pell{\'o}, R., D{\'{\i}}az, C., \& Arag{\'o}n-Salamanca, A.\ 2005,
  Revista Mexicana de Astronomia y Astrofisica Conference Series, 24, 268

\bibitem[P{\'e}rez-Gonz{\'a}lez et al.(2003)]{perez03}
  P{\'e}rez-Gonz{\'a}lez, P.~G., Zamorano, J., Gallego, J., Arag{\'o}n-Salamanca, A., \& Gil de Paz, A.\ 2003,
  \apj, 591, 827 
 
\bibitem[Rodriguez-Eugenio et al.(2006)]{rodriguez06}
  Rodriguez-Eugenio, N., Noeske, K.~G., Acosta-Pulido, J., Barrena, R., 
  Prada, F., Manchado, A., \& EGS Teams 2006, in press (astro-ph/0604027)

\bibitem[Schechter(1976)]{schechter76}
  Schechter, P.\ 1976, \apj, 203, 297 
 
\bibitem[Schlegel, Finkbeiner, \& Davis(1998)]{schlegel98}
  Schlegel, D.~J., Finkbeiner, D.~P., \& Davis, M.\ 1998, \apj, 500, 525 

\bibitem[Shimasaku et al.(2003)]{shima03}
  Shimasaku, K., et al.\ 2003, \apjl, 586, L111 
 
\bibitem[Shimasaku et al.(2004)]{shima04}
  Shimasaku, K., et al.\ 2004, \apjl, 605, L93 
 
\bibitem[Shimasaku et al.(2006)]{shima06}
  Shimasaku, K., et al.\ 2006, \pasj, 58, 313

\bibitem[Somerville et al.(2004)]{somerville04}
  Somerville, R.~S., Lee, K., Ferguson, H.~C., Gardner, J.~P., Moustakas, L.~A., \& Giavalisco, 
  M.\ 2004, \apjl, 600, L171 

\bibitem[Spergel et al.(2003)]{spergel03}
  Spergel, D.~N., et al.\ 2003, \apjs, 148, 175 
 
\bibitem[Spergel et al.(2006)]{spergel06}
  Spergel, D.~N., et al.\ 2006, submitted (astro-ph/0603449)
 
\bibitem[Sullivan et al.(2000)]{sullivan00}
  Sullivan, M., Treyer, M.~A., Ellis, R.~S., Bridges, T.~J., Milliard, B., \& Donas, J.\ 2000, 
\mnras, 312, 442 
 
\bibitem[Taniguchi et al.(2003)]{taniguchi03}
  Taniguchi, Y., et al.\ 2003, \apjl, 585, L97 
 
\bibitem[Taniguchi et al.(2005)]{taniguchi05}
  Taniguchi, Y., et al.\ 2005, \pasj, 57, 165 

\bibitem[Teplitz et al.(2000)]{teplitz00}
  Teplitz, H.~I., et al.\ 2000, \apj, 542, 18 
 
\bibitem[Teplitz et al.(2003)]{teplitz03}
  Teplitz, H.~I., Collins, N.~R., Gardner, J.~P., Hill, R.~S., \& Rhodes, J.\ 2003, \apj, 589, 704
 
\bibitem[Tresse \& Maddox(1998)]{TM98}
  Tresse, L., \& Maddox, S.~J.\ 1998, \apj, 495, 691 
 
\bibitem[Tresse et al.(2002)]{tresse02}
  Tresse, L., Maddox, S.~J., Le F{\`e}vre, O., \& Cuby, J.-G.\ 2002, \mnras, 337, 369 
 
\bibitem[Treyer et al.(1998)]{treyer98}
  Treyer, M.~A., Ellis, R.~S., Milliard, B., Donas, J., \& Bridges, T.~J.\ 1998, \mnras, 300, 303 
 
\bibitem[Treyer et al.(2005)]{treyer05}
  Treyer, M., et al.\ 2005, \apjl, 619, L19

\bibitem[Umeda et al.(2004)]{umeda04}
  Umeda, K., et al.\ 2004, \apj, 601, 805 
 
\bibitem[Wilson et al.(2002)]{wilson02}
  Wilson, G., Cowie, L.~L., Barger, A.~J., \& Burke, D.~J.\ 2002, \aj, 124, 1258 
 
\bibitem[Wyder et al.(2005)]{wyder05}
  Wyder, T.~K., et al.\ 2005, \apjl, 619, L15

\bibitem[Yan et al.(1999)]{yan99}
  Yan, L., McCarthy, P.~J., Freudling, W., Teplitz, H.~I., Malumuth, E.~M., Weymann, R.~J., \&
  Malkan, M.~A.\ 1999, \apjl, 519, L47 

\bibitem[Yip et al.(2004)]{yip04}
  Yip, C.~W., et al.\ 2004, \aj, 128, 585 

\end{thebibliography}
\end{document}